\documentclass[aps,pre,twocolumn,superscriptaddress,showpacs,amsmath,amssymb,footinbib,longbibliography]{revtex4-2}
\usepackage[english]{babel}
\usepackage{amssymb,amsbsy,amsmath}
\usepackage{mathtools}
\usepackage{graphicx}% Include figure files
\usepackage{dcolumn}% Align table columns on decimal point
\usepackage{bm}% bold math
\usepackage{subfigure}
\usepackage{here}
\usepackage{color}
\usepackage{textcomp}
\usepackage{braket}%Bra-Ket-Schreibweise
\usepackage[colorlinks,bookmarks=false,citecolor=blue,linkcolor=red,urlcolor=blue]{hyperref}
\newcommand{\op}[1]{%
    \fontdimen12\textfont3=2pt\fontdimen12\scriptfont3=1.4pt%
    \!\null\mathop{\vphantom{#1}\smash{#1}}\limits_{\sim}\null\!}
\newcommand{\xref}[1]{\protect\ref{#1}}
\newcommand{\figref}[1]{Fig.~\protect\ref{#1}}
\newcommand{\fmref}[1]{(\protect\ref{#1})}
\def\bra#1{\langle \, {#1} \, | \,}
\def\ket#1{\, | \, {#1} \, \rangle}
\renewcommand{\eqref}[1]{Eq.~(\protect\ref{#1})}

\newcommand{\mat}[1]{\mathbf{#1}}

\newcommand{\vecops}[1]{\op{\vec{s}}_{#1}}

\newcommand{\vecoptau}[1]{\op{\vec{\tau}}_{#1}}
\begin{document}
\title{Studies of decoherence in strongly anisotropic spin triangles with toroidal or general non-collinear easy axes}

\author{Kilian Irl\"ander}
\email{ORCID: 0000-0002-0223-6506}
\author{J\"urgen Schnack}
\email{ORCID: 0000-0003-0702-2723, jschnack@uni-bielefeld.de}
\affiliation{Fakult\"at f\"ur Physik, Universit\"at Bielefeld, Postfach 100131, D-33501 Bielefeld, Germany}

\date{\today}

\begin{abstract}
Magnetic molecules are investigated with respect to their usability
as units in future quantum devices. In view of quantum computing, 
a necessary prerequisite is a long coherence time of superpositions 
of low-lying levels. In this article,
we investigate by means of numerical simulations whether a toroidal structure 
of single-ion easy anisotropy axes is advantageous as often conjectured.
Our results demonstrate that there is no general advantage of 
toroidal magnetic molecules, but that arrangements of tilted anisotropy 
axes perform best in many cases.
\end{abstract}

%\pacs{75.10.Jm,75.50.Xx,75.40.Mg} 

\maketitle

%%%%%%%%%%%%%%%%%%%%%%%%%%%%%%%%%%%%%%%%%%%%%%%%%%%%%%%%%%%%%%%%%%%%%%%%
\section{Introduction}

Molecular spins are being investigated as one prospective platform for quantum computation \cite{MST:PRL06,ARM:PRL07,Wer:NM07,TTT:N11,KWW:PRB14,SKD:N16,FPC:NC16,GFB:PRL17,GLH:NC19,AtS:JACS19,CSC:PRR20,CZC:APL21,PCW:npjQI21,LMU:NP21,CPC:JPCL22}. In order to reduce decoherence effects, clock transitions have been established as
promising processes with long lasting coherence \cite{WTG:NN13,KWM:NC14,SKD:N16,McI:NC22,KCH:arXiv21}.
Clock transitions are spin transitions made up of two eigenstates having close to the same expectation value 
of the magnetic moment and thus the same slope of their Zeeman energies as function of applied field.
The energy difference which provides the frequency of the transition is thus rather stable against 
field fluctuations.
One preferred constellation is to have zero slope of the Zeeman levels at all, either at the extreme
points of Zeeman curves of parabolic shape \cite{SKD:N16} 
or for Zeeman curves that belong to zero moments and 
are thus totally flat \cite{VoS:PRB20,BYW:SA18}.

Somewhat along these lines, toroidal magnetic states of molecular nanomagnets \cite{THM:ACIE06,ARS:N07,SoC:PRB08,SFM:JPCM08,SoC:PRB10,GLZ:IC12,ULH:JACS12,WSL:CS12,XCZ:IC12,DVG:CAEJ15,GGM:ACIE18,VLS:ACIE18,CrA:PRB18,LVG:DT19,RAS:PB20,Pav:PRB20,ZZQ:M20,ABV:EJIC21,HyS:MC22,YUC:CC22} are considered to be promising candidates 
of quantum computation since they do not possess magnetic moments for ideally symmetric cases \cite{CUS:ACIE08}.
Several recent publications echo this hypothesis, e.g., \cite{VSL:NC17,CrA:PRB18,ULT:CSR14}, however, to our knowledge 
no decoherence calculations or systematical experiments have actually been performed for such systems. 
The argument that toroidal quantum states are promising for quantum computing rests on the
-- maybe reasonable -- assumption that they as well are insensitive to weak fluctuating homogeneous magnetic fields \cite{TSL:PRB12,VSL:NC17}. 
%NEW
It was, however, shown in e.g.\ \cite{VoS:PRB20,VSS:NJP21} that a true many-body
treatment of the interacting system and bath spins goes beyond the mean-field picture of a fluctuating field 
and that the reason for decoherence is entanglement between system and bath.

In the following we are going to investigate a spin triangle as a typical
representative of toroidal magnetic molecules \cite{CUS:ACIE08}.
We study its decoherence while immersed in a bath of nearby spins with 
mutual dipolar interactions 
%NEW
between system and bath spins and among bath spins.
The time evolution of the combined systems is described by the time-dependent 
Schr{\"o}dinger equation and thus unitary as in \cite{VoS:PRB20}. We thus refrain from mean field assumptions
or assumptions about transition matrix elements necessary for a description 
in terms of Lindblad master equations 
%NEW
or perturbation/scattering theory \cite{Lin:CMP76,Tab:AP17}.
The Hamiltonian is designed to model the impact of either electronic or
nuclear bath spins since nuclear spins are often the source of decoherence in
molecular insulators \cite{PrS:RPP00,CHS:JPCL20,CJS:JPCL20,KCH:arXiv21} whereas electronic spins are the source 
of decoherence for molecules 
deposited on metallic surfaces, see e.g.\ \cite{VoS:PRB20,BYW:SA18}.
No approximations as in e.g.\ \cite{CHS:JPCL20,CJS:JPCL20,CCZ:PCCP22}
are made concerning 
the dipolar interaction; it is fully anisotropic.

\begin{figure}[ht!]
\centering
\includegraphics[clip, width=0.67\columnwidth]{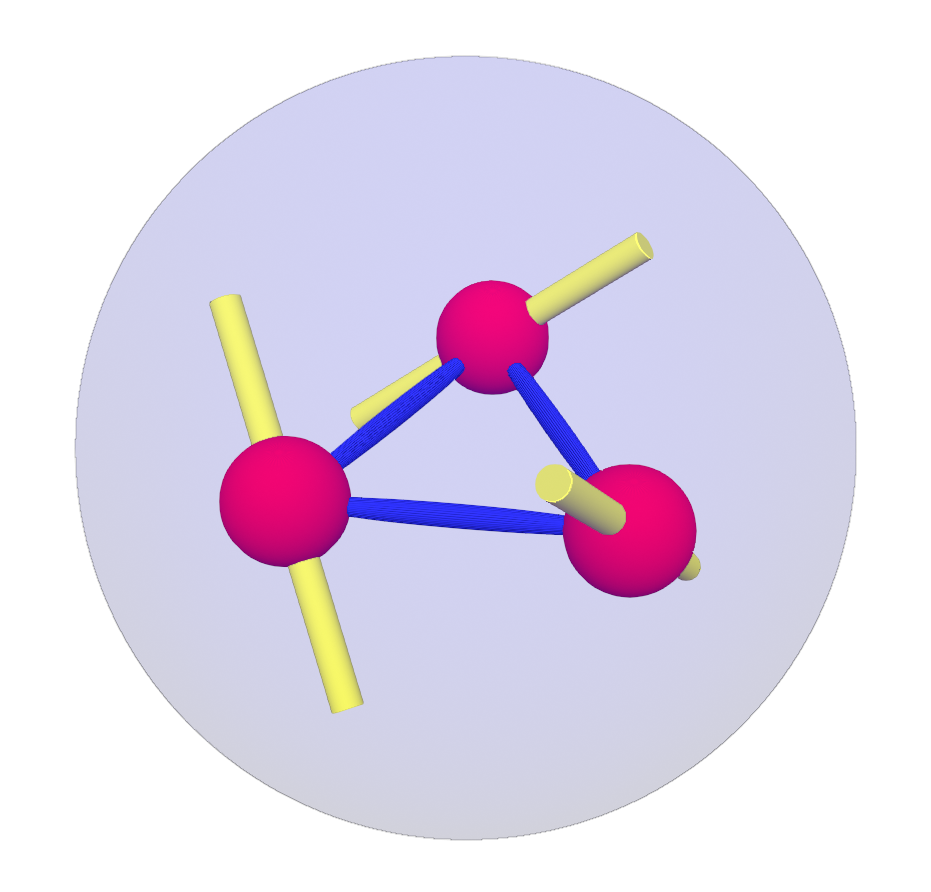}
\caption{System of three spin-1 particles in red, Heisenberg coupling in blue and anisotropy axes tilted by $\theta_1=\theta_2=\theta_3=\theta$ (here: $\theta=50^{\circ}$) in yellow with a varying number of bath spins $n_{\text{bath}}$ (chosen between 4 and 10 in this paper) located on the surface of the surrounding purple sphere. Bath spins are coupled to each other and to the system spins via dipole-dipole interactions.}
\label{system}
\end{figure}

We decided to choose electronic bath spins since the sole purpose of our study, 
which hopefully initiates further investigations,
is to demonstrate that there is no simple correlation between toroidicity and
extended coherence times. To this end, we provide examples and systematic studies
for numerous arrangements of the surrounding bath of decohering spins. 
In contrast to widespread expectations it turns out that a tilted toroidal arrangement
of easy anisotropy axes as displayed in \figref{system} yields 
the longest coherence times in most of our simulations except for some cases
with rather weak anisotropy discussed in Sec.~\xref{sec5} as well as in the appendix.

The paper is organized as follows. In Sec. \xref{sec2} we introduce the 
employed spin Hamiltonian as well as the necessary technicalities. 
Section~\xref{sec3} presents numerical examples whereas Sec.~\xref{sec4}, \xref{sec5}, 
\xref{sec6}, and \xref{sec7}, discuss dependencies of coherence times
on parameters of the spin system systematically.
A summary is given in Sec.~\xref{sec8}.

%%%%%%%%%%%%%%%%%%%%%%%%%%%%%%%%%%%%%%%%%%%%%%%%%%%%%%%%%%%%%%%%%%%%%%%%
\section{Methods}
\label{sec2}

We calculate decoherence times of superpositions in a strongly anisotropic triangle of three spins $s=1$ with easy-axis anisotropy coupled to a spin bath using exact diagonalization and unitary time evolution. To see if there is anything inherently beneficial about toroidal (superposition) states that makes them more stable towards external perturbations in the context of decoherence, 
we evaluate 
%NEW
for which Hamiltonians there are superpositions among the low-lying spectrum with longer or shorter coherence times.

The system considered is displayed in \figref{system}. The three system spins ($s=1$) are 
antiferromagnetically coupled and there is a tangential easy-axis anisotropy on each site, 
i.e. the anisotropy axes are perpendicular to the altitude of the equilateral triangle formed 
by the sites of these three system spins
%NEW
reminiscent of e.g.\ Dy$_3$ toroidal molecules \cite{THM:ACIE06,CUS:ACIE08}.
We consider a bath of ($s=\frac{1}{2}$) spins that are coupled to each other and to the system spins via dipole-dipole interactions. The number of bath spins is chosen between 4 and 10. This small number is sufficient for our purpose.
The system spins are located at a distance of $r_s=1$ from the origin, while bath spins are placed randomly on the sphere around the origin with radius $r_b=2$ (in arbitrary units, see below).

The complete Hamiltonian
\begin{align}
\op{H}=\op{H}_{\text{S}}+\op{H}_{\text{SB}}+\op{H}_{\text{B}}
\end{align}
is made up of the system Hamiltonian $\op{H}_{\text{S}}$, the system-bath interaction Hamiltonian $\op{H}_{\text{SB}}$ and the bath-bath interaction Hamiltonian $\op{H}_{\text{B}}$.
The system Hamiltonian $\op{H}_{\text{S}}$ is comprised of three terms, the Heisenberg exchange interaction 
\begin{align}
\op{H}_{\text{Heisenberg}}=-2J\cdot \sum_{i=0}^2 \vec{\op{s}}_i\cdot \vec{\op{s}}_{i+1}
\end{align}
with $\vecops{3}  \equiv \vecops{0}$, the single-spin (single-ion) anisotropy
\begin{align}
\op{H}_{\text{anisotropy}} = \sum_{i=0}^2 \vec{\op{s}}_i\cdot \mat{D} \cdot \vec{\op{s}}_i
\end{align}
with the anisotropy tensors $\mat{D}_i$ as well as a Zeeman interaction term. All calculations were performed with a weak magnetic field $B_z=0.05$~T acting only on the central spin system. This splits the states that are degenerate at $B=0$ in order to be able to distinguish them numerically. We have numerically verified that while there is some quantitative effect on coherence times there is no significant qualitative difference to the situation with $B=0$.

The system-bath Hamiltonian $\op{H}_{\text{SB}}$ is defined as follows
\begin{align}
\op{H}_{\text{SB}}=\sum_{\substack{i=0,\\j=3}}^{2,N}\frac{A_1}{r_{ij}^3}\left(\vecops{i} \cdot\vecops{j}-\frac{3\left(\vecops{i}\cdot\vec{r}_{ij}\right)\left(\vecops{j}\cdot\vec{r}_{ij}\right)}{r_{ij}^2}\right)
\end{align}
with
\begin{align}
A_1=\frac{\mu_0 g_{\text{S}} \mu_{\text{S}}g \mu}{4\pi} \ .
\end{align}
The bath Hamiltonian $\op{H}_{\text{B}}$ reads
\begin{align}
\op{H}_{\text{B}}=&\sum_{3\leq i<j}^{N}\frac{A_2}{r_{ij}^3}\left(\vecops{i} \cdot\vecops{j}-\frac{3\left(\vecops{i}\cdot\vec{r}_{ij}\right)\left(\vecops{j}\cdot\vec{r}_{ij}\right)}{r_{ij}^2}\right)
\end{align}
with
\begin{align}
A_2=\frac{\mu_0 (g \mu)^2}{4\pi} \ .
\end{align}
Here $\mu_0$ is the vacuum permeability while $g\mu$ and $g_{\text{S}} \mu_{\text{S}}$ denote the magnetic interaction strength of the bath and system spins, respectively. We do not specify them, and we do not specify the unit of length
since we take $A_1$ and $A_2$ as adjustable parameters of our investigation 
%NEW
that would enable us to switch between electronic and nuclear bath spins.
%NEW
We would like to emphasize that we do not approximate the dipolar interactions by their diagonal,
i.e.\ Heisenberg-like part, as often done, see e.g.\ \cite{CHS:JPCL20,CJS:JPCL20,CCZ:PCCP22} as typical examples.
If the dipolar interaction is approximated by terms of the form
$2 \op{s}_i^z\op{s}_j^z + \op{s}_i^x\op{s}_j^x + \op{s}_i^y\op{s}_j^y$, then total $\op{S}^z$ is a conserved 
quantity for this interaction as it is for the Heisenberg part of the Hamiltonian.
We are truly convinced that the symmetry-breaking anisotropic parts of the dipolar interaction play an
important role for decoherence since they allow many more transitions like for instance flip-flop
transitions which are often discussed in a perturbation picture of decoherence, compare e.g.\ \cite{WTG:NN13}.

In Ref.~\cite{VoS:PRB20}, the effects of the magnitude of system parameters $A_1$ and $A_2$ are illustrated. To summarize, $A_1$ has a significant effect on the time scale of decoherence for all superpositions. This is because the 
%NEW
many-body energy eigenstates of the full system
get less and less energetically isolated as $A_1$ is increased and the original Zeeman structure of the system is lost when adding the bath and considering the full system. $A_2$ controls the relative differences in coherence times between different superpositions but affects them a lot less than $A_1$. 
%NEW
In an approximate mean-field picture a strong $A_1$ would lead to stronger time-dependent 
detunings of the effective magnetic field at the site of the system spins, thus destroying 
the coherence of the transition.
Our findings align with these general statements, 
and $A_1$ and $A_2$ are chosen to have fixed values of 0.1 K for all following calculations (for other values of $A_1$ and $A_2$ see the Appendix). 
This somewhat arbitrary choice is justified as the aim of these investigations is to find 
relative differences in coherence times, not calculating them accurately for realistic systems.
%More examples for different choices of parameters are provided in the appendix.

The decoherence times are calculated via time evolution based on exact diagonalization. To perform the time evolution of the whole system with a time-independent magnetic field, let \{$\ket{a_i}$\} be the eigenbasis of the initial ($t=0$) Hamiltonian of the system $A$. Initialize $A$ into an initial state $\ket{\Psi_A(t=0)}$. Most often, this will be a superposition of two non-degenerate states $n$ and $p$
\begin{align}\label{superpos}
\ket{\Psi_A(t=0)}=\frac{1}{\sqrt{2}}\left(\ket{a_n}+\ket{a_p}\right) \ .
\end{align}
Then define a random initial state $\ket{\Psi_B(t=0)}$ for the bath $B$ and form a product state
\begin{align}\label{initial}
\ket{\Psi_0} = \ket{\Psi(t=0)} = \ket{\Psi_A(t=0)} \otimes \ket{\Psi_B(t=0)} \ .
\end{align}
Let \{$\ket{m}$\} be the product basis of the whole system ($A\otimes B$) 
and \{$\ket{\varphi_l}$\} the eigenbasis of the Hamiltonian of the whole system with eigenvalues $E_l$. 
Then a change into this eigenbasis
\begin{align}
\ket{\Psi_0} = \sum_{l=1}^{\text{dim}(\mathrm{H})} \braket{\varphi_l|\Psi_0}\cdot\ket{\varphi_l}
\end{align}
yields for the time-evolved state
\begin{align}
\ket{\Psi(t)} = \sum_{l=1}^{\text{dim}(\mathrm{H})} 
e^{i\cdot E_l\cdot t} \braket{\varphi_l|\Psi_0}\cdot\ket{\varphi_l} \ .
\end{align}
Changing back into the product basis gives
\begin{align}
\ket{\Psi(t)} = \sum_{l,m=1}^{\text{dim}(\mathrm{H})} e^{i\cdot E_l\cdot t} \braket{\varphi_l|\Psi_0}\cdot\braket{m|\varphi_l}\cdot\ket{m} \ .
\end{align}
All terms in this equation are known to machine precision by exact diagonalization.

In order to quantify the decoherence of the superposition, we employ the reduced density matrix, denoted by $\rho$, as 
a quantifier. There are various options for the quantification of decoherence but in our context it does not really matter which one is chosen, and we just consider the absolute value of the off-diagonal element of $\rho$. 
If the initial state was a superposition as defined in \eqref{superpos}, this quantity can simply be calculated as 
\begin{align}
|\rho_{n,p}| = |\braket{a_n|\op{\rho}|a_p}| \ .
\label{deco}
\end{align}

\begin{figure}[H]
\centering
\includegraphics[width=0.45\columnwidth]{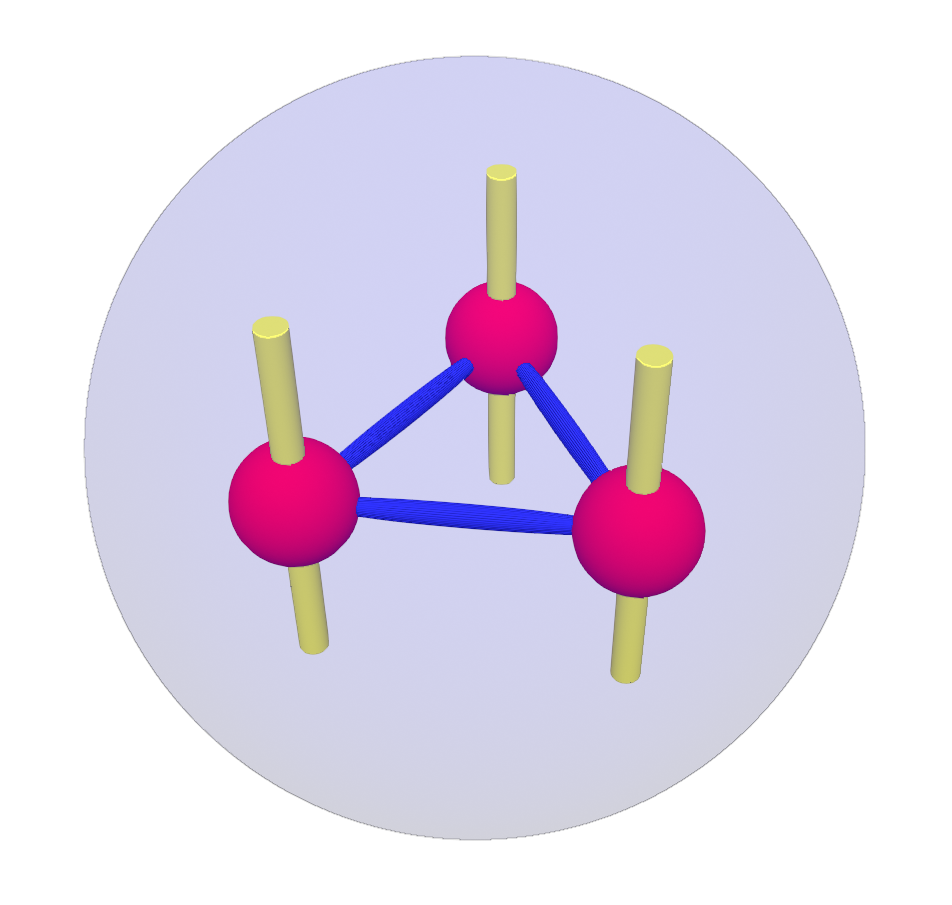}
\qquad
\includegraphics[width=0.45\columnwidth]{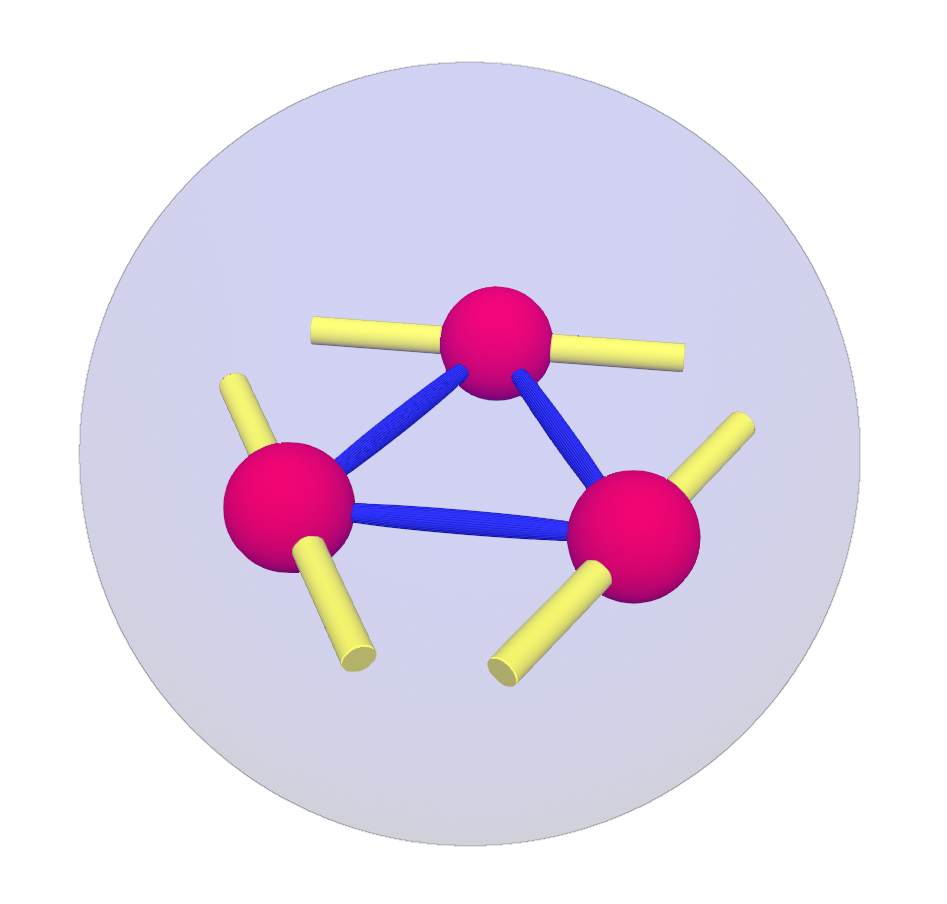}
\caption{\label{SMM_tor}LHS: System for $\theta=0^{\circ}$; this would be a typical SMM configuration. 
RHS: System for $\theta=90^{\circ}$; this corresponds to a perfect toroidal configuration.
For angles inbetween see \figref{system}.}
\end{figure}

In this paper we investigate the dependence of decoherence rates on the tilting of all anisotropy axes along the global $\theta$ direction where $\theta=0^{\circ}$ corresponds to a perfect alignment of easy anisotropy axes for a 
single molecule magnet (SMM) and $\theta=90^{\circ}$ represents a perfect toroidal configuration, see \figref{SMM_tor}. 
%NEW
The configuration with $\theta=90^{\circ}$ will however not be considered in this paper as it is then impossible to initialize a toroidal moment with a magnetic field in $z$-direction because the field direction is perpendicular to the one of the easy axes.
We will also need a definition of the toroidal magnetic moment of a spin triangle in order to evaluate its relevance for decoherence rates. This is given by 
\begin{align}
\vecoptau \ =\  g\cdot \mu_B \sum_{i=0}^{2} \vec{r}_i \times  \vecops{i} \ ,
\label{tau}
\end{align}
with $\vec{r}_i$ being the classical positions of the spins contributing to the sum.
%NEW
Then $\tau_z$, as used in later figures, is defined as the expectation value with respect to the initial state,
compare \fmref{initial},
\begin{align}
\tau_z = \bra{\Psi_A(t=0)}\op{\tau}_z\ket{\Psi_A(t=0)}
\ .
\label{tauexp}
\end{align}

\section{Examples of decoherence calculations}
\label{sec3}

\begin{figure}[H]
\centering
\includegraphics[width=0.48\columnwidth]{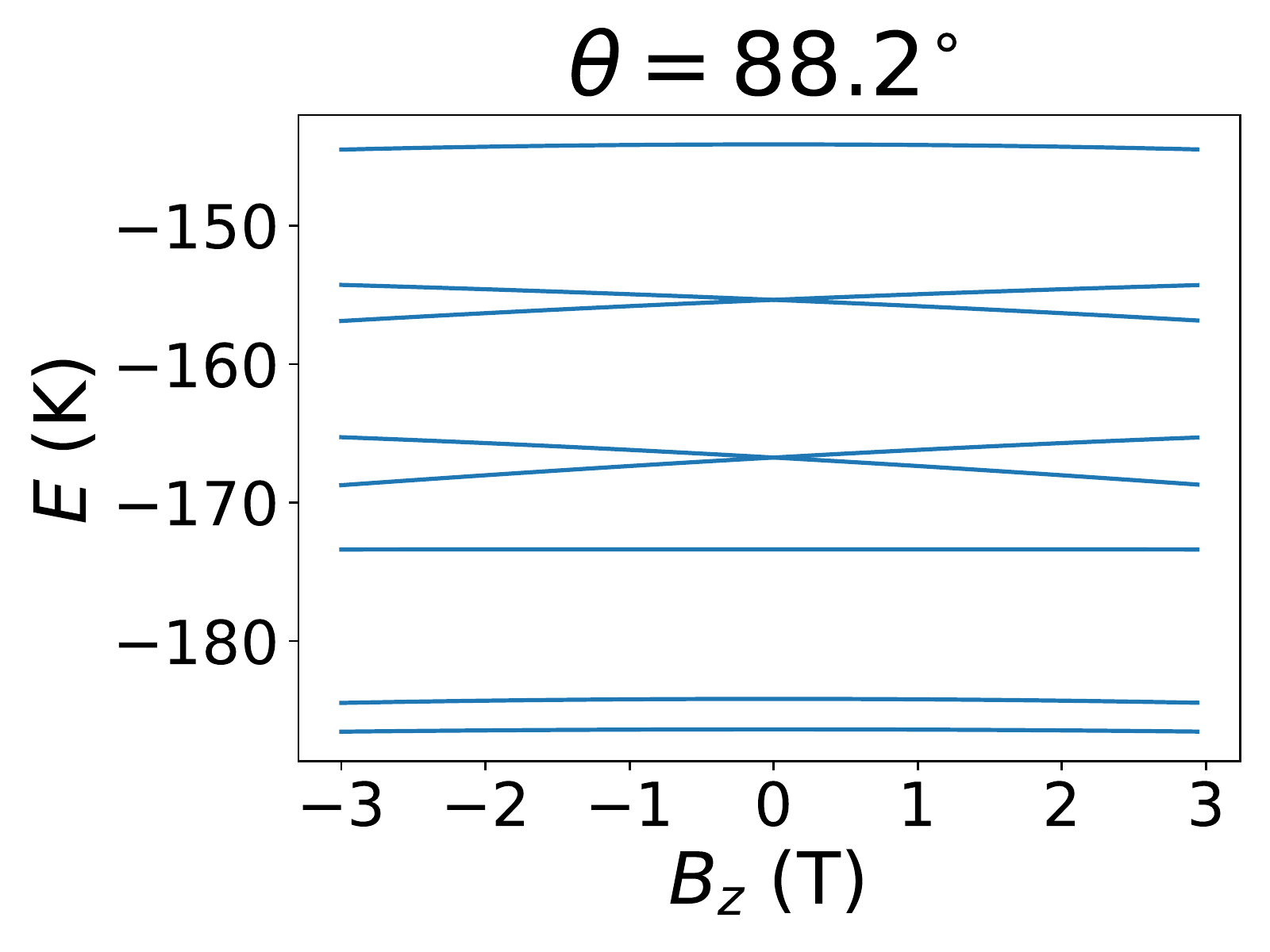}
\includegraphics[width=0.48\columnwidth]{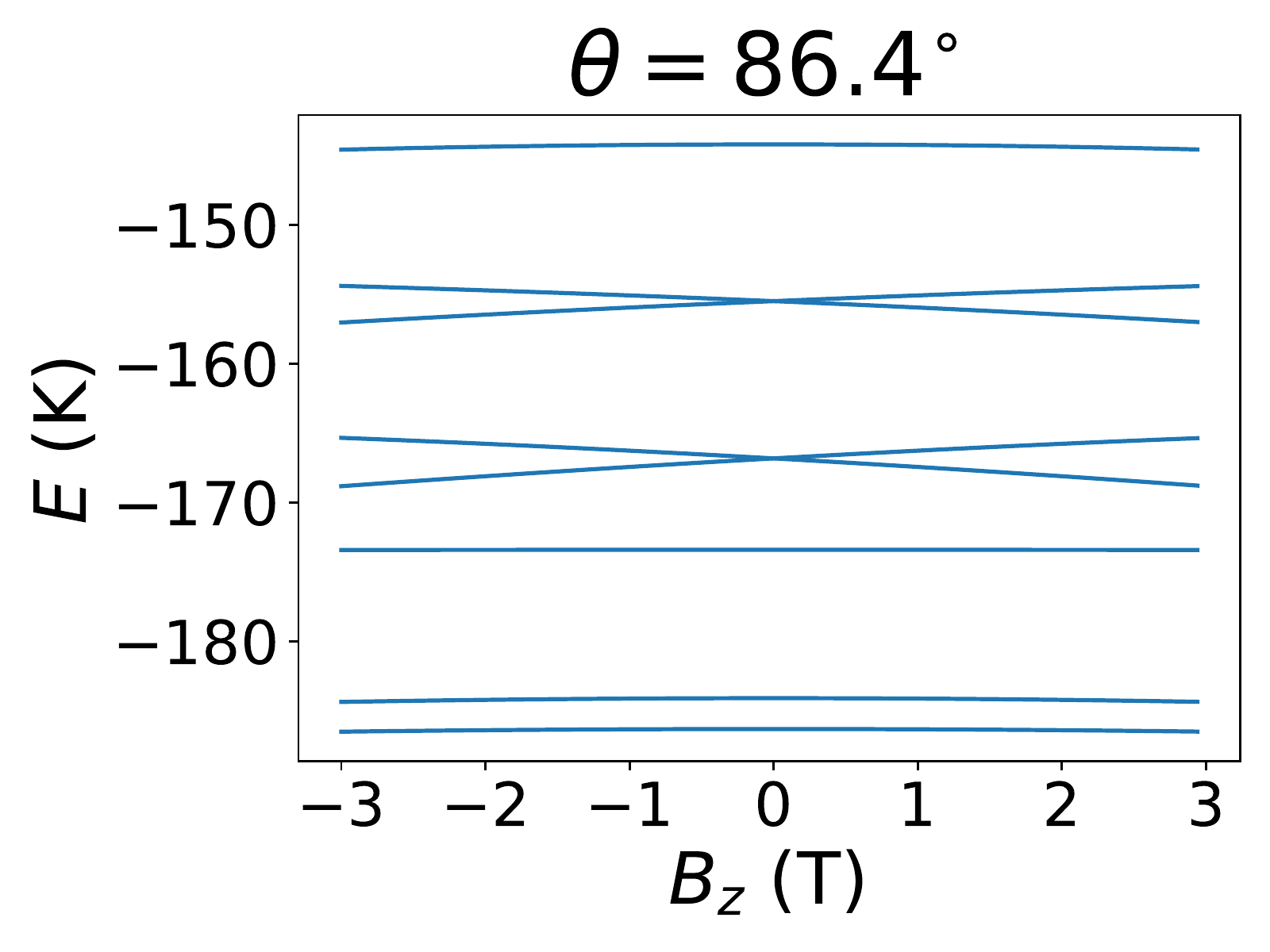}\\
\includegraphics[width=0.48\columnwidth]{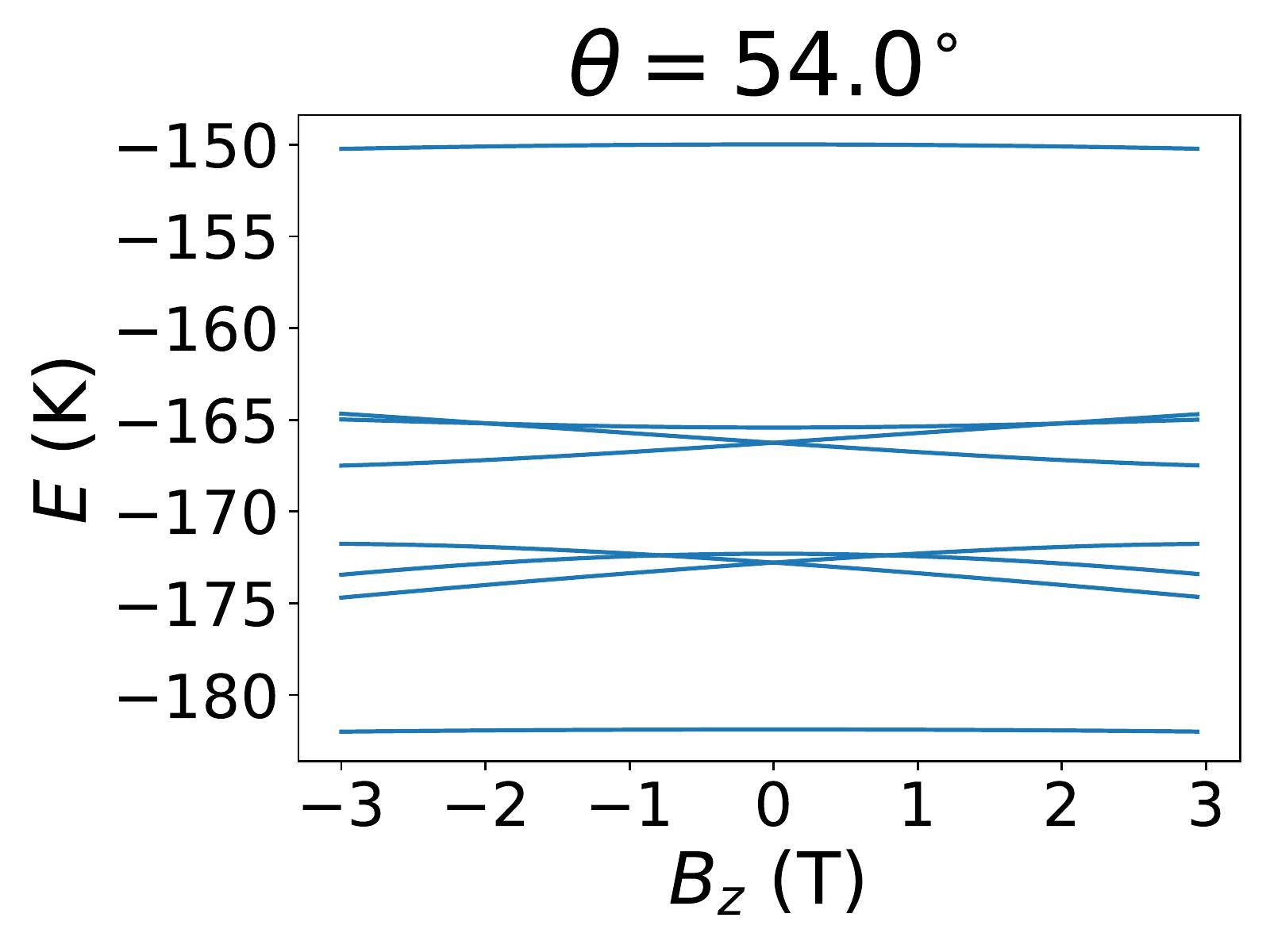}
\includegraphics[width=0.48\columnwidth]{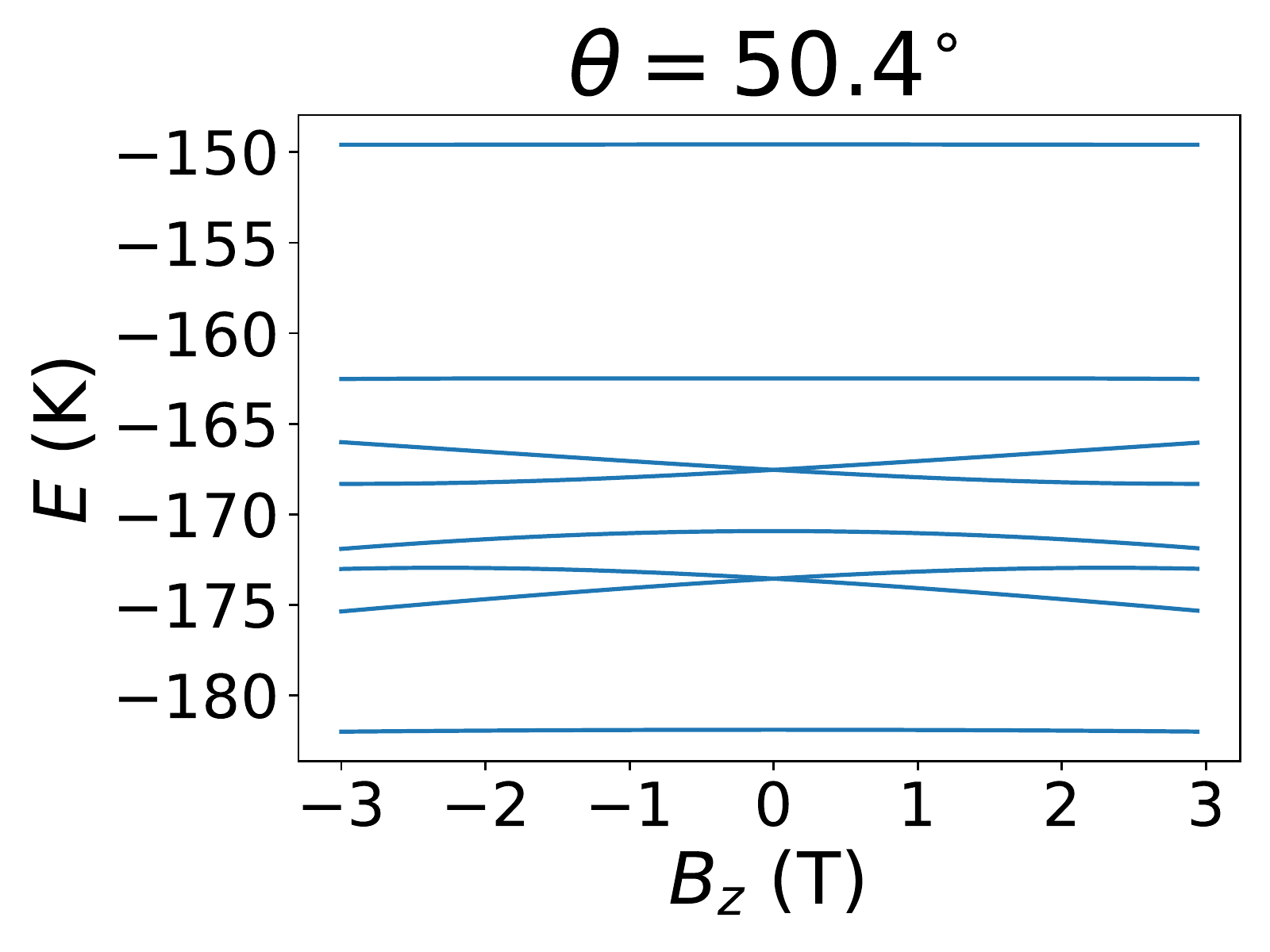}\\
\includegraphics[width=0.48\columnwidth]{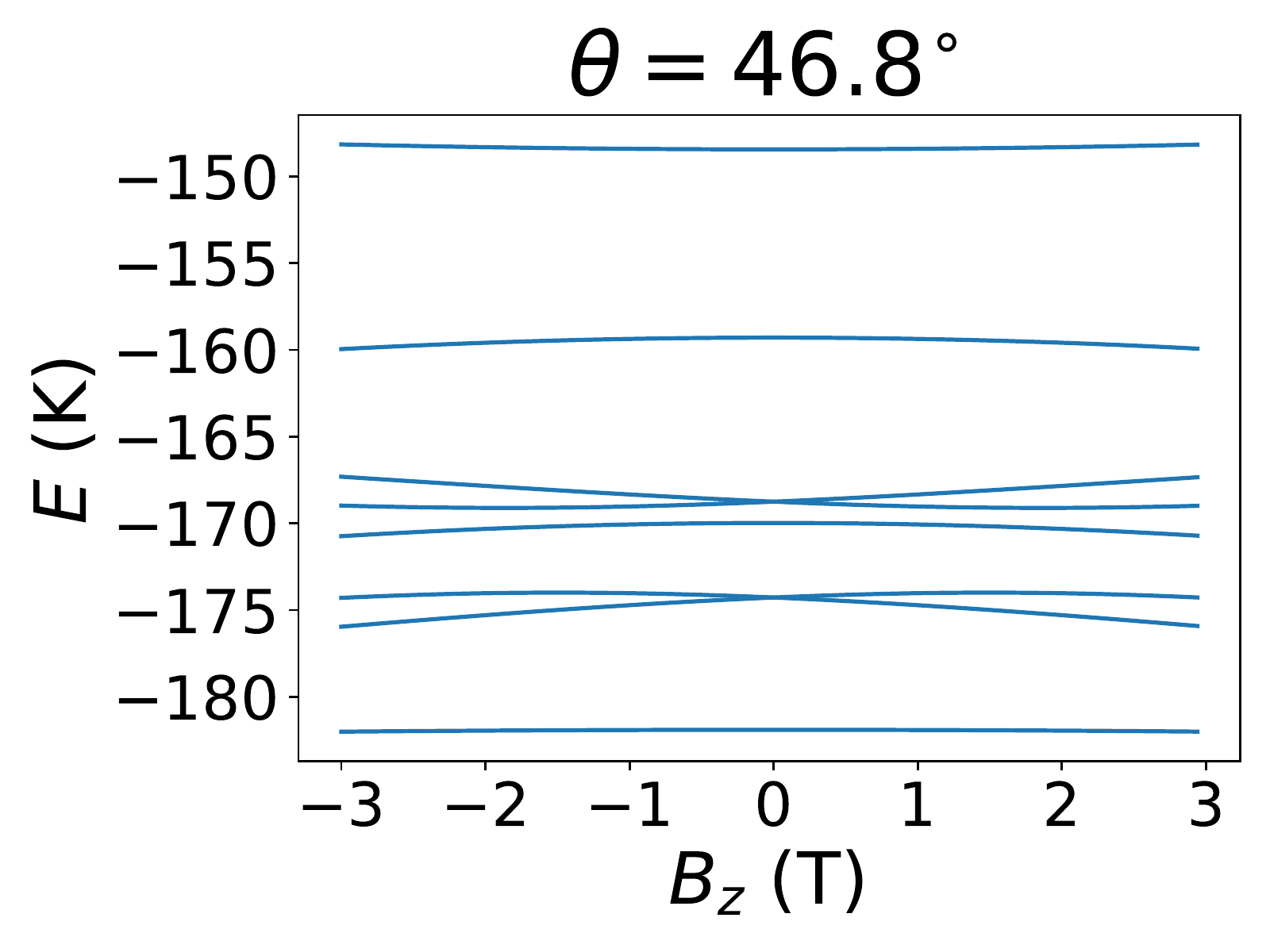}
\includegraphics[width=0.48\columnwidth]{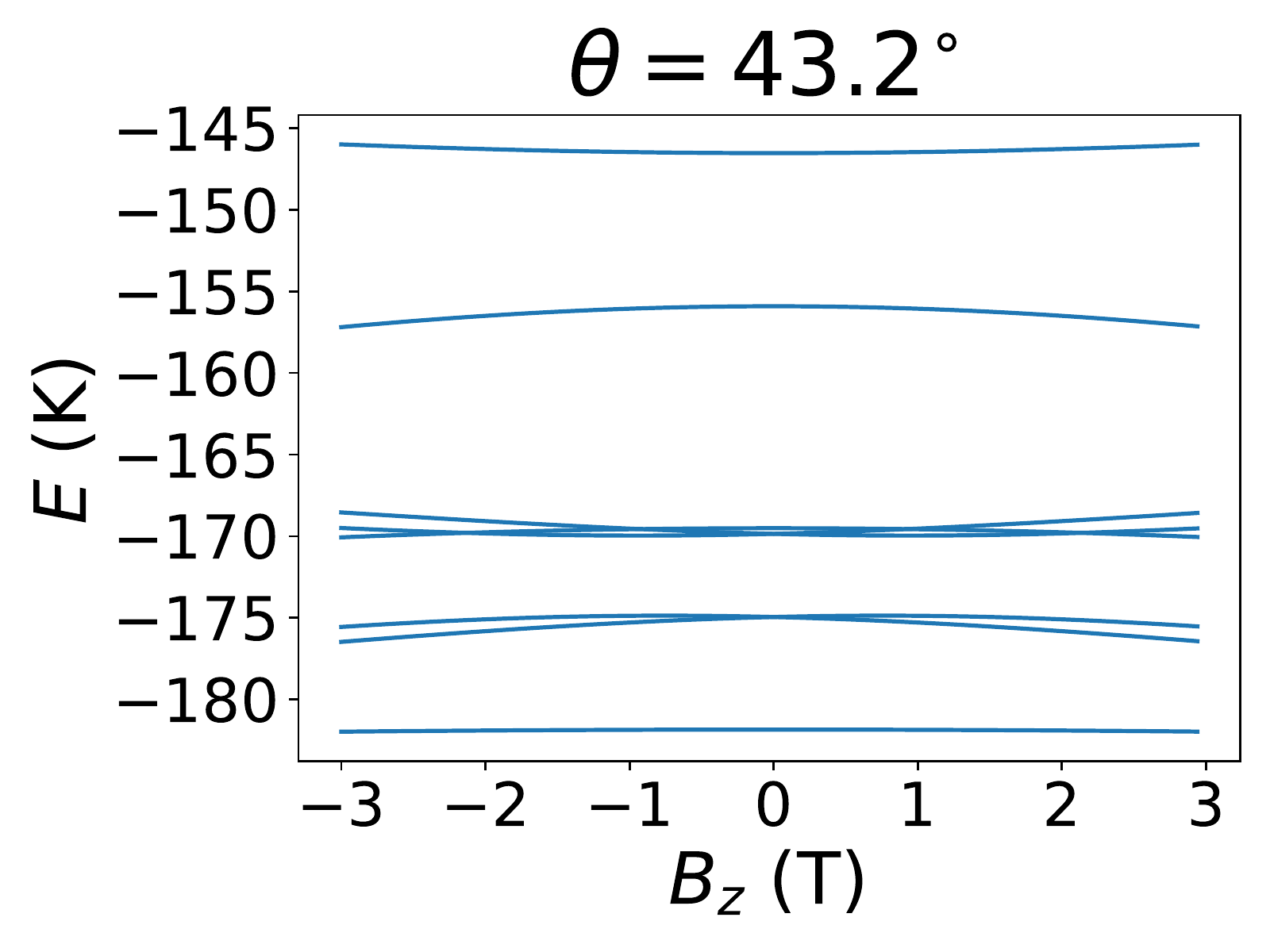}\\
\includegraphics[width=0.48\columnwidth]{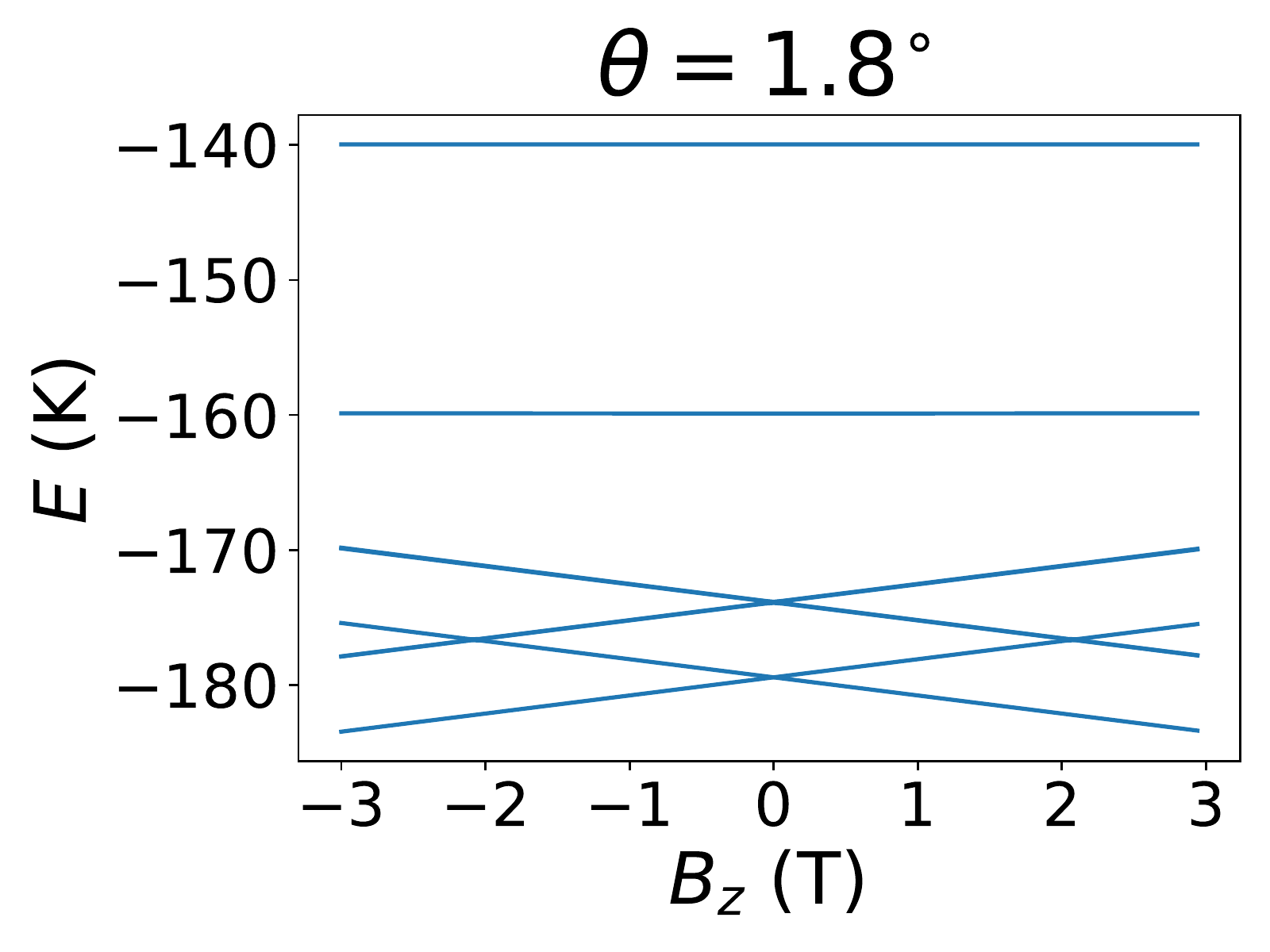}
\includegraphics[width=0.48\columnwidth]{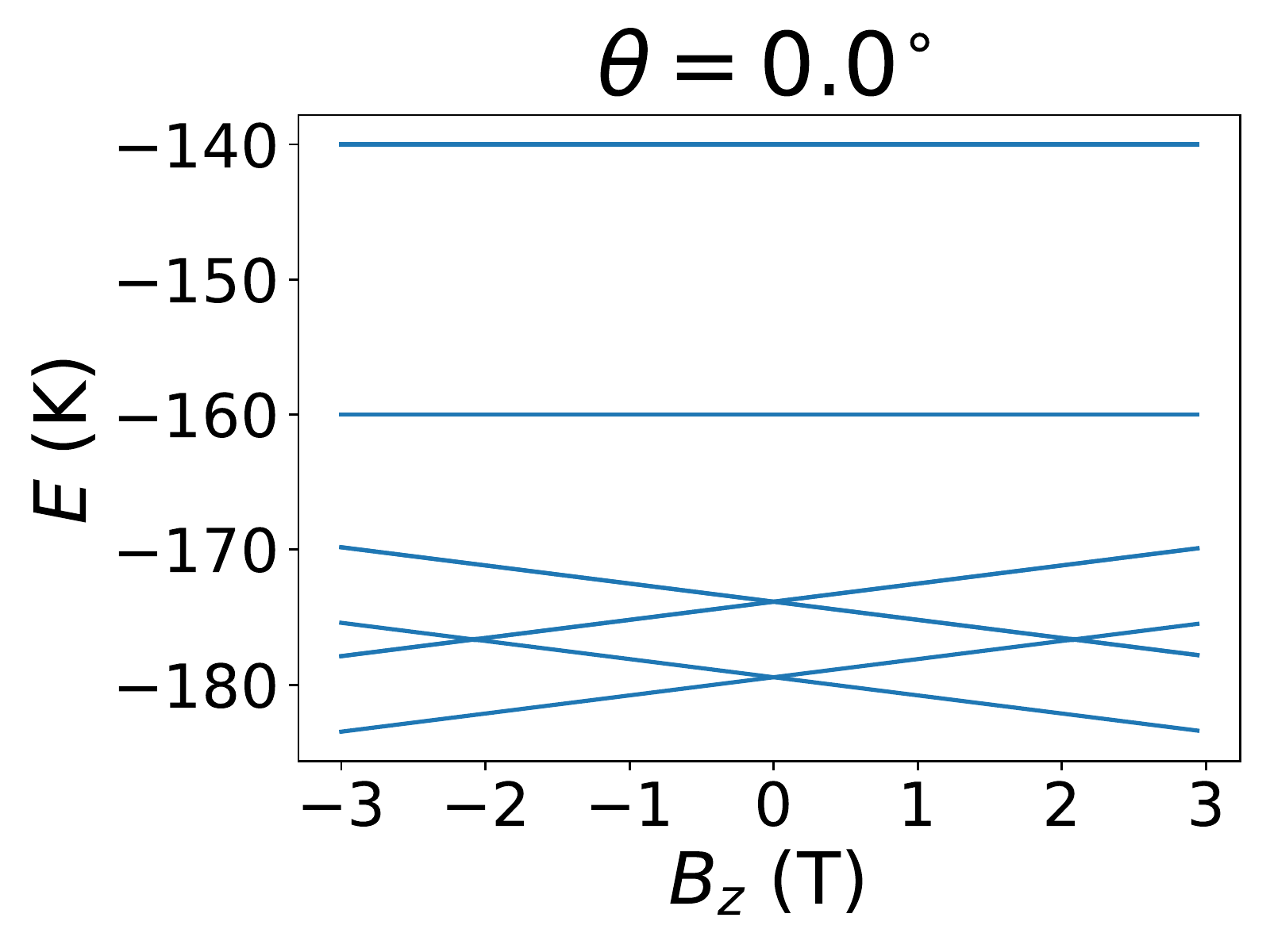}
\caption{\label{zeeman}Zeeman diagrams of the eight lowest-lying states of the system without bath for different values for the tilting angle of the anisotropy axes $\theta$. For a given $B_z$ (typically $B_z=0.05$~T) the states are enumerated
$0,1,2,3,\dots$ from below.}
\end{figure}

To model a strongly anisotropic system, $J=-10$ K and $D=-50$ K are chosen in order to work with 
%NEW
(order of magnitude) typical numbers. A few examples using different parameters are provided in 
the appendix. 
Some of the resulting Zeeman diagrams as a function of $B_z$ displaying the lowest 
%NEW
eight eigenstates are shown in \figref{zeeman}.

\begin{figure}[ht!]
\centering
\includegraphics[width=0.9\columnwidth]{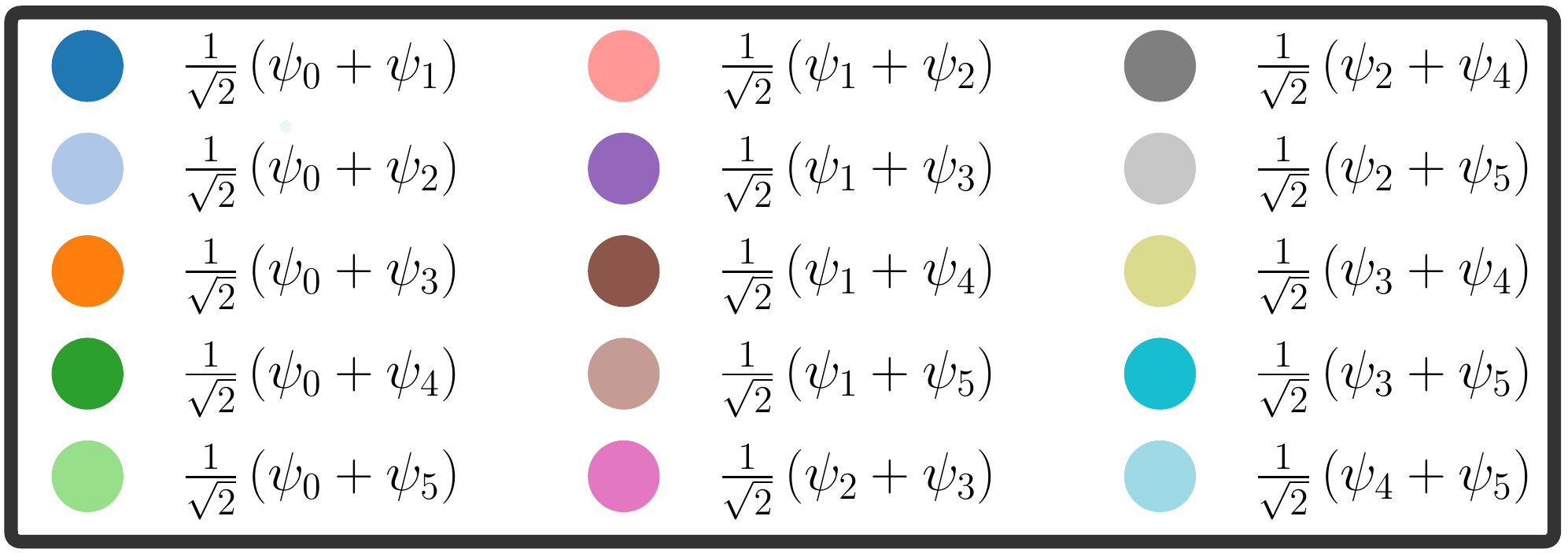}
\caption{\label{legend}Legend to show which superposition is represented by which color.}
\end{figure}

In order to test which system configurations and which superpositions show the longest coherence times, we consider all possible two-state superpositions of the six energetically lowest eigenstates for different values of $\theta$ at $B_z=0.05$ T and perform a time evolution as laid out above. We ignore possible experimental difficulties concerning the initialization of these superpositions
%NEW
\footnote{We speculate that superpositions of toroidal states can be obtained e.g.\ when molecules are deposited on surfaces using
pulses of the STM tunneling current \cite{BPC:S15} or by microwave pulses in cases when the anisotropy axes are tilted from the molecular 
plane since the magnetization along the molecular axis is proportional to the toroidal moment of a given state.} 
as we aim to identify characteristics of those superpositions which display long coherence times. In order to avoid duplicate plot legends, the color code for the superpositions is displayed once in \figref{legend}.

\begin{figure}[H]
\centering
\includegraphics[width=0.48\columnwidth]{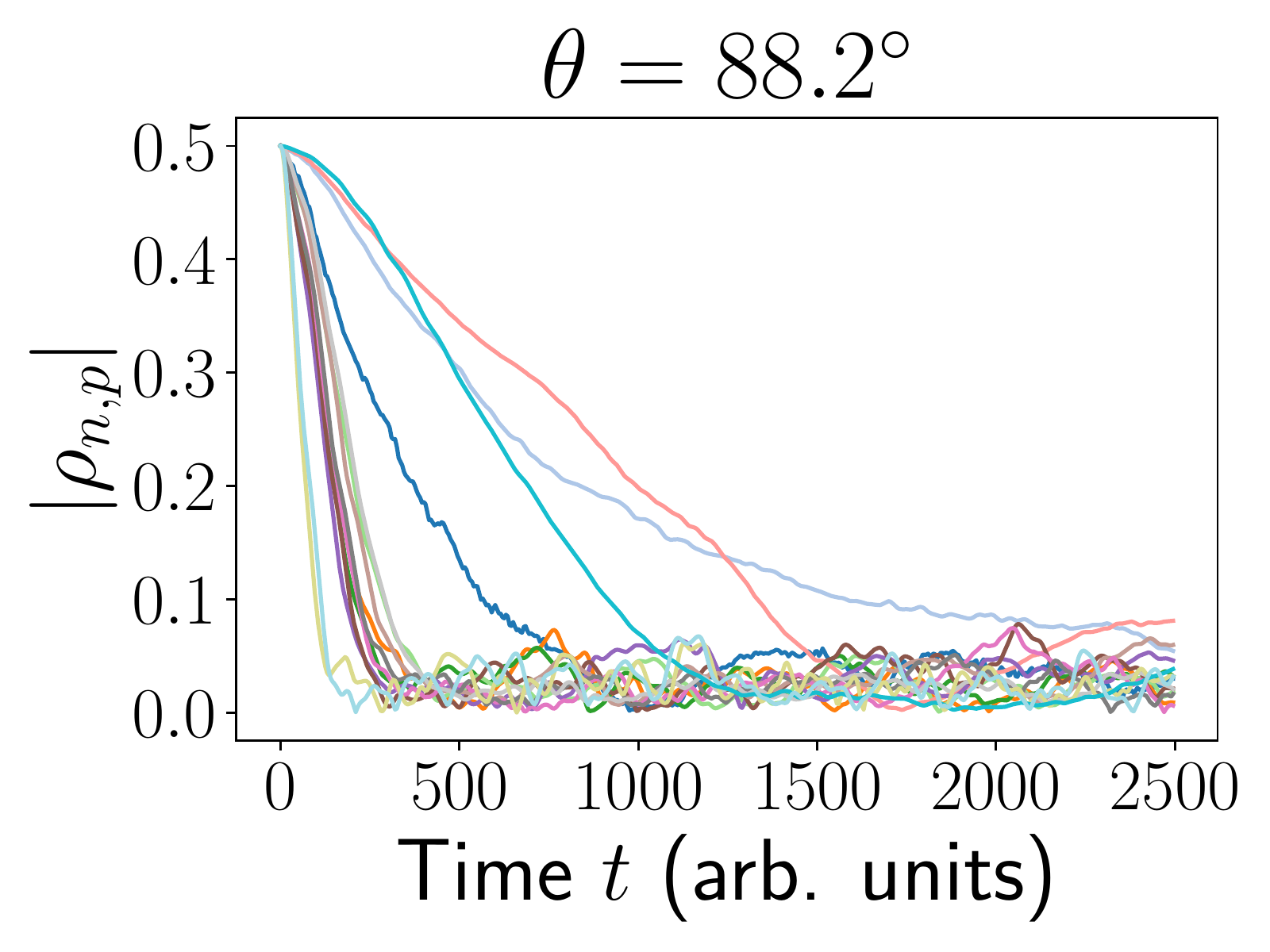}
\includegraphics[width=0.48\columnwidth]{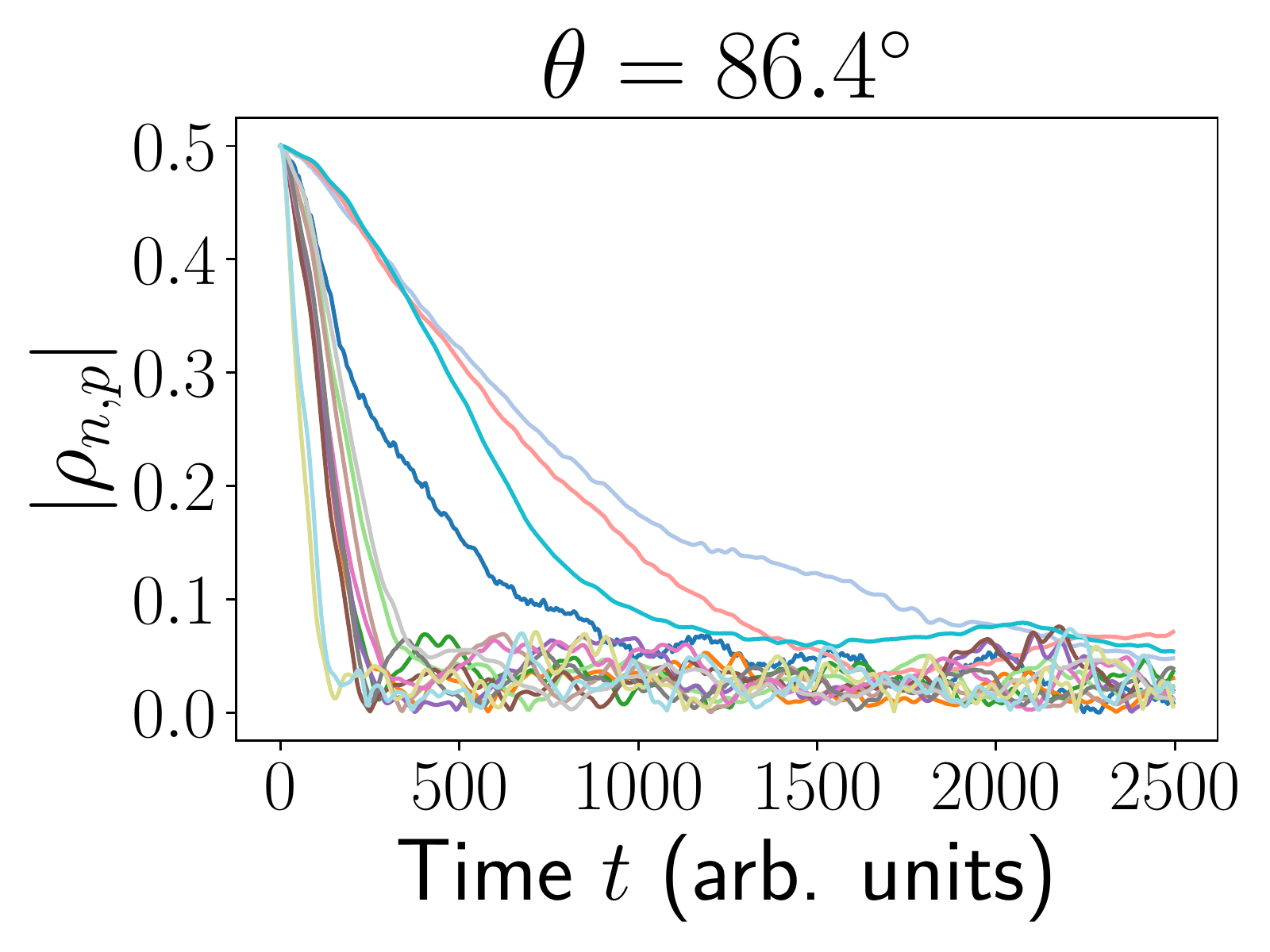}
\includegraphics[width=0.48\columnwidth]{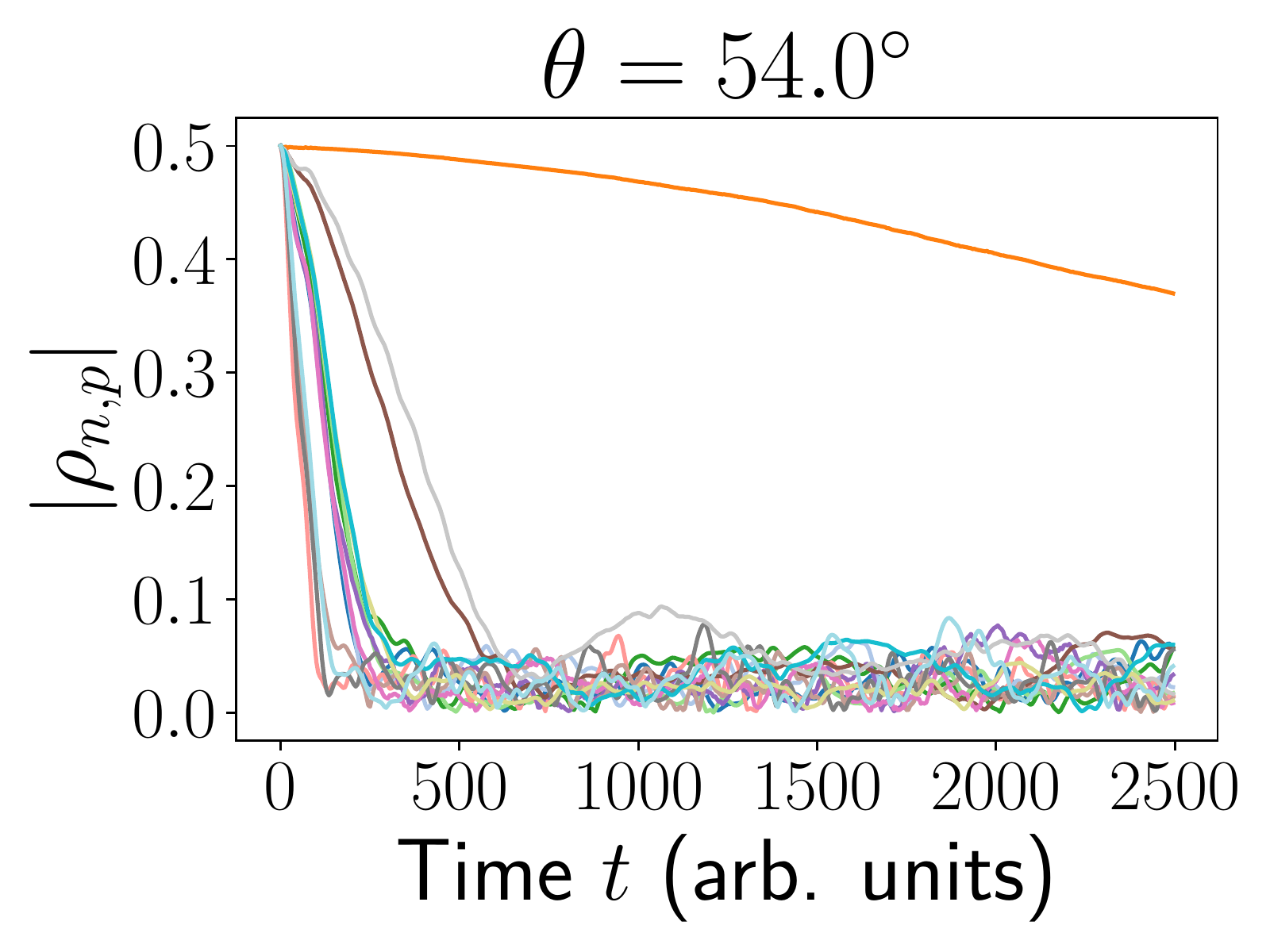}
\includegraphics[width=0.48\columnwidth]{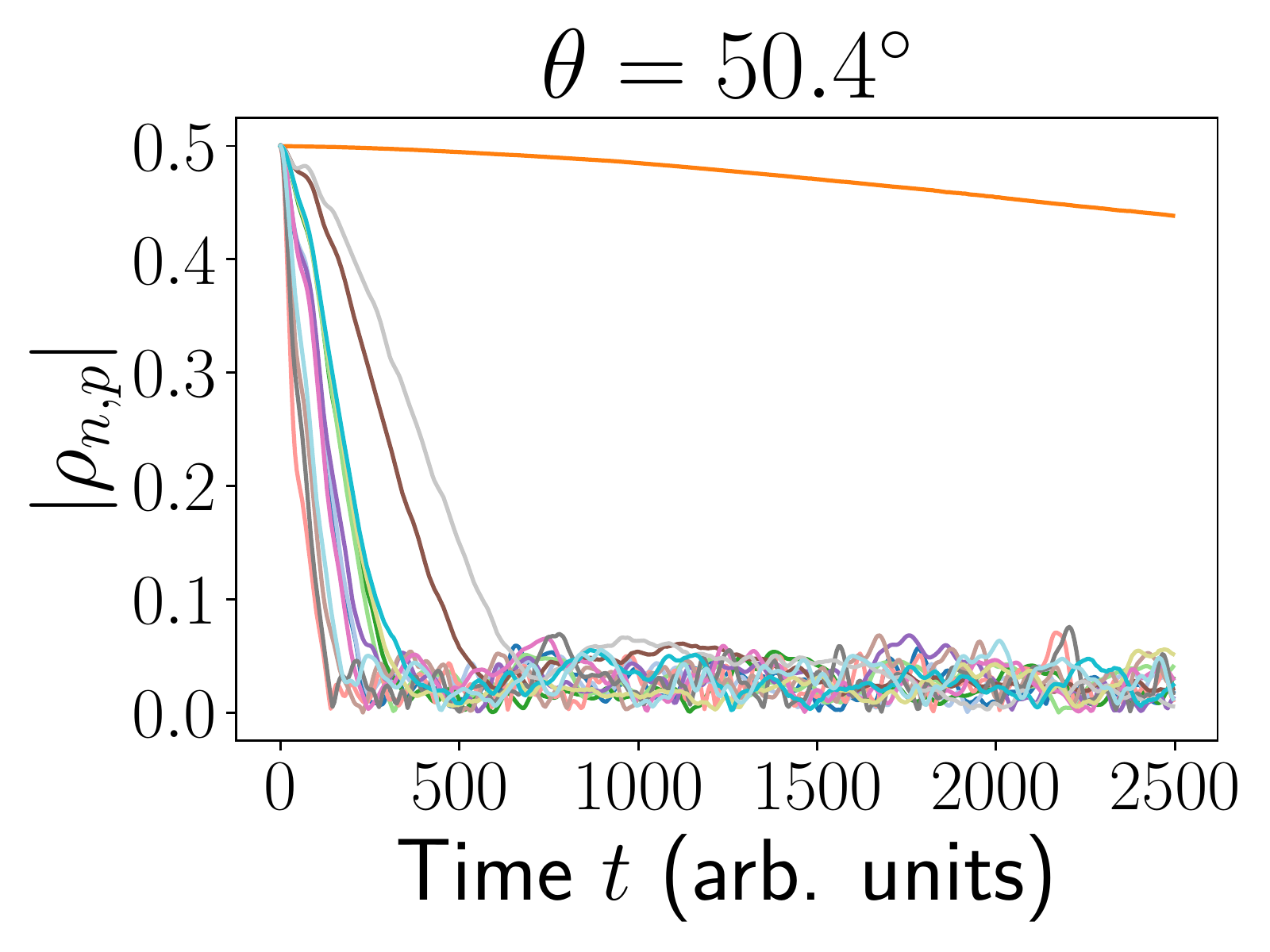}
\includegraphics[width=0.48\columnwidth]{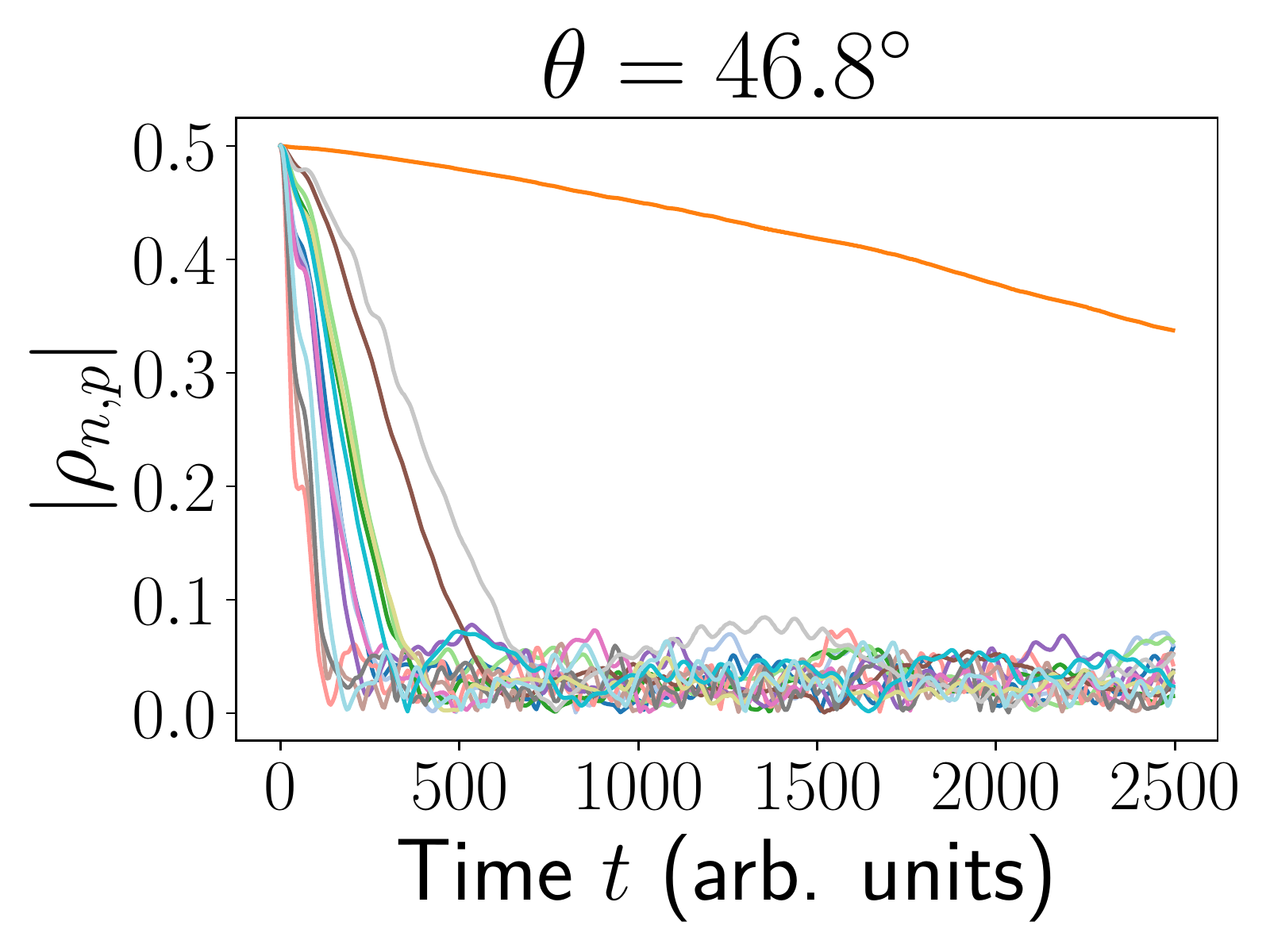}
\includegraphics[width=0.48\columnwidth]{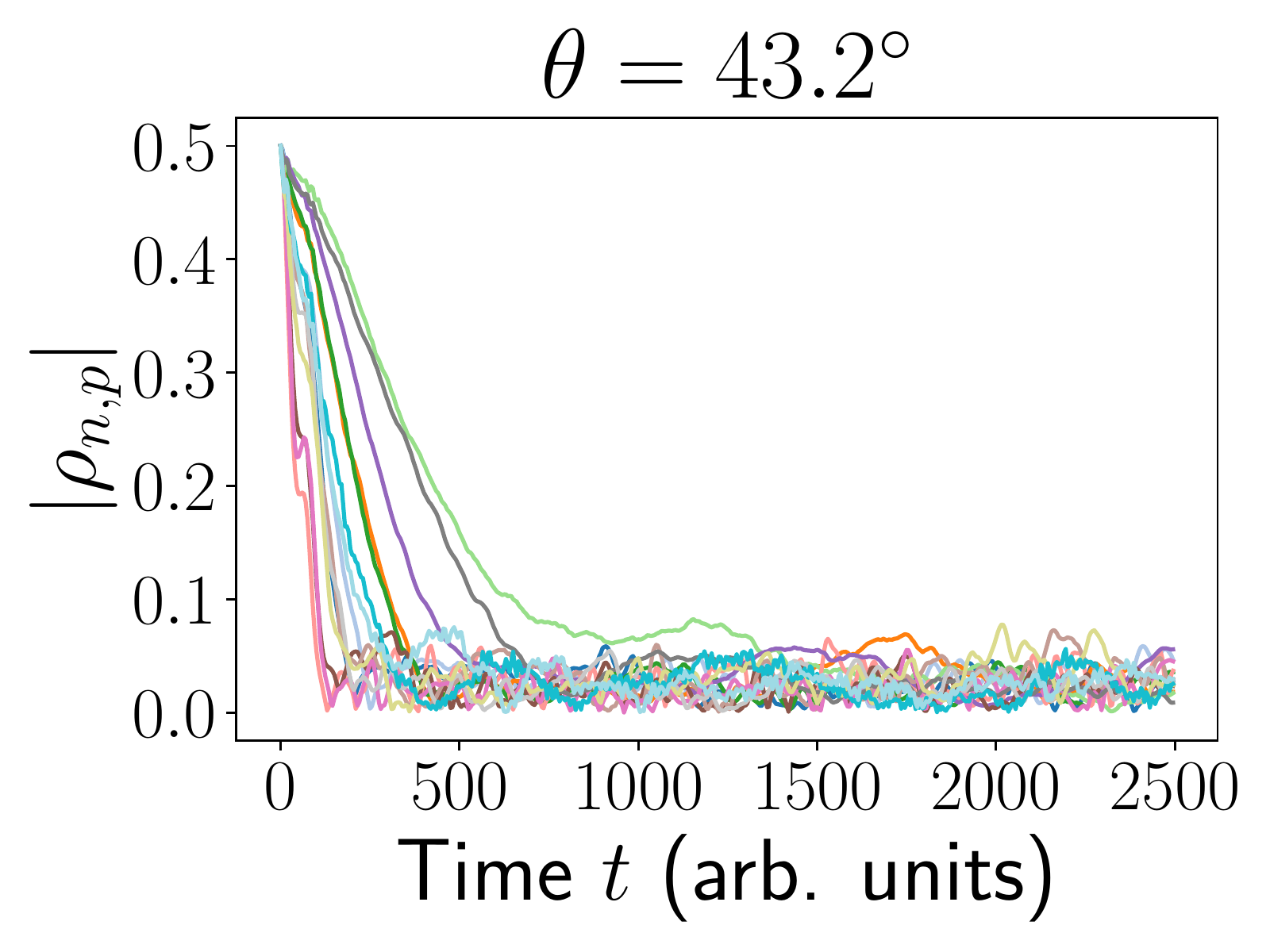}
\includegraphics[width=0.48\columnwidth]{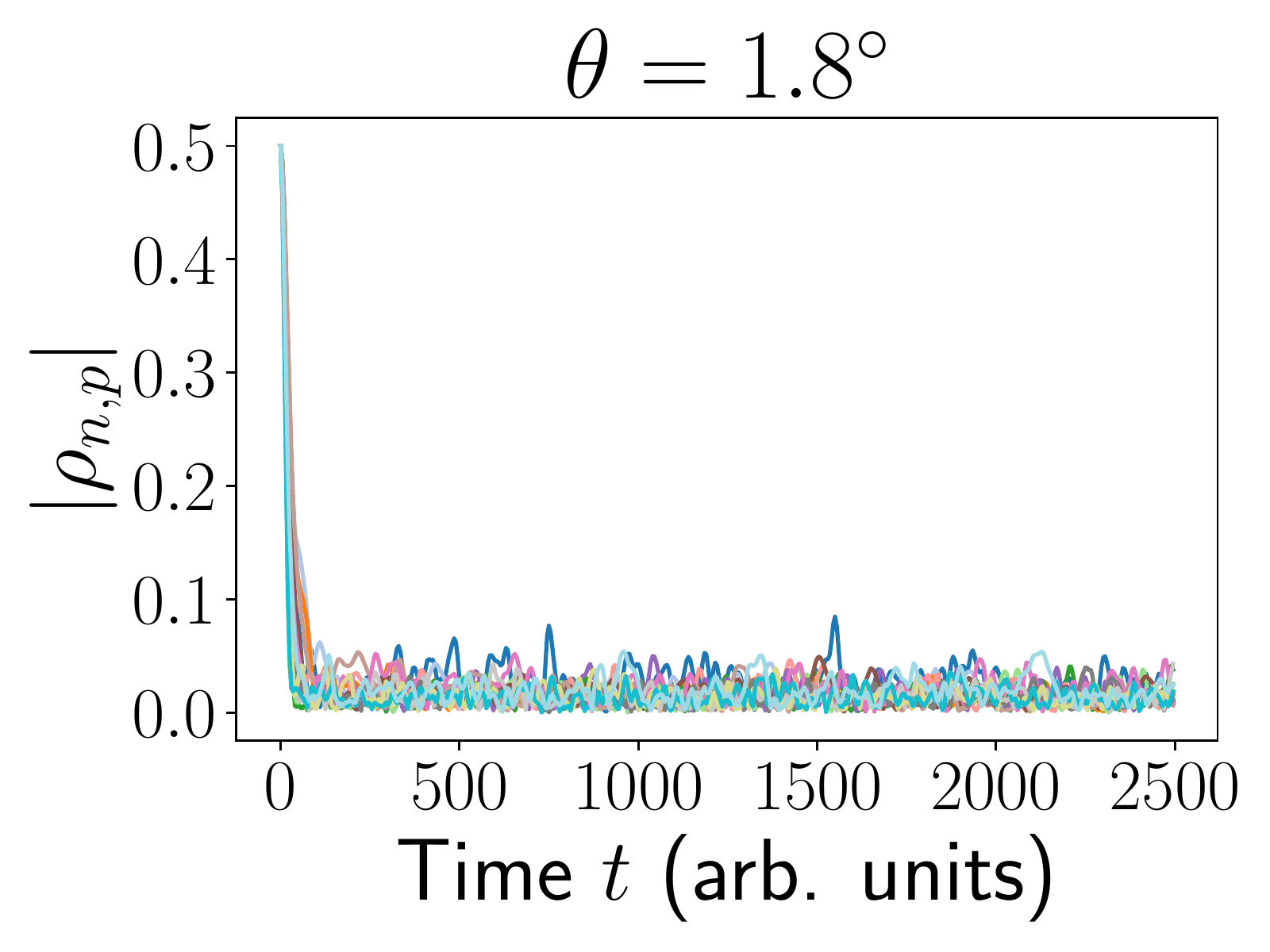}
\includegraphics[width=0.48\columnwidth]{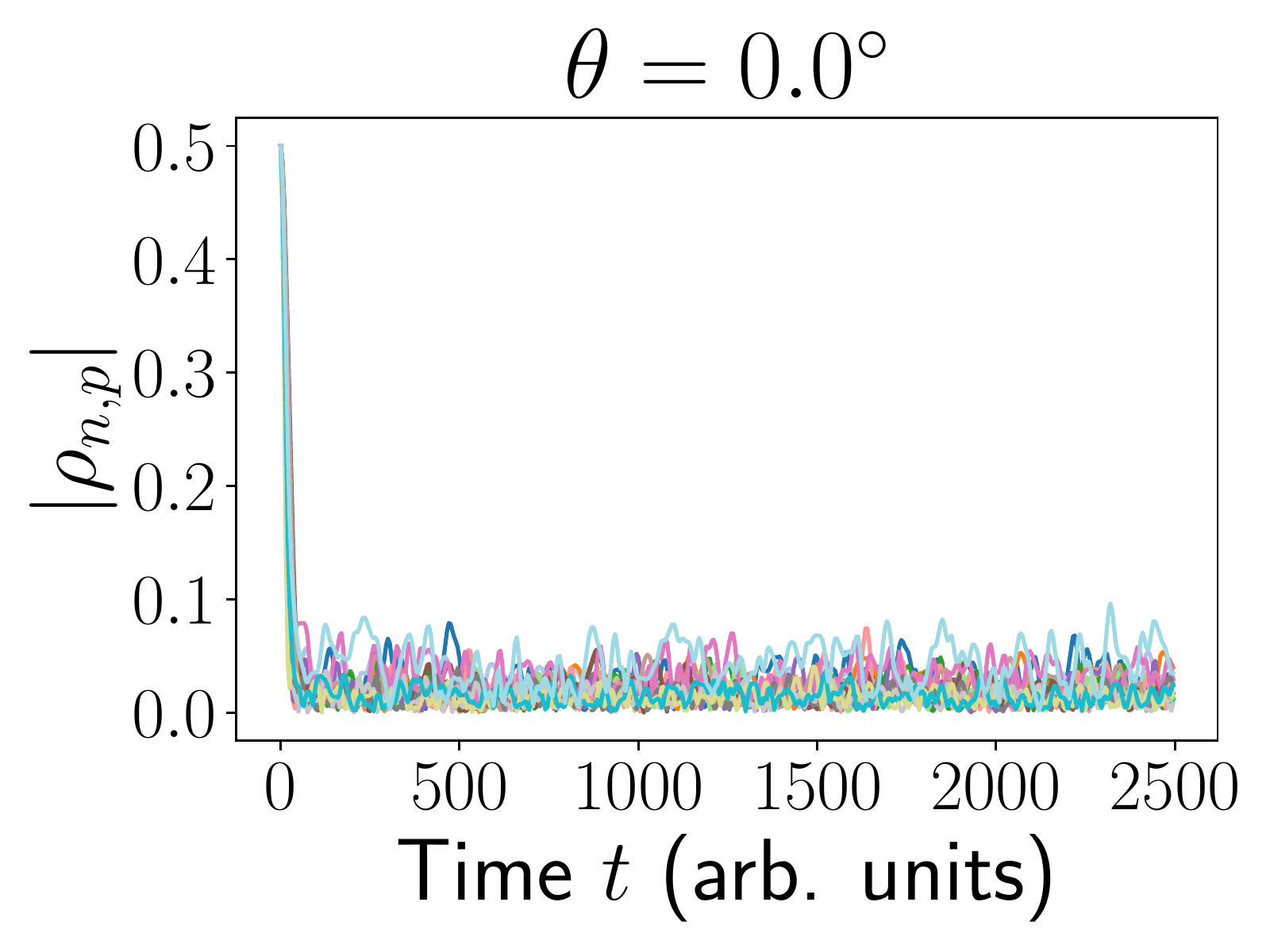}

\caption{Decoherence (see \eqref{deco}) over time of all two-state superpositions of the six lowest-lying energy eigenstates for a fixed random bath with $n_{\text{bath}}=8$; angles $\theta$ correspond to \figref{zeeman}. Legend is displayed in \figref{legend}.} 
\label{decodiagrams} 
\end{figure}

Some sample results corresponding to the Zeeman diagrams from \figref{zeeman} are shown in \figref{decodiagrams}. While many superpositions decohere almost instantly, others show significantly longer coherence times. In the SMM configuration ($\theta=0^{\circ}$), all superpositions decohere quickly, while for the almost toroidal configuration ($\theta=88.2^{\circ}$), there are some states which survive a decent amount of time. For the parameter configuration chosen here, however, middle-sized angles $\theta \approx 50^{\circ}$ perform best in most cases.

\begin{figure}[ht!]
\centering
\includegraphics[width=0.675\columnwidth]{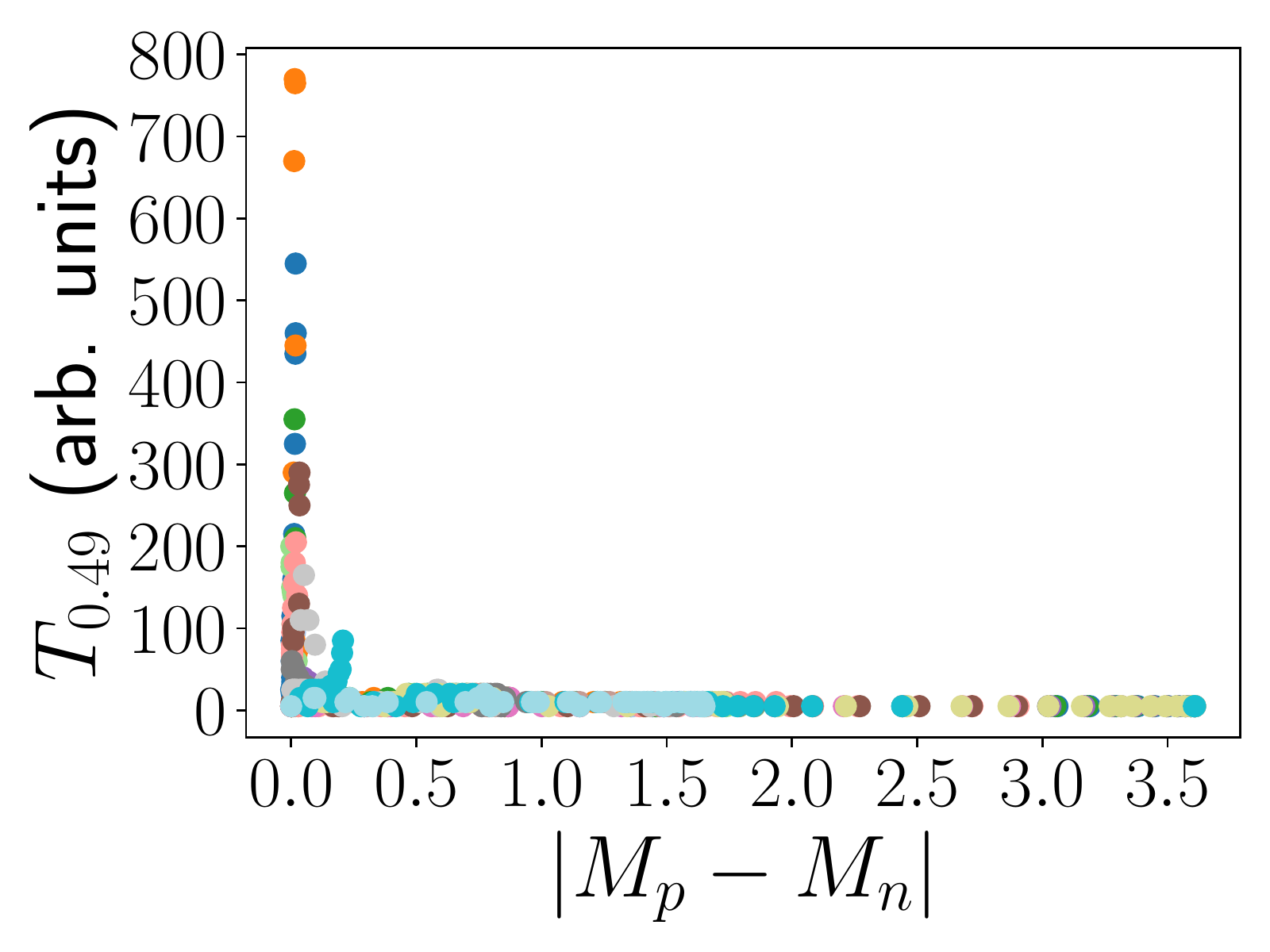}
\includegraphics[width=0.675\columnwidth]{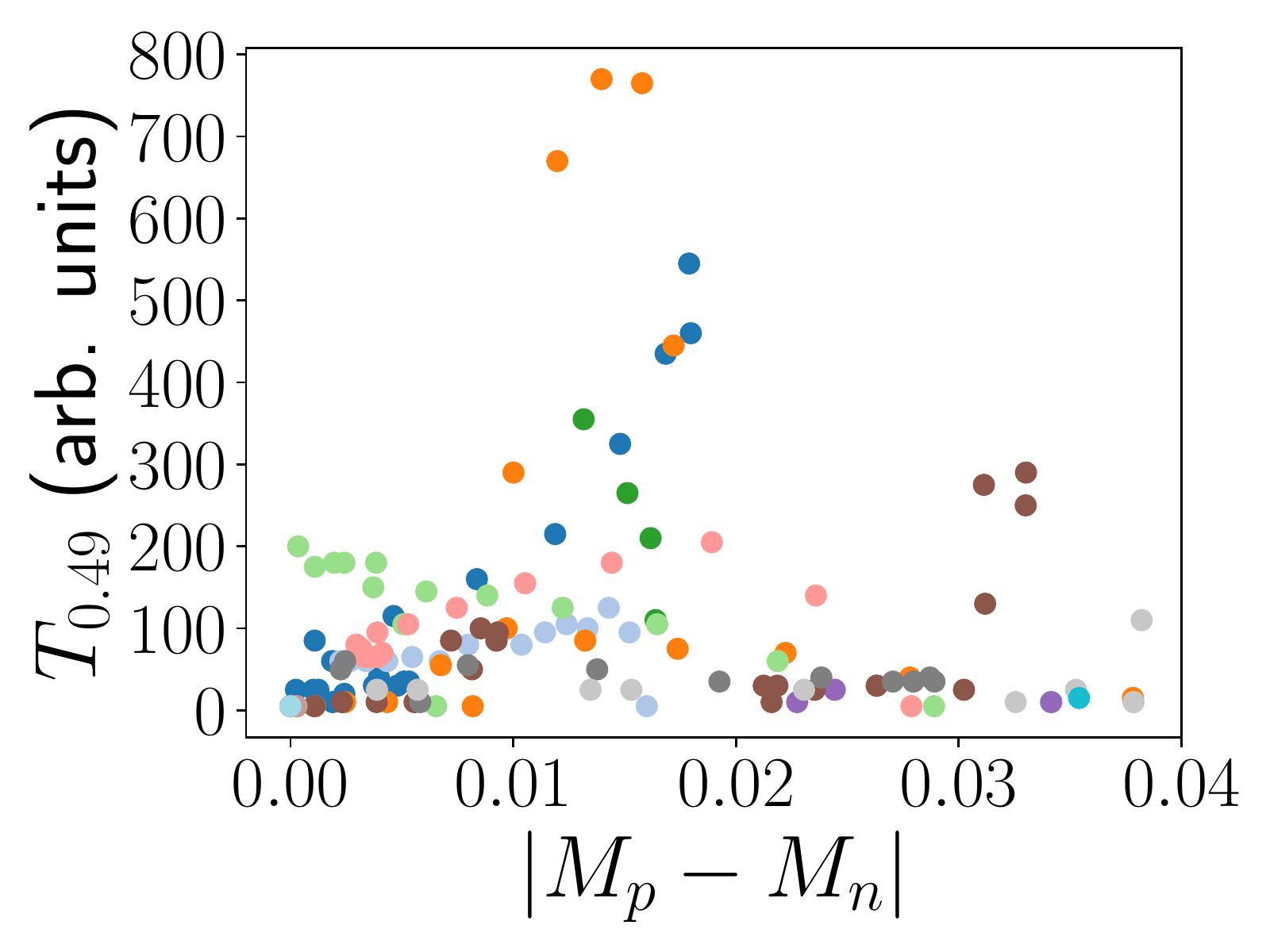}
\caption{Coherence times vs$.$ the absolute value of the difference in the expectation value for $S_z$ of the eigenstates in each superposition ($|M_p-M_n|$) for 50 values of $\theta$ between $0^{\circ}$ and $88.2^{\circ}$ at $B_z=0.05$~T for a random bath with $n_{\text{bath}}=8$. Top: Only superpositions $|M_p-M_n| \approx 0$ have comparatively long coherence times. Bottom: Zoomed-in view. While $|M_p-M_n| \approx 0$ is necessary, however, it is not sufficient to predict long coherence times. Legend is displayed in \figref{legend}.} 
\label{slope}
\end{figure}

To make system configurations and superpositions comparable, we choose to consider the time it takes for the absolute value of the off-diagonal element to fall below the more or less arbitrarily chosen value of 0.49 for the first time for a given superposition. This time is denoted by $T_{0.49}$. In \figref{slope}, we first verify the claim that superpositions of states with near-identical magnetic moments (clock transitions) indeed survive the longest. To this end, we calculate the absolute value of the difference in the magnetic moment $M_{\nu}$ for each superposition and compare it to the corresponding coherence time. Here, the magnetic moment of a state is defined as the negative expectation value of the $\op{S}_z$ operator acting on the system without bath times the gyromagnetic factor $g=2$ times the Bohr magneton $\mu_B$. 
Figure \xref{slope}~(top) shows that only states with small differences in magnetic moment can form superpositions expressing long coherence times. 
However, this property of clock transitions is not sufficient to predict well-performing superpositions 
as can be deduced from \figref{slope}~(bottom). 
There must be additional factors at play in order to explain why some superpositions 
perform poorly despite $|M_p-M_n|$ being very small.

% \comment{Alles folgende weg bis I do not ...}
% As a sidenote, this cannot be explained by the second derivative of the energy difference with respect to the magnetic field either. When considering the second derivative instead of the first and making a plot like the one shown in \figref{slope}~(bottom), they look almost exactly the same for the first and second derivatives. The only real difference is a scaling of the $x$-axis by a factor of 0.05 which is the strength of $B_z$. This can easily be explained by the observation that the superpositions shown in \figref{slope}~(bottom) almost all have $M_p^0-M_n^0=0$ for $B=0$. Then, choosing $B_z$ so small that a linear approximation holds and with 
% \begin{align}
% \Delta M_i \coloneqq M_i^{B_z}-M_i^{0}\ \text{for}\ i \in \{n,p\}\ ,
% \end{align}
% such a linear approximation gives
% \begin{align}
% \frac{\Delta M_p}{\Delta B_z}-\frac{\Delta M_n}{\Delta B_z} &\approx \frac{\partial M_p^0}{\partial B_z}-\frac{\partial M_n^0}{\partial B_z}\\
% \Leftrightarrow \frac{\Delta M_p-\Delta M_n}{\Delta B_z} &\approx \frac{\partial (M_p^0-M_n^0)}{\partial B_z}\\
% \Leftrightarrow M_p^{B_z}-M_n^{B_z}&\approx B_z\cdot\frac{\partial (M_p^0-M_n^0)}{\partial B_z}\ .
% \end{align}
% This means that the first and second derivatives of the energy differences yield almost the same information and are just scaled by $B_z$ as claimed above.
% \comment{I do not understand this remark. $\rightarrow$ Clearer now?}

%NEW
As a sidenote, we checked that this cannot be explained by the second derivative of the energy difference 
with respect to the magnetic field either. Rather, it is probably related to the way a certain clock transition couples
to the decohering bath.

\begin{figure}[ht!]
\centering
\includegraphics[width=0.675\columnwidth]{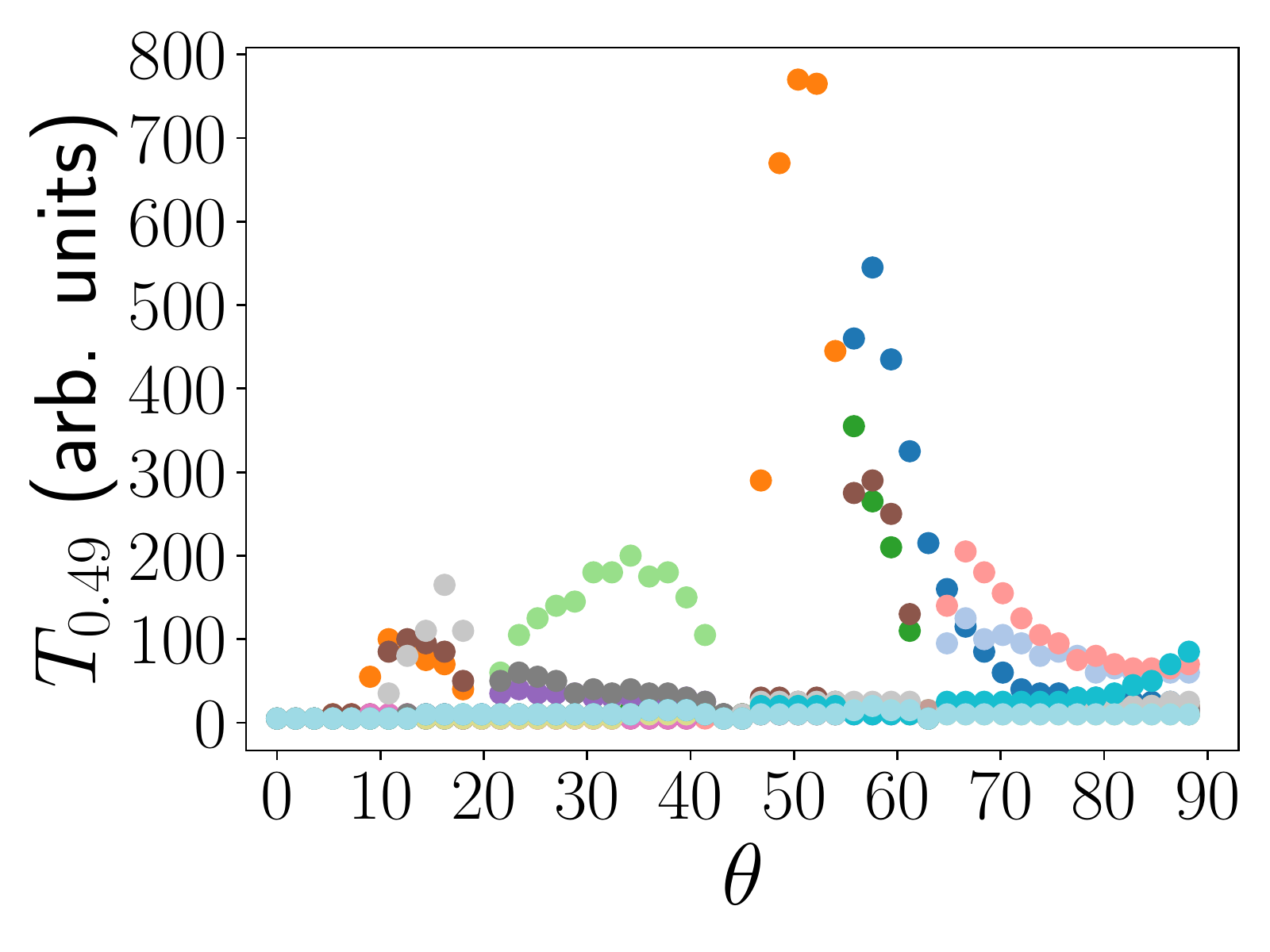}
\includegraphics[width=0.675\columnwidth]{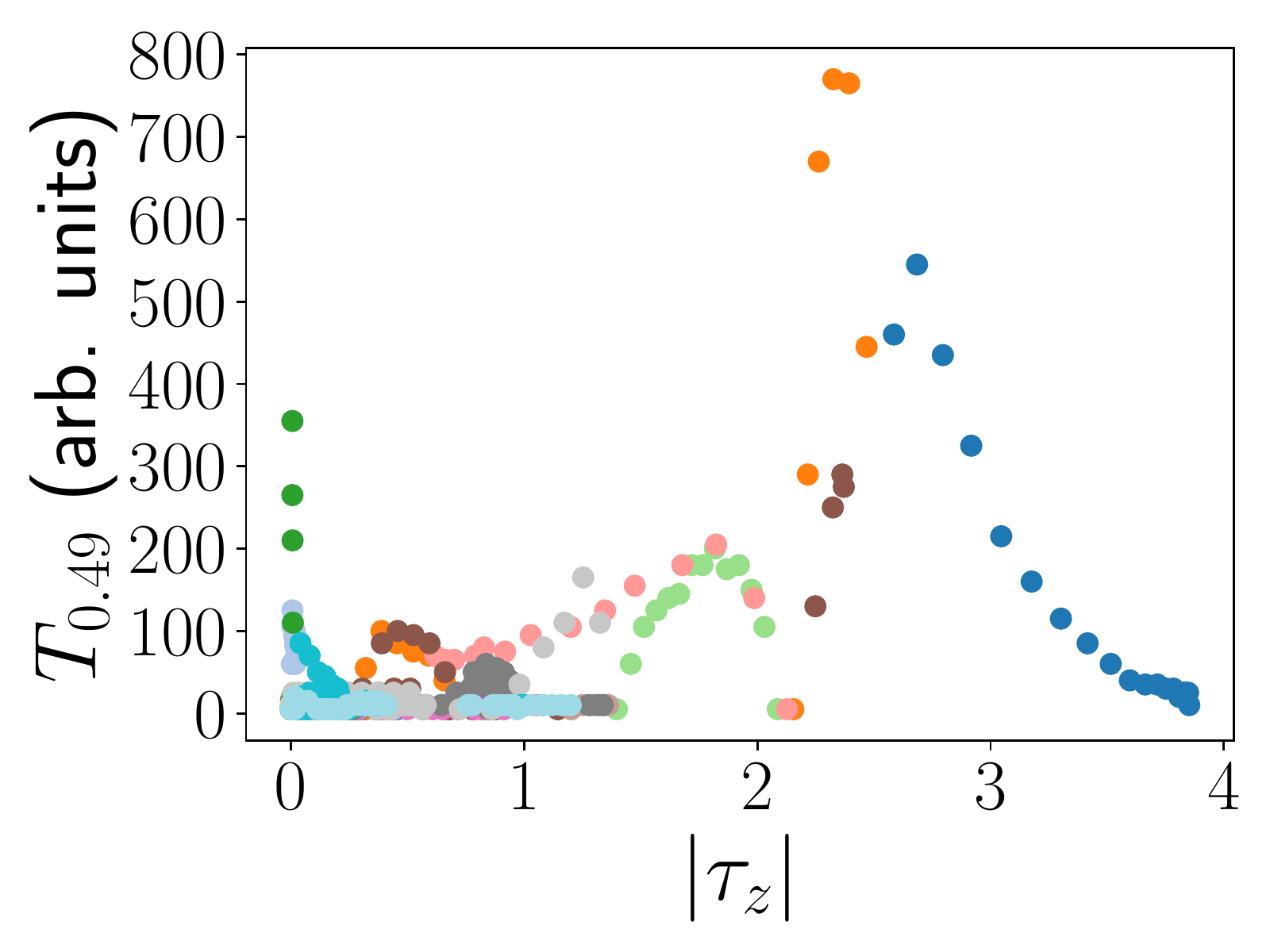}
\caption{Coherence times vs$.$ tilting angle (top) and absolute value of the $z$-component 
of the toroidal moment of superpositions (bottom, see \eqref{tau}) for 50 values of $\theta$ between $0^{\circ}$ and $88.2^{\circ}$ at $B_z=0.05$~T for a random bath
with $n_{\text{bath}}=8$. Best coherence times are observed for mid-sized angles and 
toroidal moments. Legend is displayed in \figref{legend}.
There is no one-to-one relation between $\theta$ and the expectation value of the toroidal moment;
the latter may simply vanish for certain energy eigenstates or their superpositions even if $\theta\neq 0$.} 
\label{anglevsbreak}
\end{figure}

Figure \xref{anglevsbreak}~(top) provides an impression that systems with strong SMM ($\theta\approx 0^{\circ}$) or toroidal ($\theta\approx 90^{\circ}$) orientation of the anisotropy axes both do not contain low-lying states to form 
long-living clock transitions. Rather, mid-sized angles seem to be most promising for the given parameter configuration.
This is a strong indication against the simple idea that superpositions with a larger toroidal moment 
should display longer coherence times. 
And, \emph{vice versa}, there is also no indication of superpositions of states forming a clock transition 
having large toroidal moments which is, although not sufficient on its own, 
obviously the deciding factor for long coherence times.

\begin{figure}[ht!]
\centering
\includegraphics[width=0.675\columnwidth]{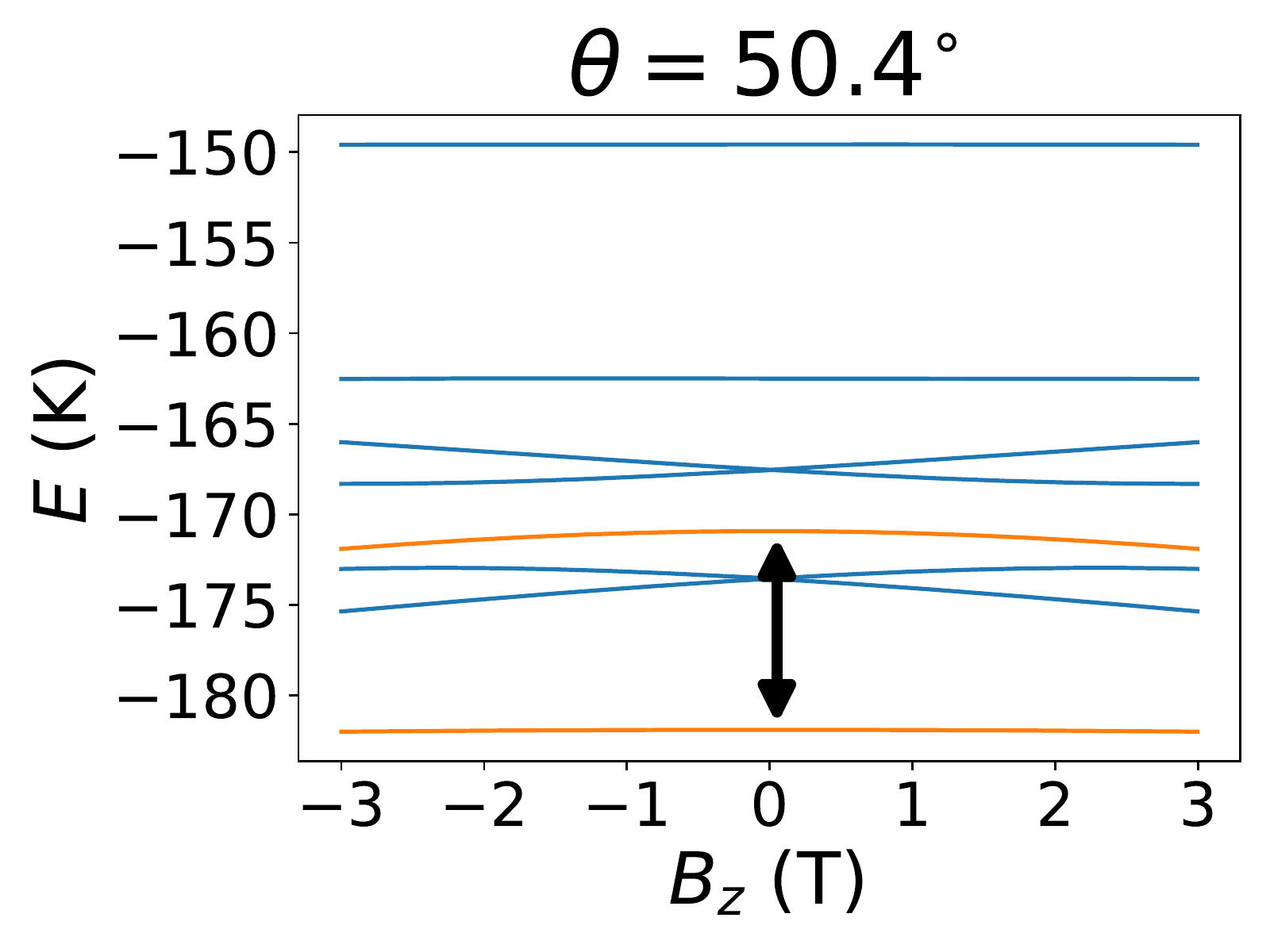}
\caption{Zeeman diagram of the six lowest-lying states of the system without bath for $\theta=50.4^{\circ}$. 
The black arrow indicates the transition with the longest coherence time 
(see also \figref{diff_seeds}) 
made up of the two orange states.} 
\label{zeeman_22} 
\end{figure}

\section{Dependence of coherence times on the placement and number of bath spins}
\label{sec4}

The best performing superposition of the system with the parameters introduced above 
is found numerically around a tilting angle of $\theta=50.4^{\circ}$ for the superposition 
of the ground state and the third excited state. In the following, 
we use this as a sample system in order to illustrate our findings. 

Note that $\theta=50.4^{\circ}$ 
is not a universal ``magic angle" as the performance of superpositions is highly 
dependent on the parameter configuration and therefore the ideal angle changes 
when altering the magnitude of parameters such as $D$ and $J$ as will be shown in Sec.~\xref{sec5}. 
Figure~\xref{zeeman_22} shows the Zeeman diagram of the system without bath for 
$\theta=50.4^{\circ}$ together with the best-performing transition.

\begin{figure}[ht!]
\centering
\includegraphics[width=0.48\columnwidth]{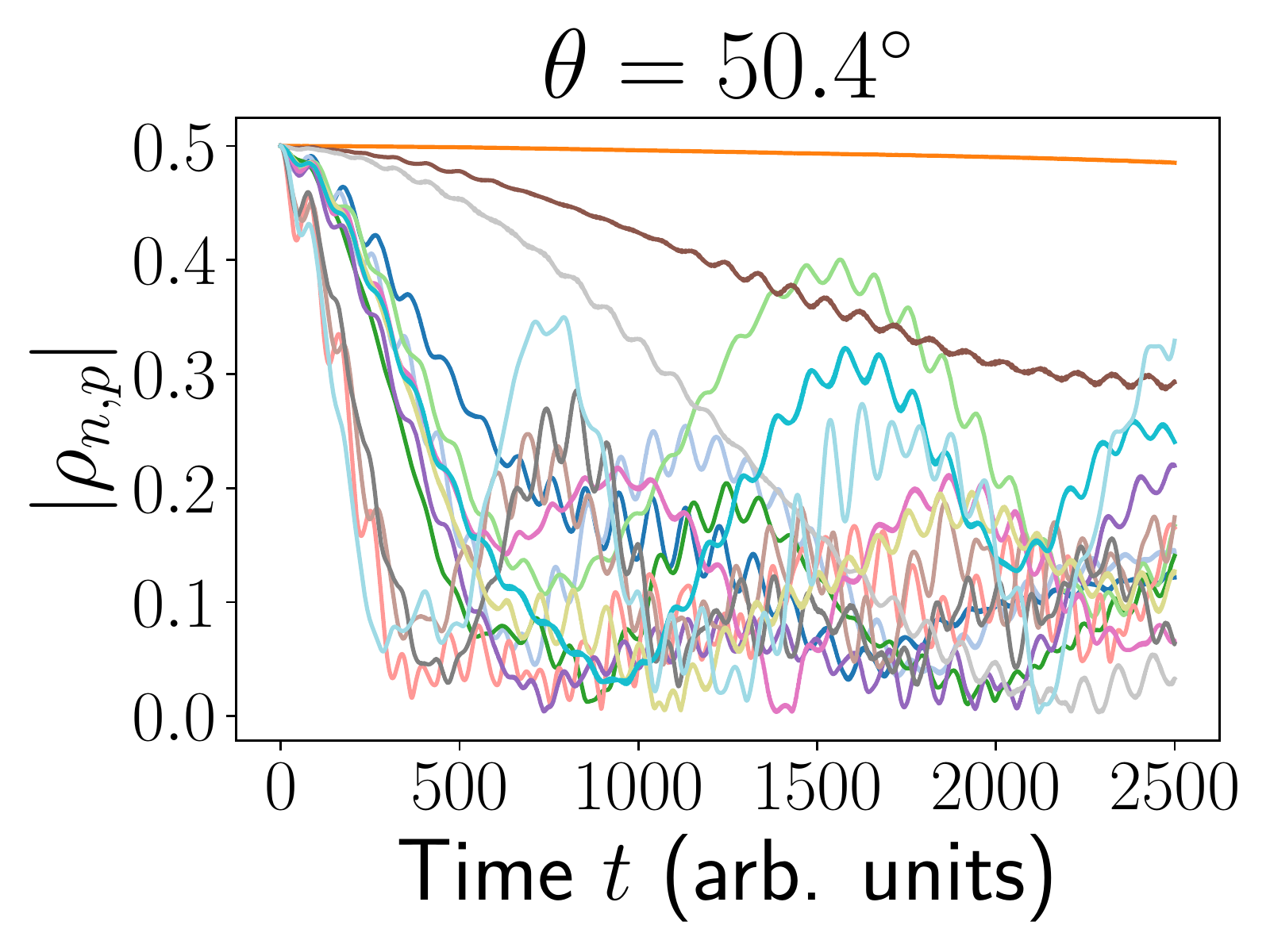}
\includegraphics[width=0.48\columnwidth]{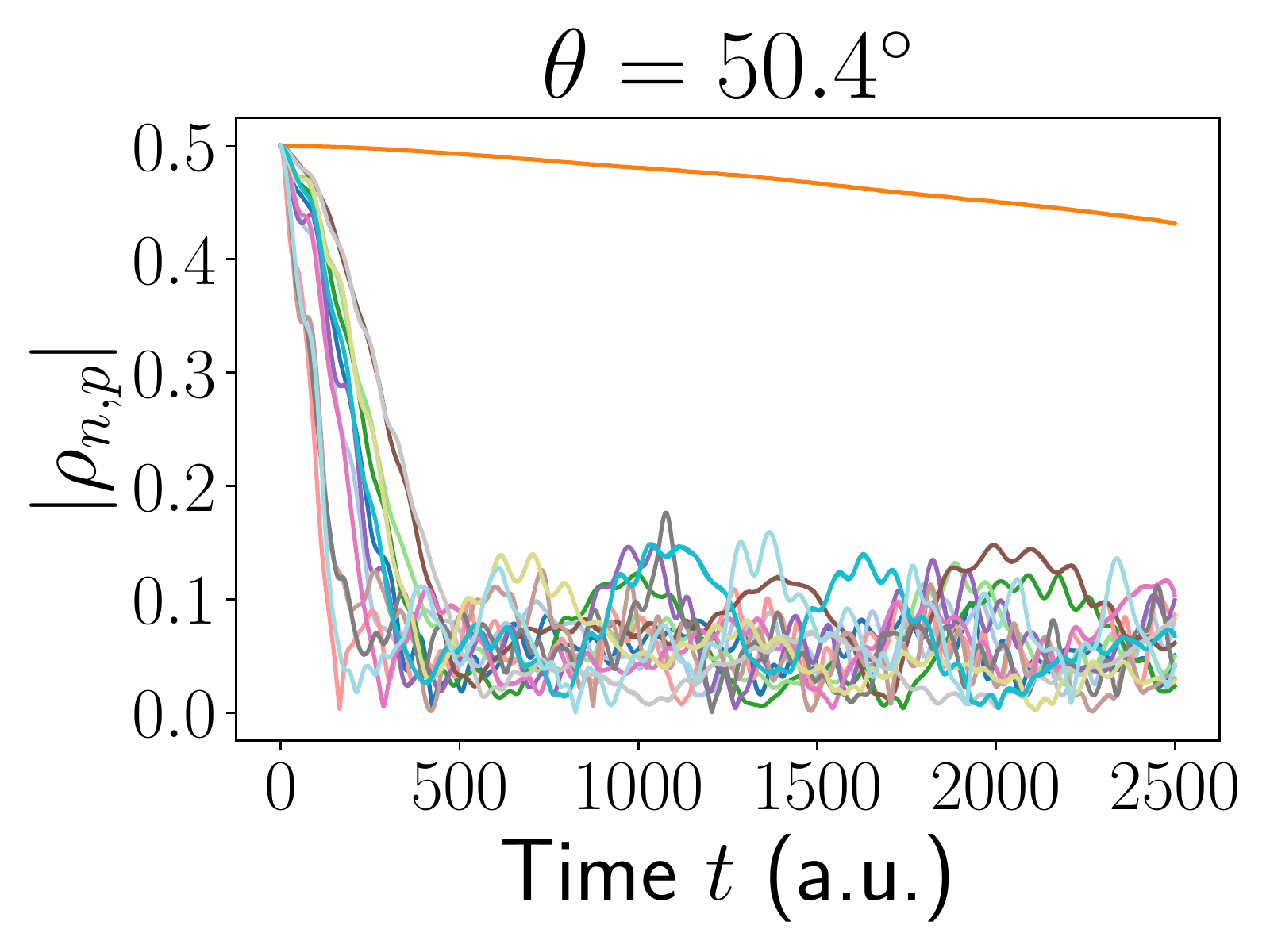}
\includegraphics[width=0.48\columnwidth]{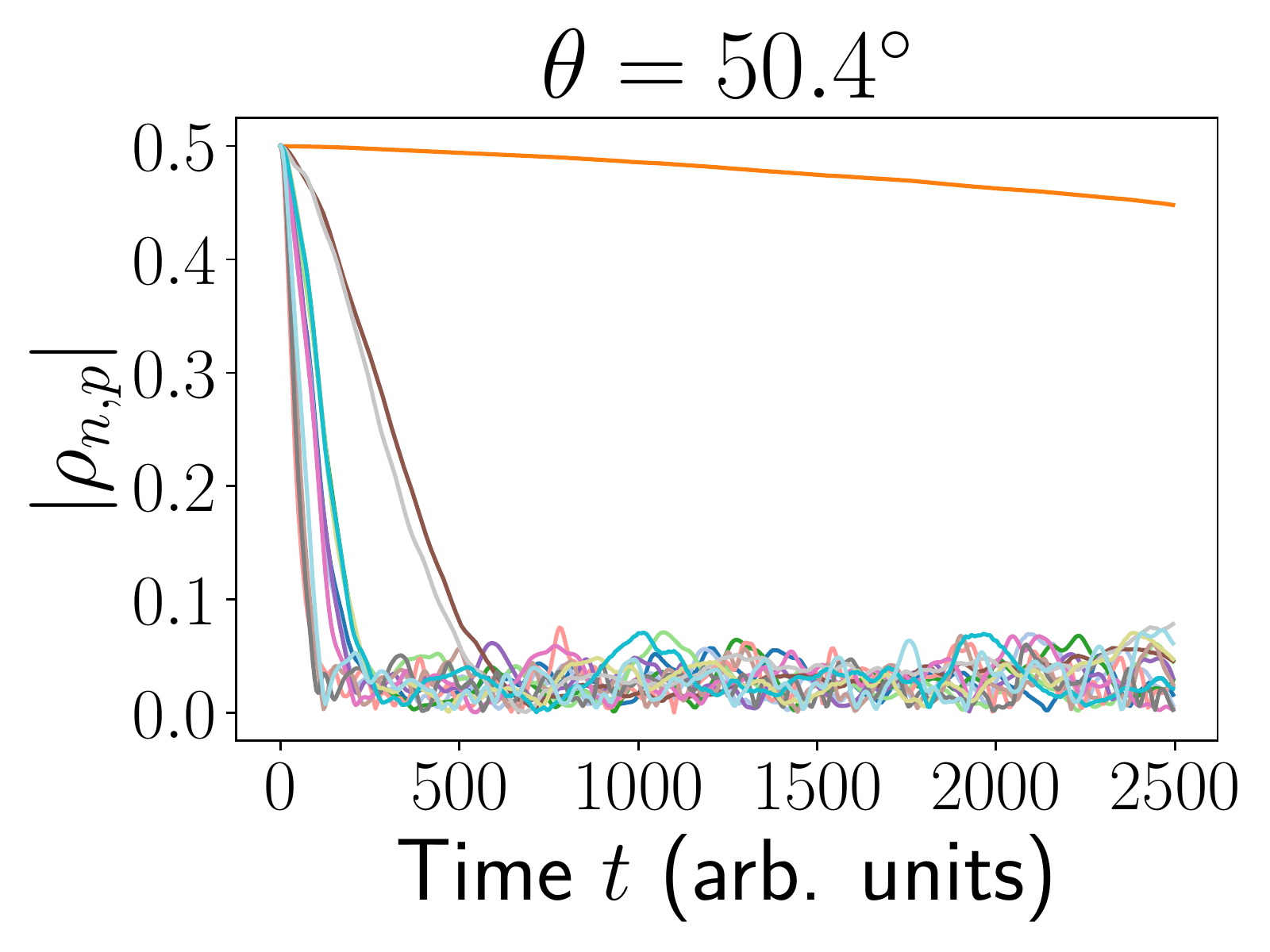}
\includegraphics[width=0.48\columnwidth]{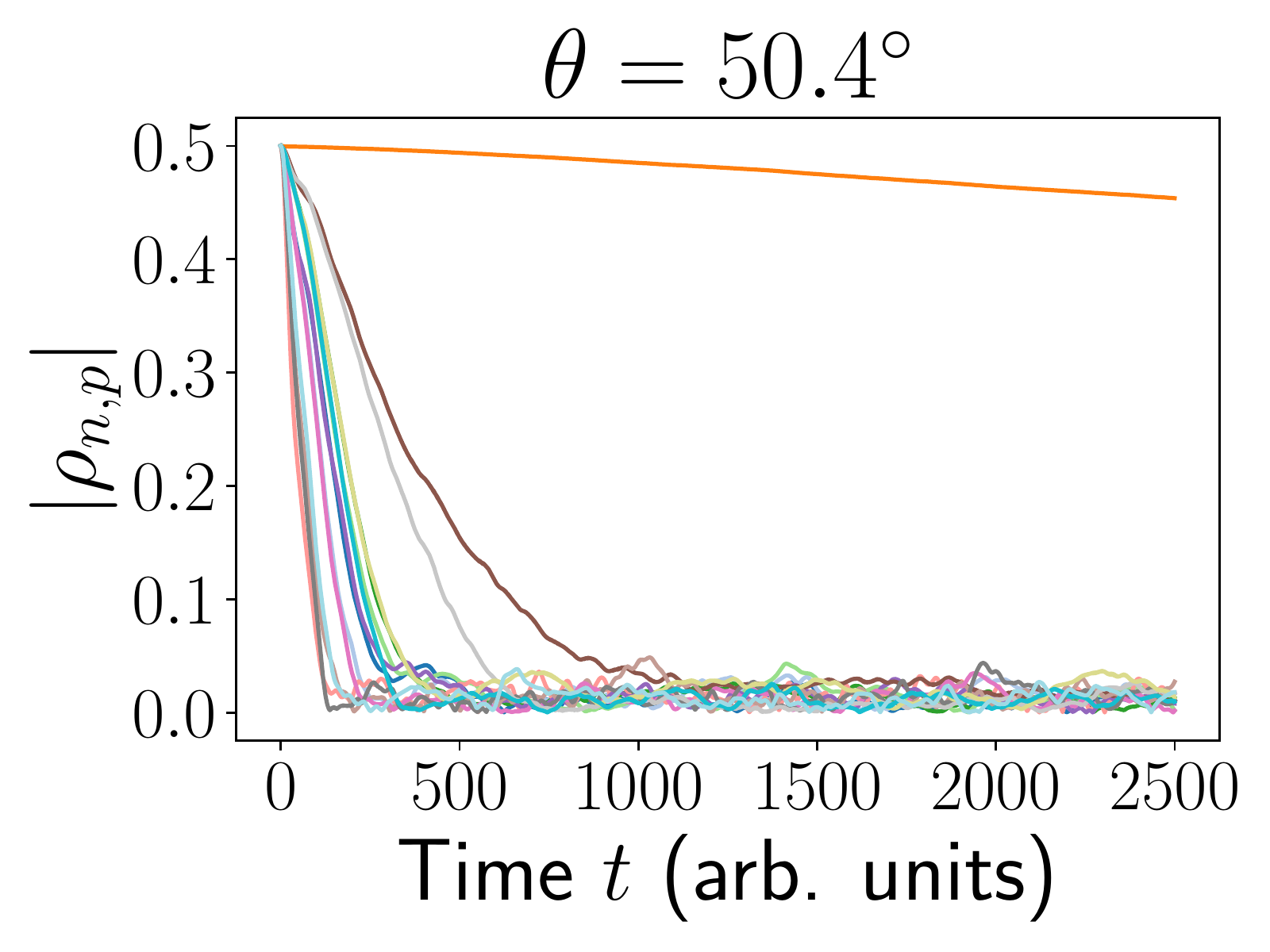}
\caption{Decoherence over time of all two-state superpositions of the six lowest-lying energy eigenstates for $n_{\text{bath}}=4$ (top left), $n_{\text{bath}}=6$ (top right), $n_{\text{bath}}=8$ (bottom left), $n_{\text{bath}}=10$ (bottom right), $B_z=0.05$, and $\theta=50.4^{\circ}$. Legend is displayed in \figref{legend}.} 
\label{diff_env} 
\end{figure}

\begin{figure}[ht!]
\centering
\includegraphics[width=0.48\columnwidth]{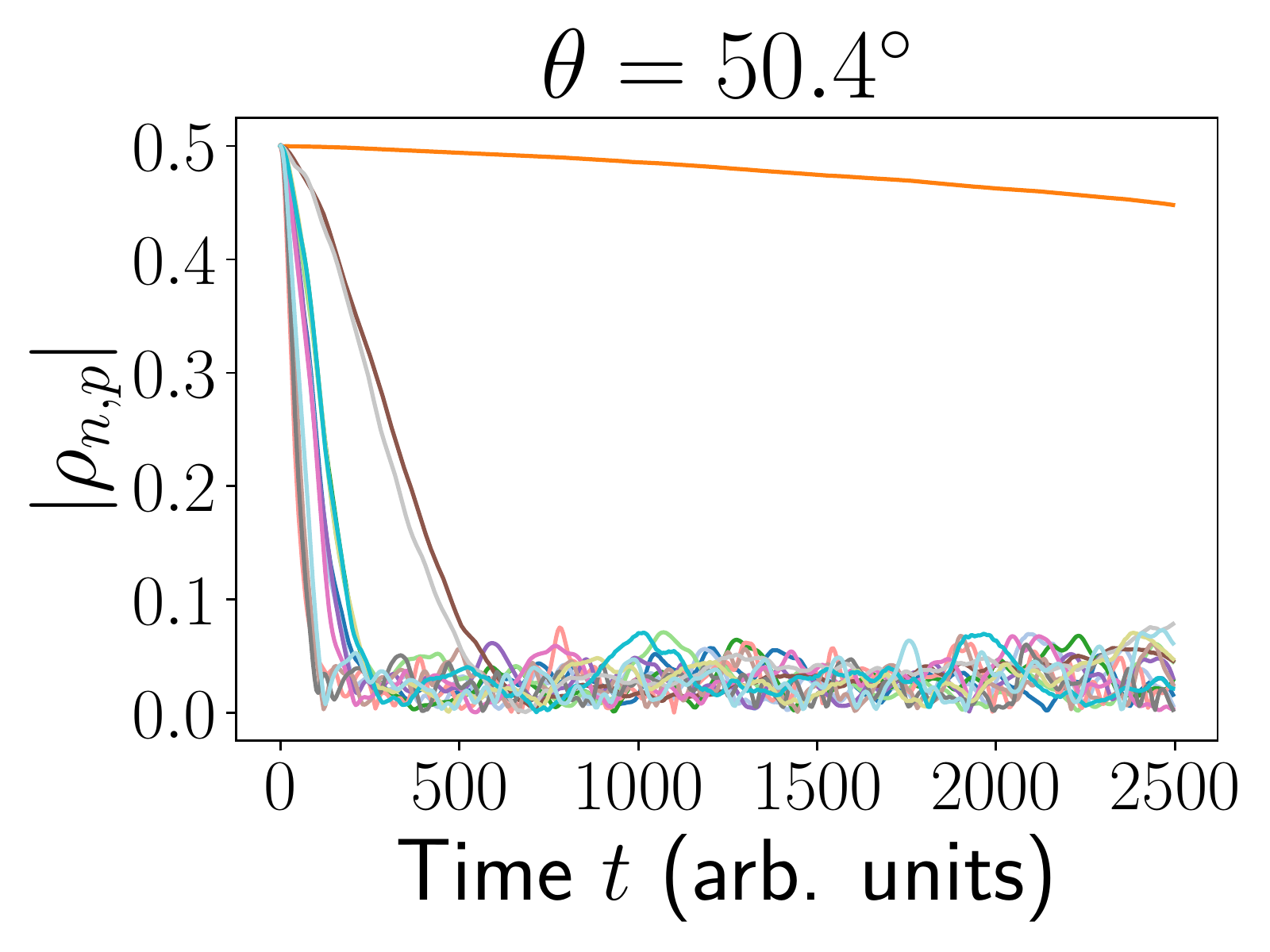}
\includegraphics[width=0.48\columnwidth]{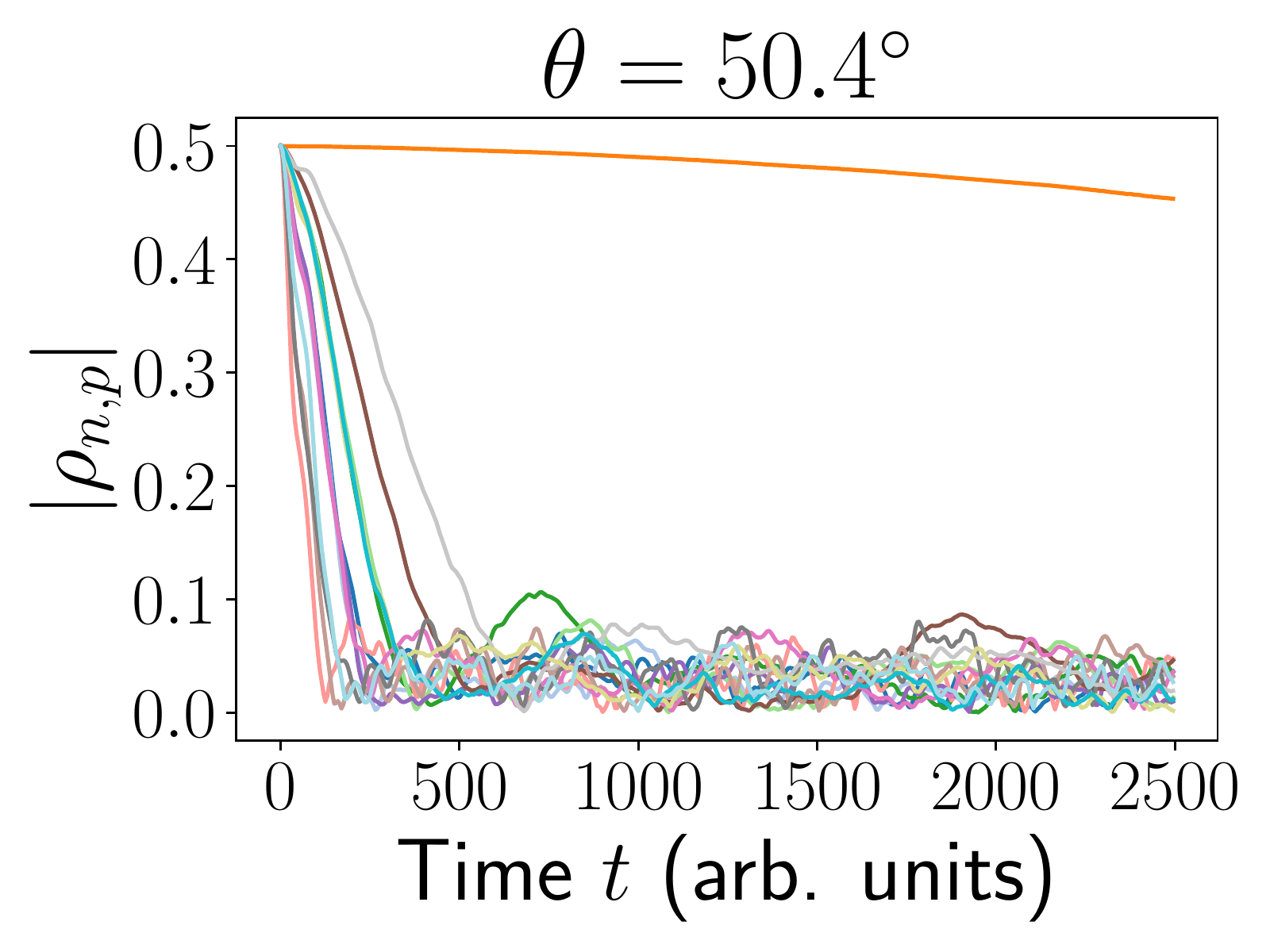}
\includegraphics[width=0.48\columnwidth]{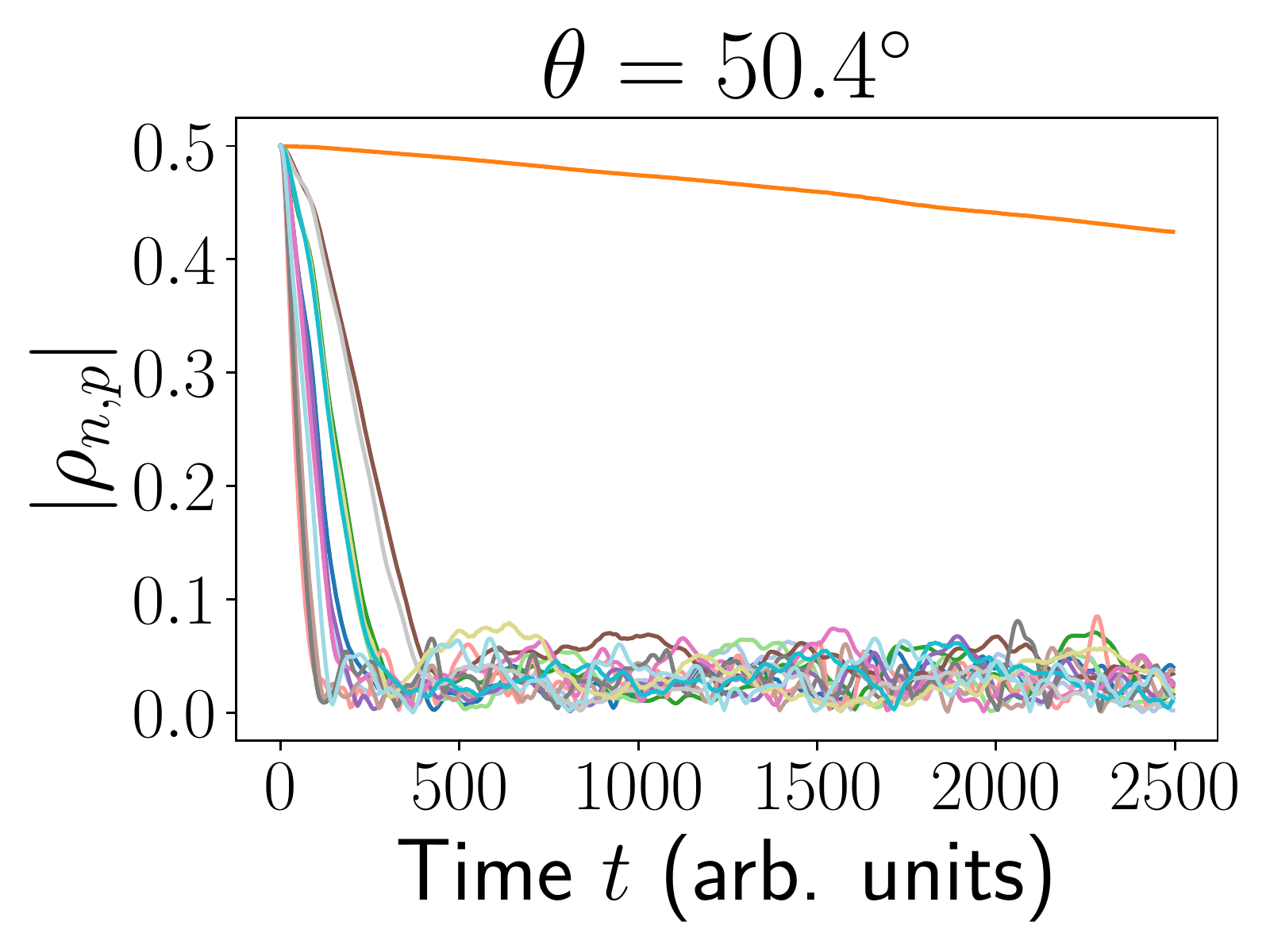}
\includegraphics[width=0.48\columnwidth]{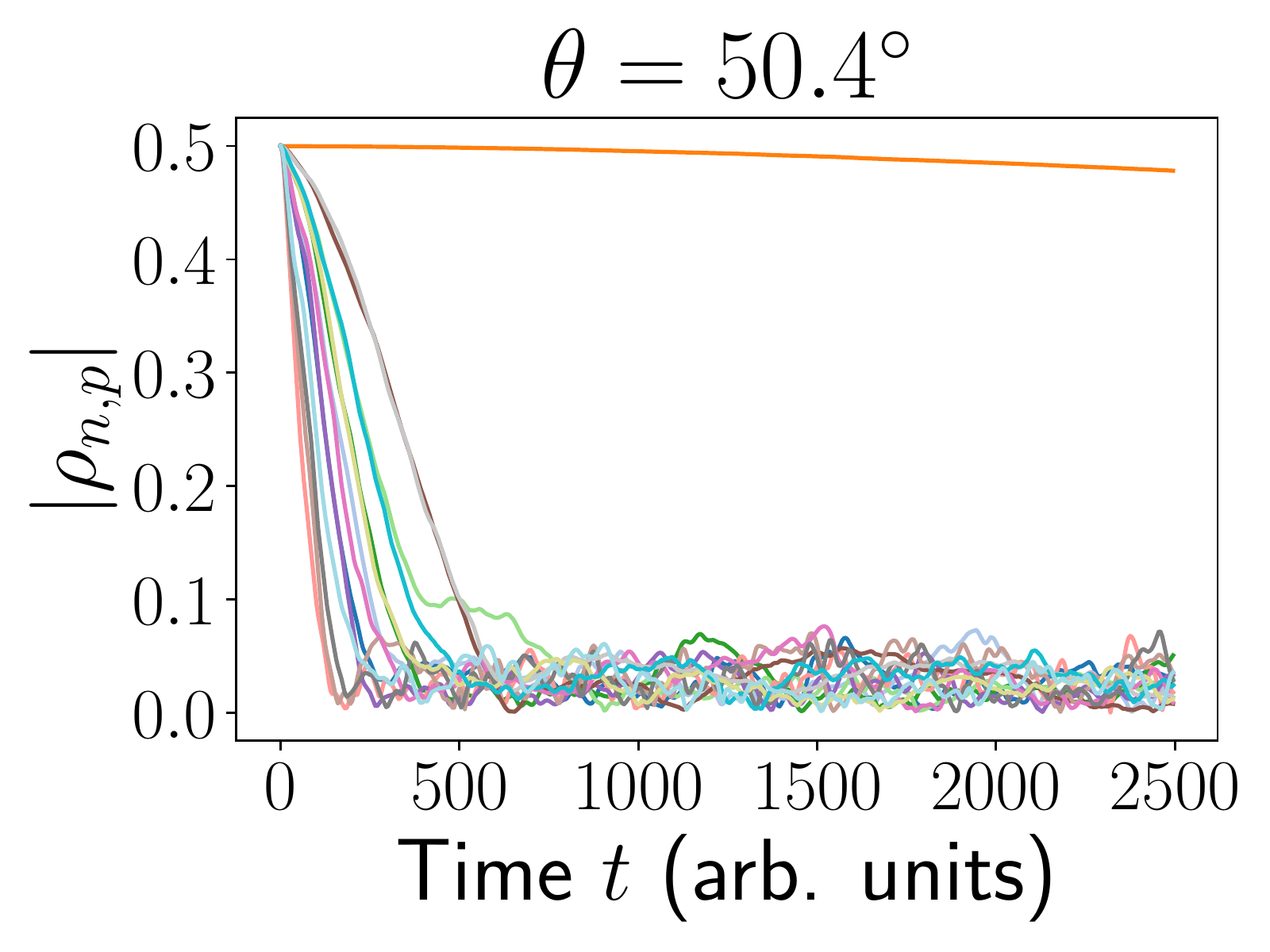}
\includegraphics[width=0.48\columnwidth]{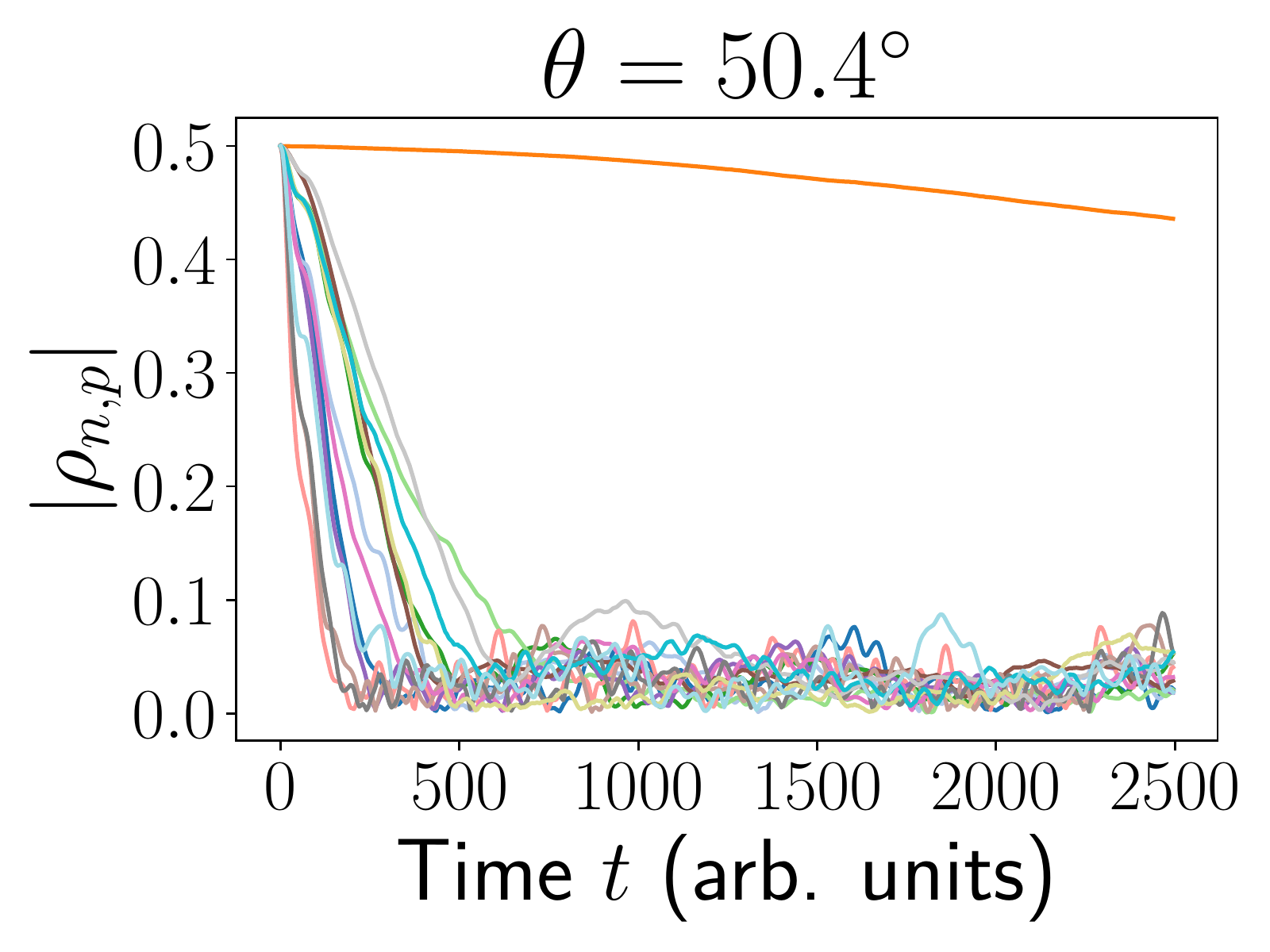}
\includegraphics[width=0.48\columnwidth]{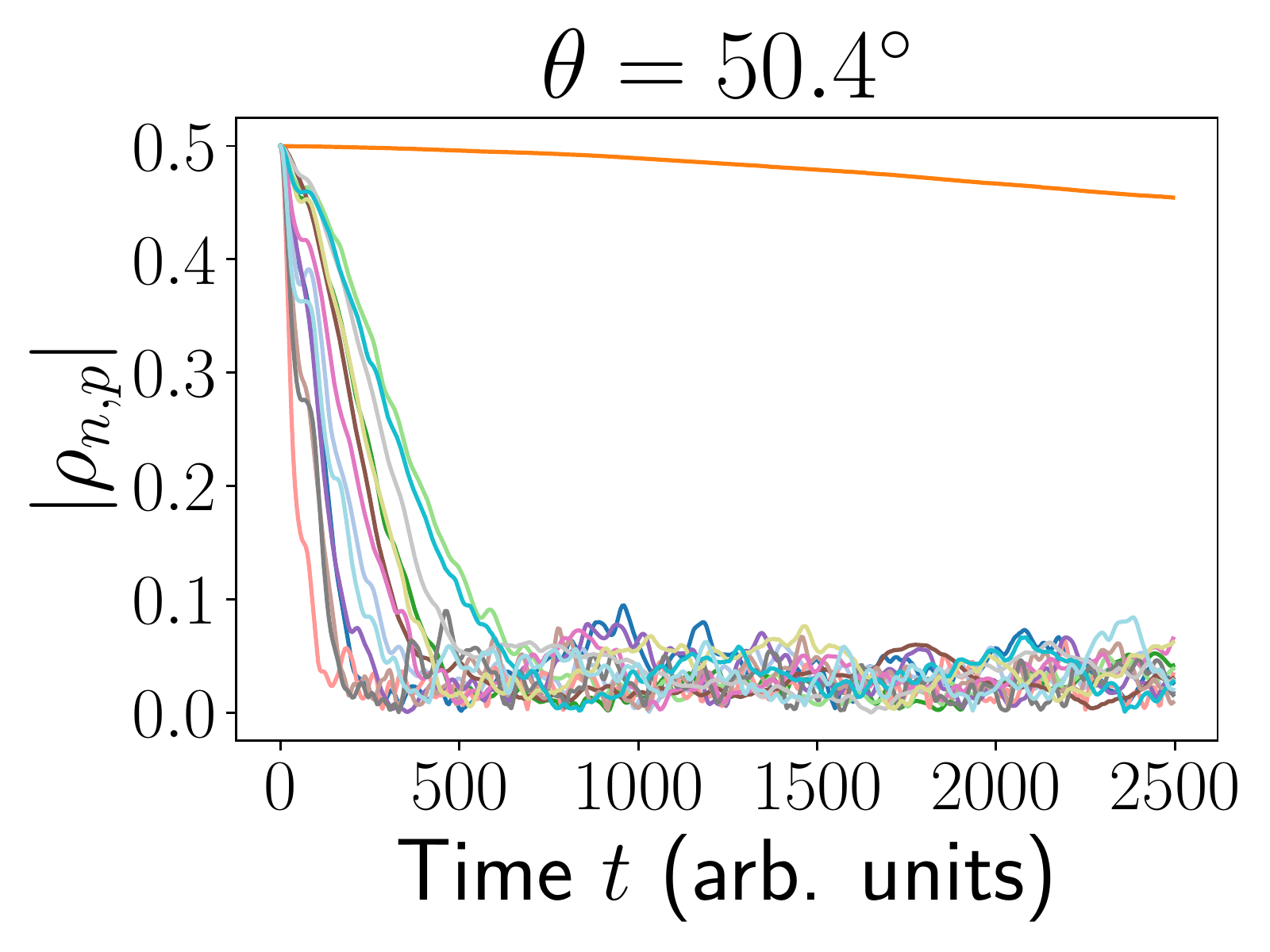}
\includegraphics[width=0.48\columnwidth]{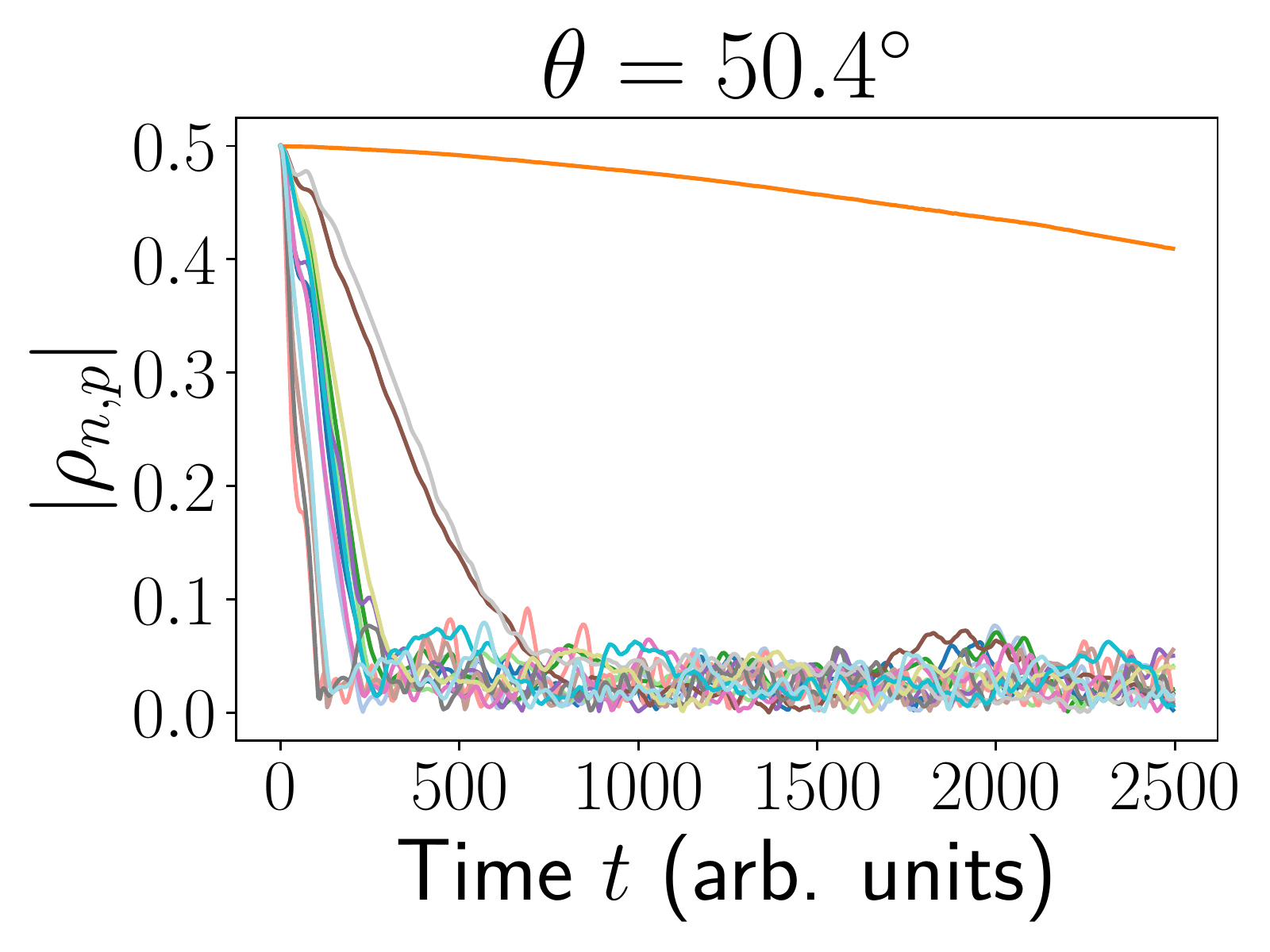}
\includegraphics[width=0.48\columnwidth]{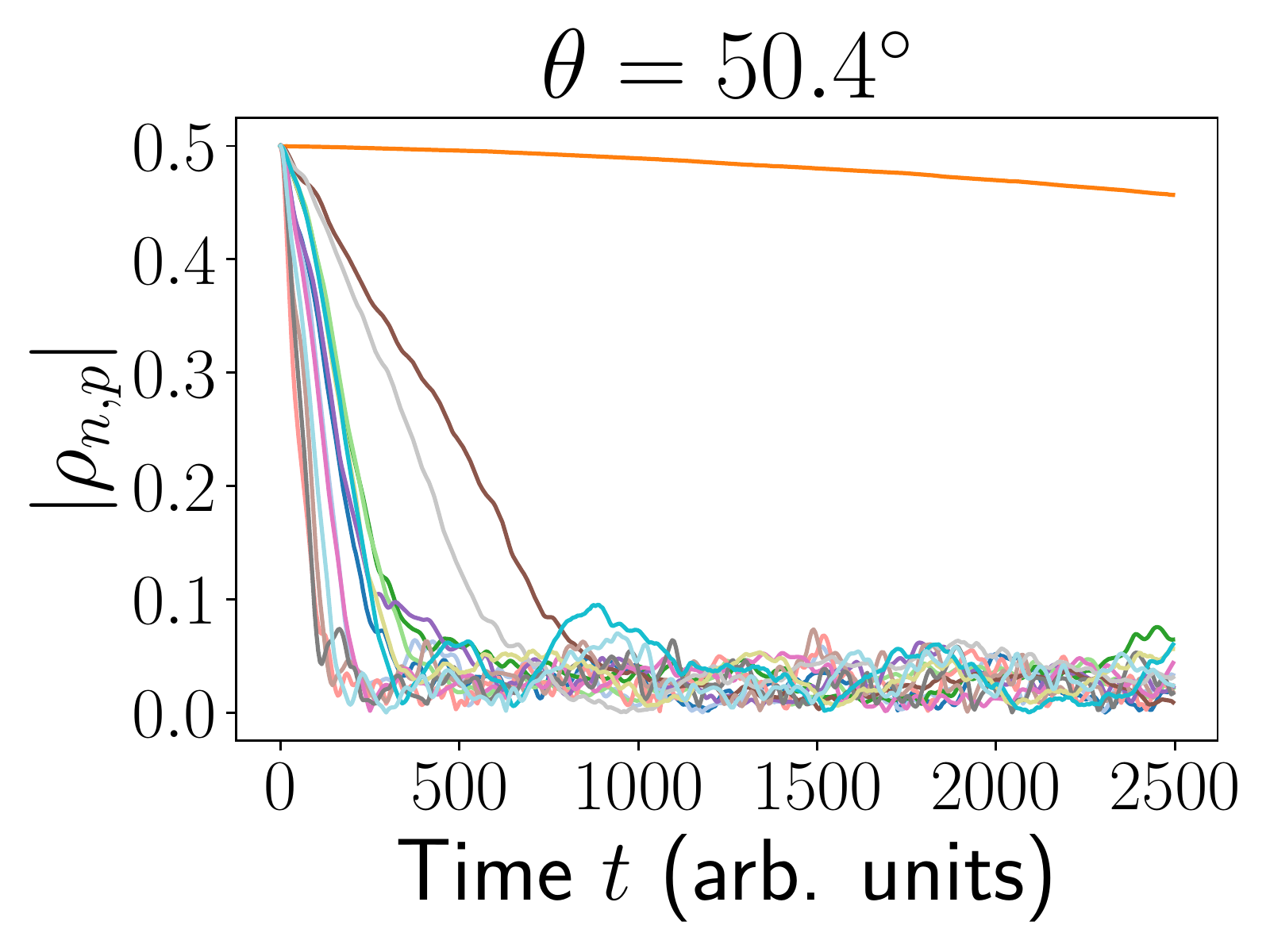}
\includegraphics[width=0.48\columnwidth]{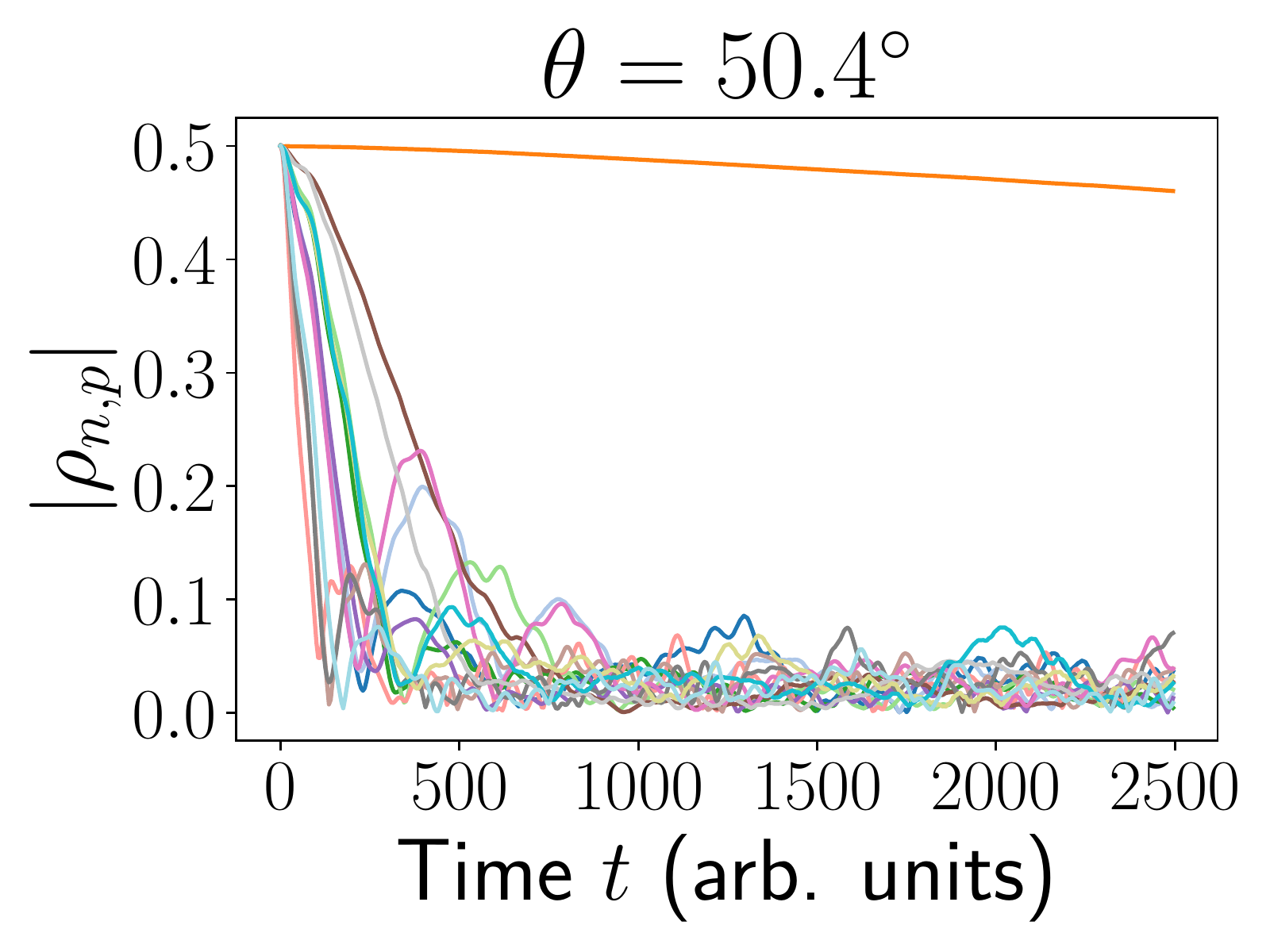}
\includegraphics[width=0.48\columnwidth]{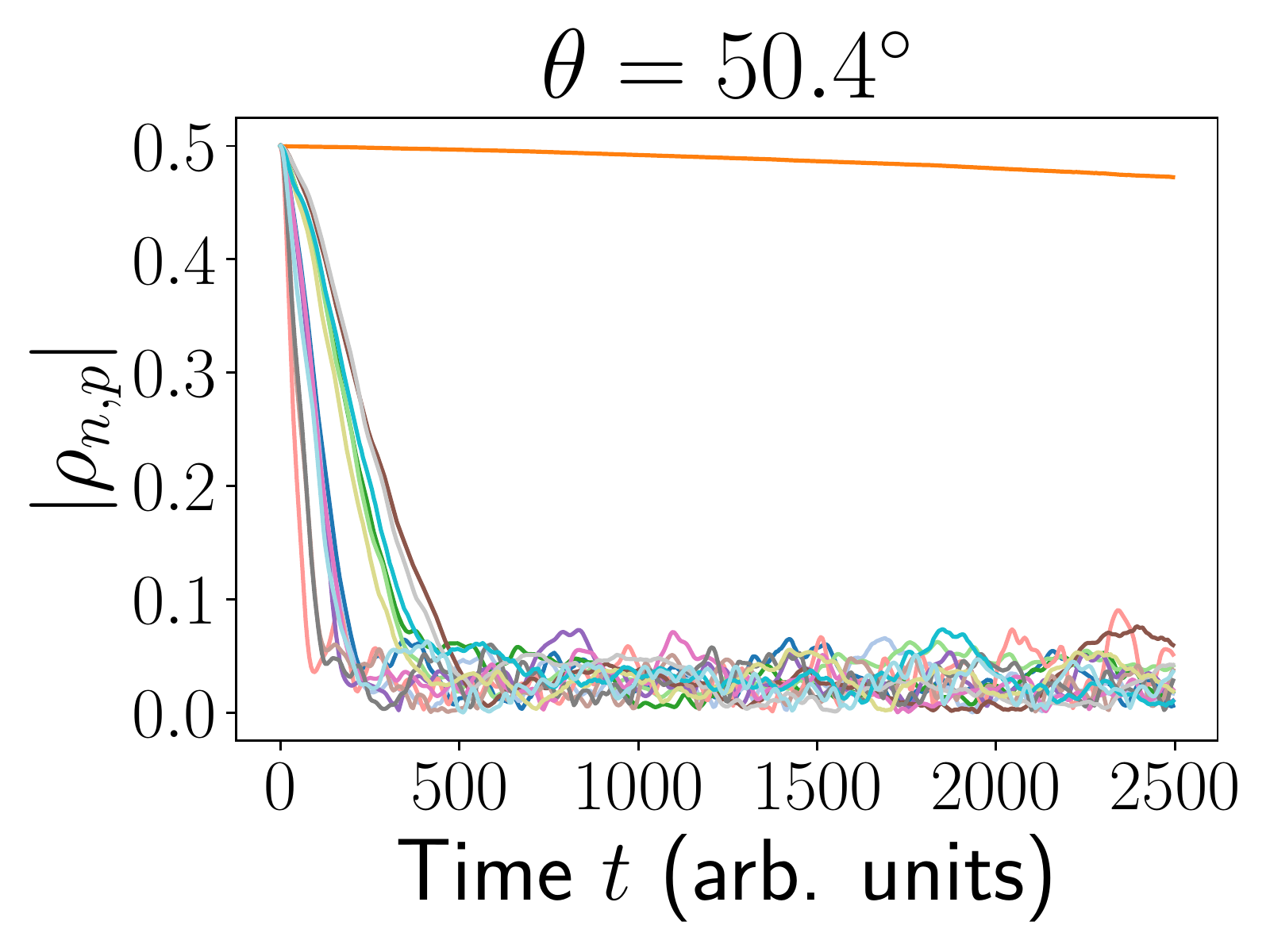}
\caption{Decoherence over time of all two-state superpositions of the six lowest-lying energy eigenstates for 10 different random baths with $n_{\text{bath}}=8$ at $B_z=0.05$ and $\theta=50.4^{\circ}$. Legend is displayed in \figref{legend}.} 
\label{diff_seeds}
\end{figure}

In order to evaluate the effect of changes of the spin bath and to eliminate 
the possibility of accidentally choosing a non-typical bath, 
the decoherence calculations were repeated 
for baths with a different number of bath spins, see \figref{diff_env},
as well as for ten different sets 
of $n_{\text{bath}}=8$ randomly placed bath spins, see \figref{diff_seeds}. 
Both \figref{diff_env} and \figref{diff_seeds} show that while coherence times are 
somewhat dependent on the nature of the spin bath, the main qualitative findings 
regarding which superpositions show long coherence times and which do not are 
largely independent of the number and placement of bath spins.
In particular, the robustness of $1/\sqrt{2}(\psi_0+\psi_3)$ (orange curves) both 
against various arrangements of the bath spins and various sizes is astonishing 
as well as encouraging in view of future applications.

We have numerically verified that these results for a tilting angle of 
$\theta = 50.4^{\circ}$ are representative for other angles. 
For simplicity, most of the following calculations have been performed 
using only a single random bath. This is justified as the concrete values 
of the coherence times are of no particular relevance to our findings and 
the relative differences between superpositions are very similar across all random baths we used.

\section{Dependence of coherence times on the magnitude of $J$ and $D$}
\label{sec5}

\begin{figure}[ht!]
\centering
\includegraphics[width=0.48\columnwidth]{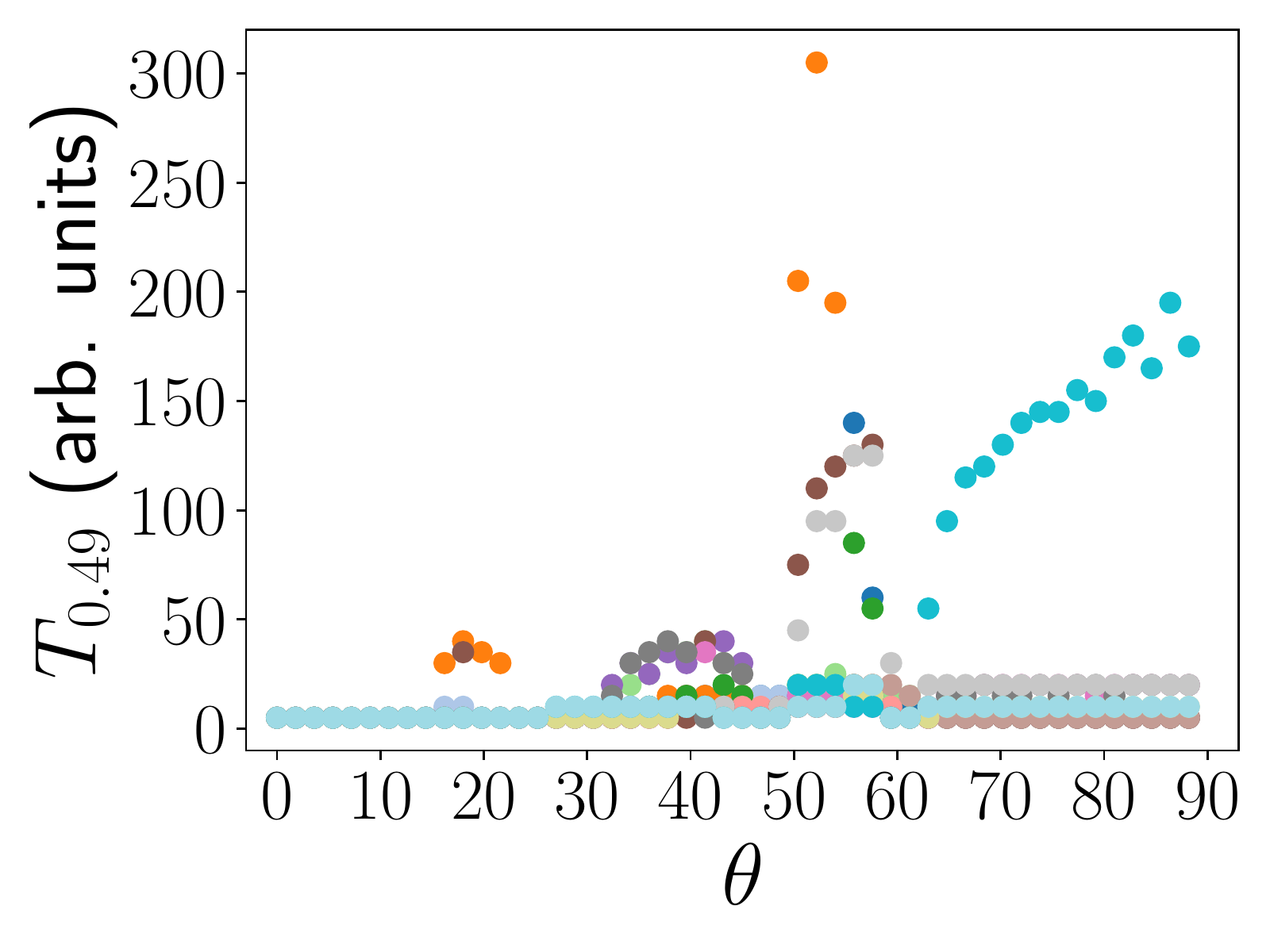}
\includegraphics[width=0.48\columnwidth]{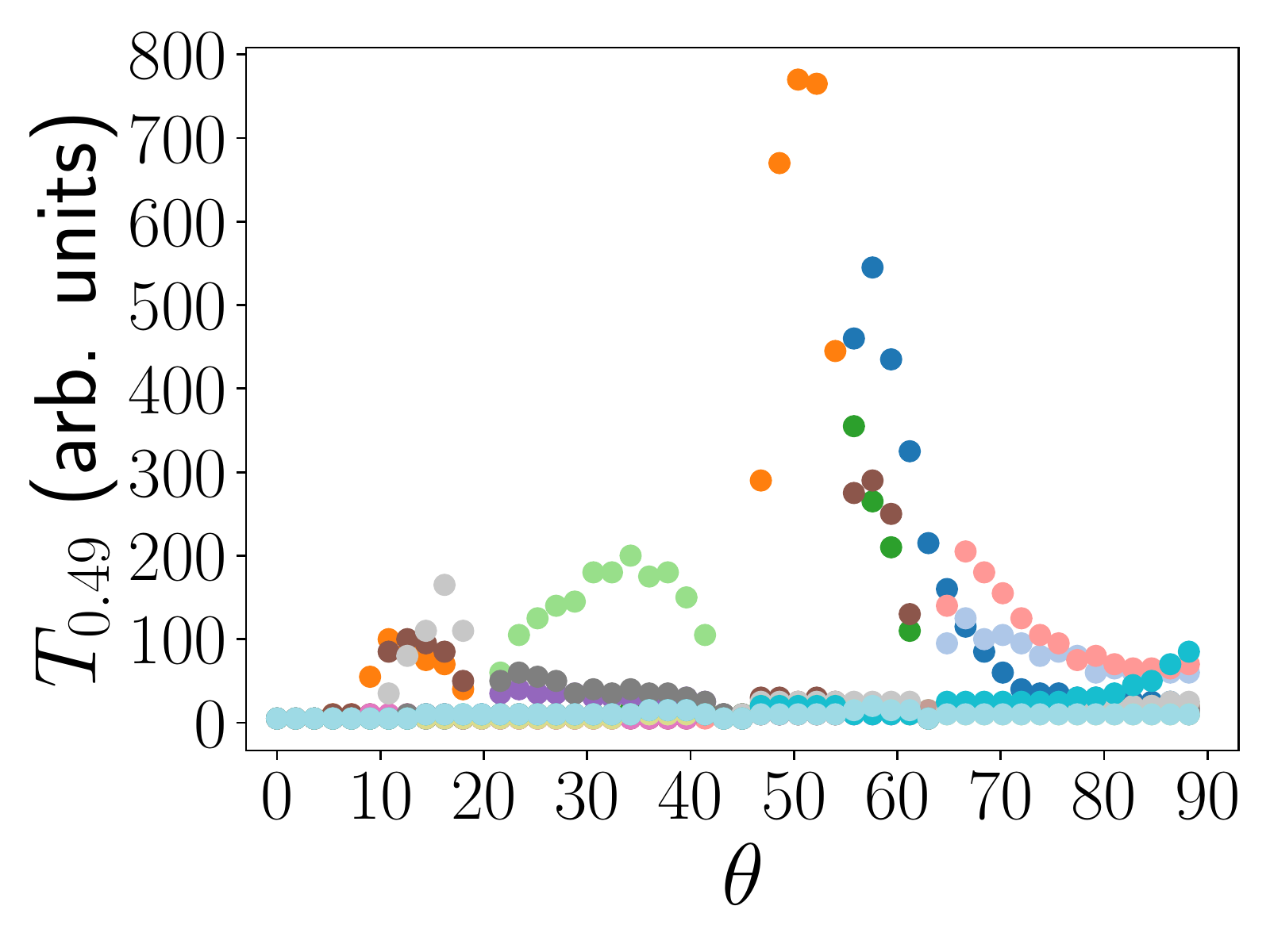}
\includegraphics[width=0.48\columnwidth]{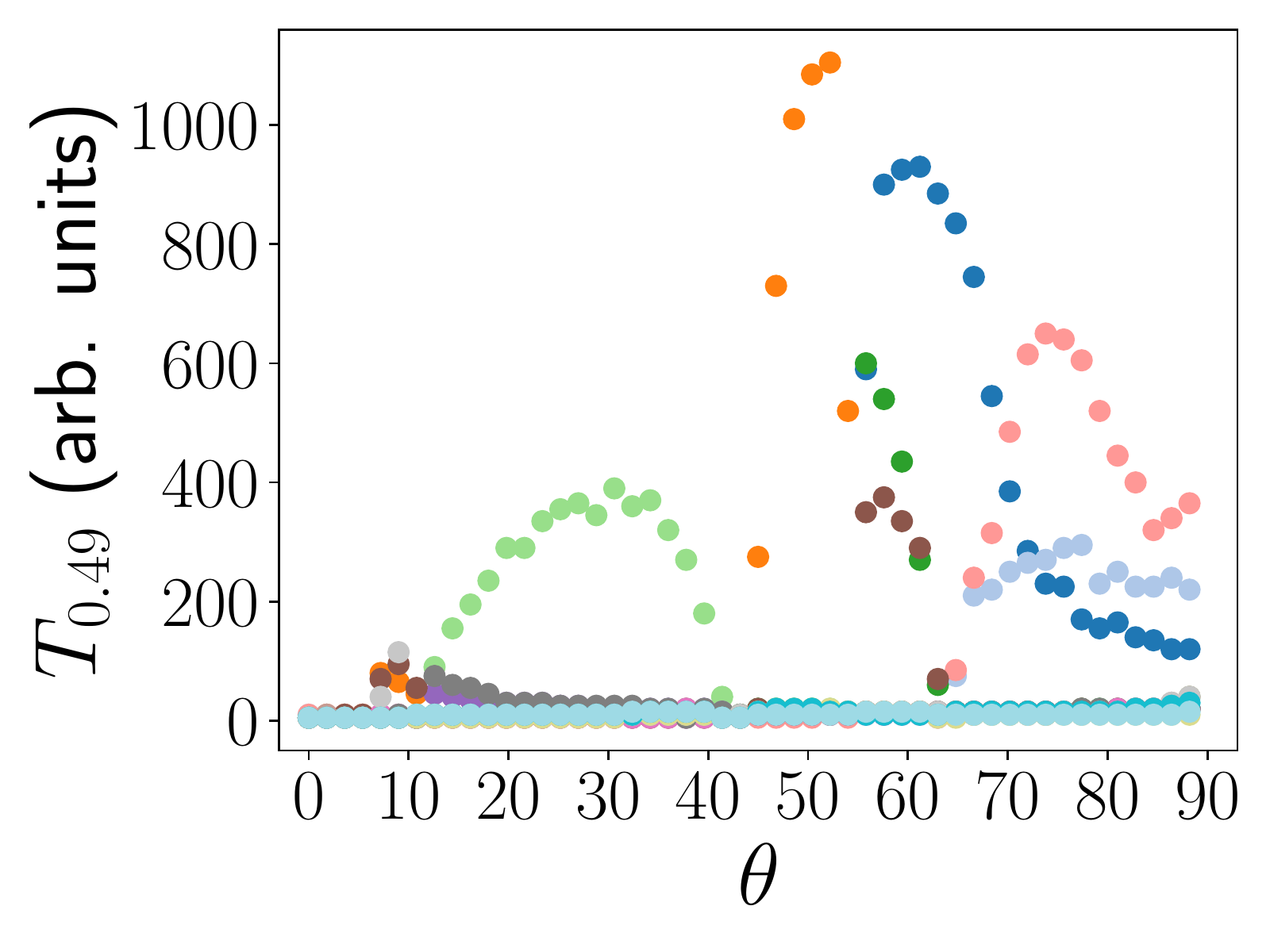}
\includegraphics[width=0.48\columnwidth]{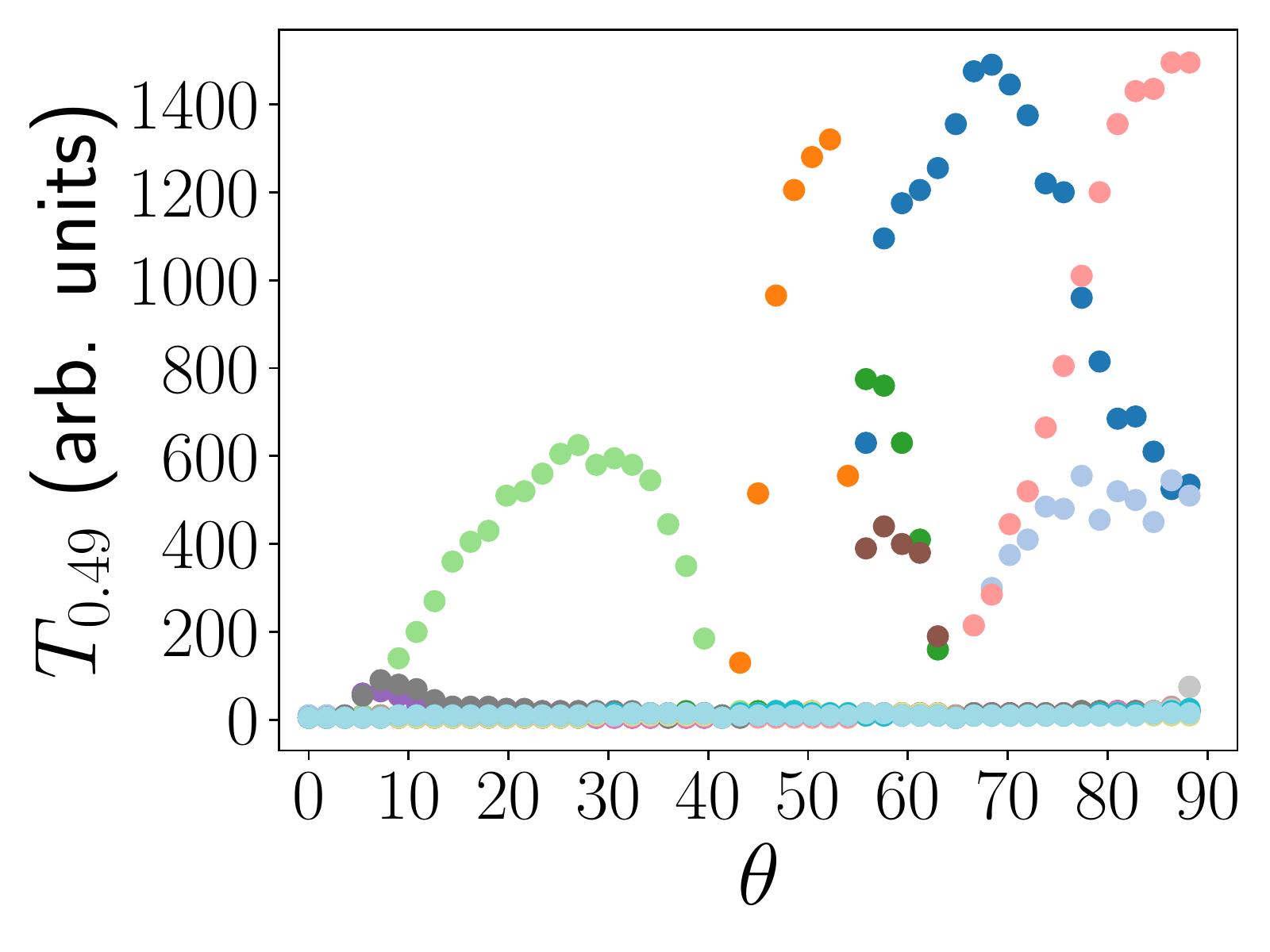}
\caption{Coherence times vs$.$ tilting angle for $D=-50$ K and (from top to bottom) $J=-5$ K, $J=-10$ K, $J=-15$ K, $J=-20$ K for 50 values of $\theta$ between $0^{\circ}$ and $88.2^{\circ}$ at $B_z=0.05$~T for a single bath with $n_{\text{bath}}=8$. Legend is displayed in \figref{legend}.} 
\label{angle_J}
\end{figure}

\begin{figure}[ht!]
\centering

\includegraphics[width=0.48\columnwidth]{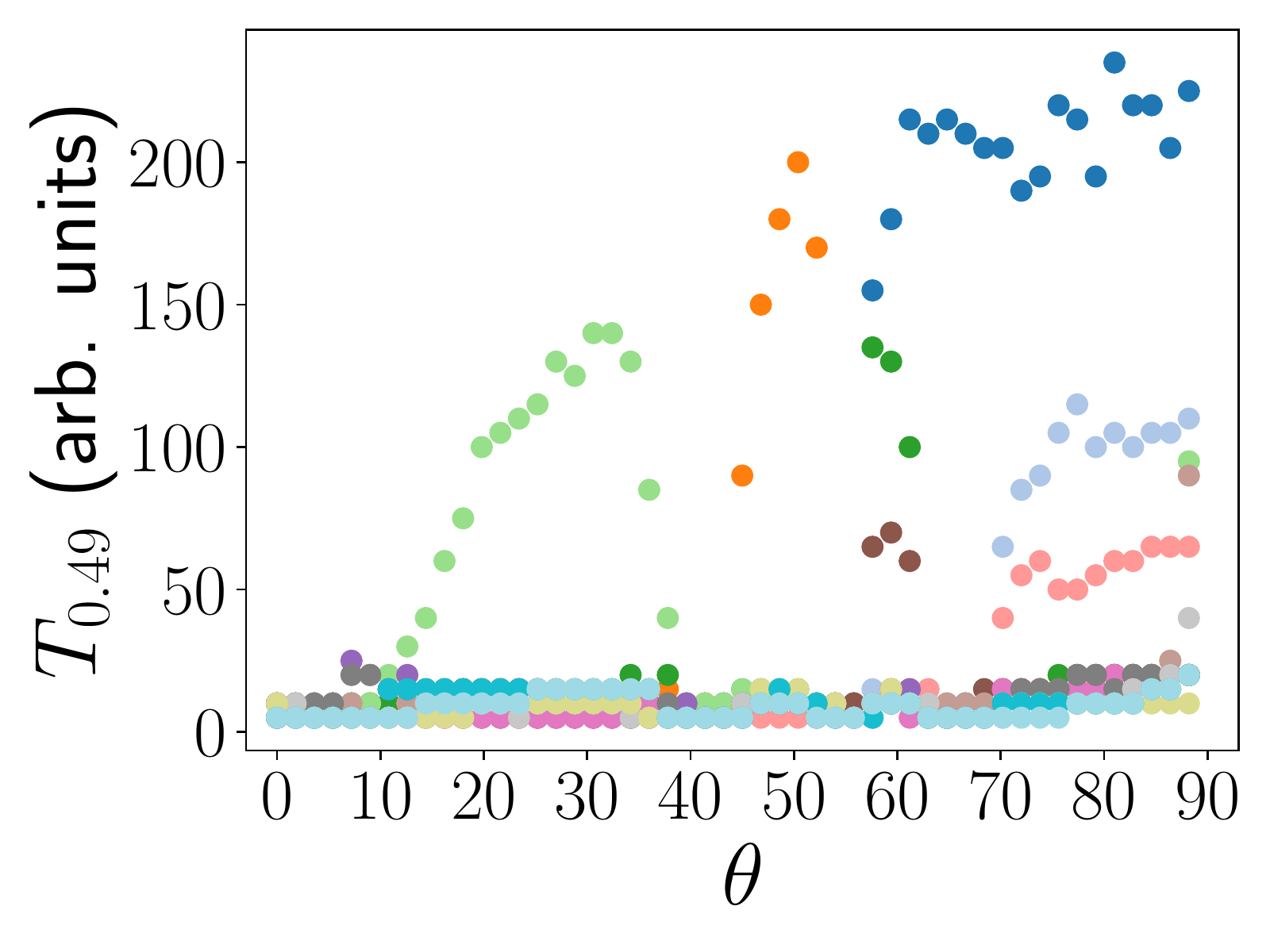}
\includegraphics[width=0.48\columnwidth]{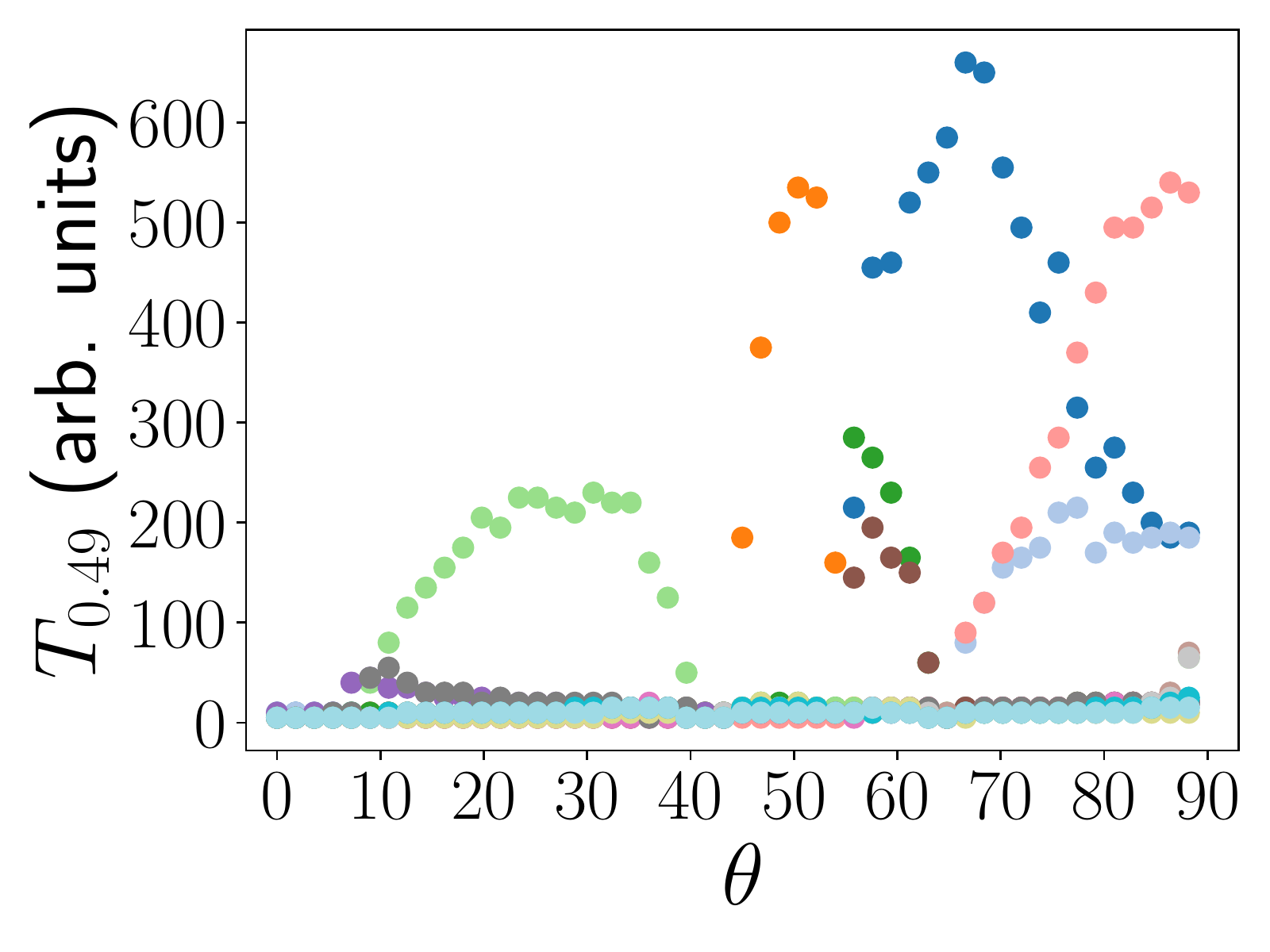}
\includegraphics[width=0.48\columnwidth]{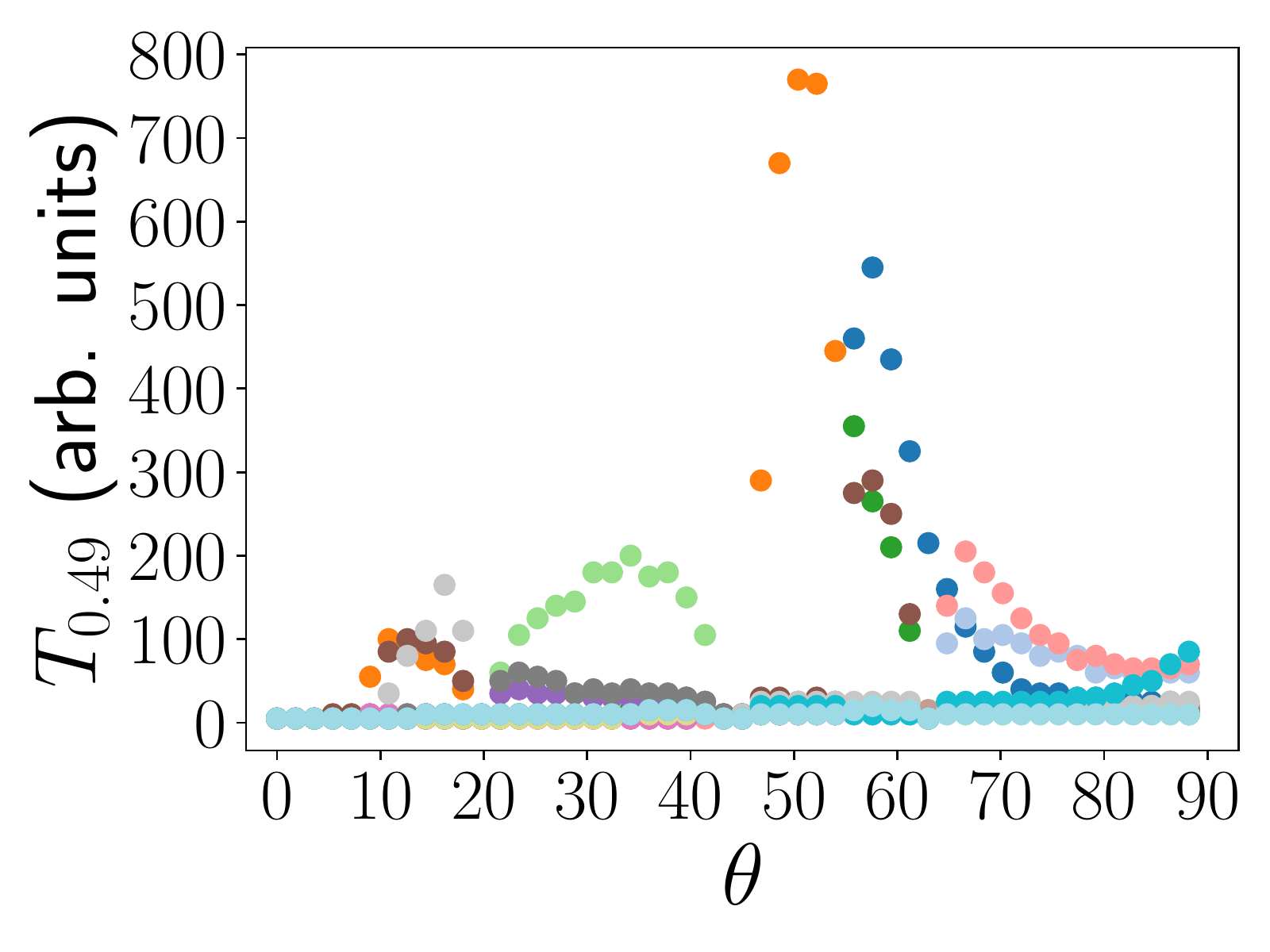}
\includegraphics[width=0.48\columnwidth]{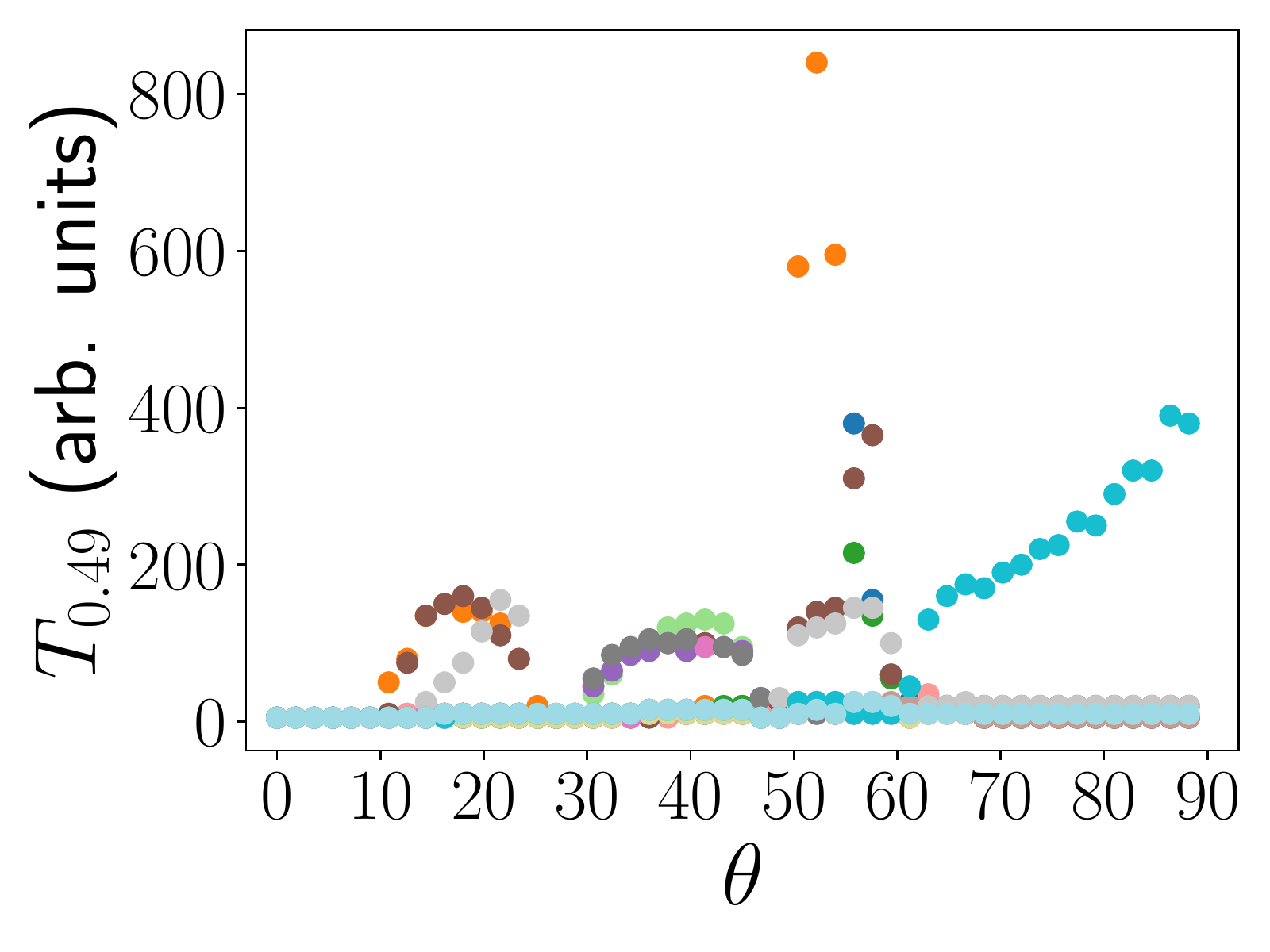}

\caption{Coherence times vs$.$ tilting angle for $J=-10$ K and (from top to bottom) $D=-10$ K, $D=-25$ K, $D=-50$ K, $D=-100$ K for 50 values of $\theta$ between $0^{\circ}$ and $88.2^{\circ}$ at $B_z=0.05$~T for a single bath with $n_{\text{bath}}=8$. Legend is displayed in \figref{legend}.} 
\label{angle_D}
\end{figure}

In this section, we aim to take a look at if and how the ideal tilting angle of $\theta = 50.4^{\circ}$ 
found for the chosen $J=-10$ K and $D=-50$ K changes with $J$ and $D$. 
Figures~\xref{angle_J} and \xref{angle_D} show the dependency in regards to $J$ and $D$, respectively. 

There are some competing trends in the data but for strongly anisotropic systems, 
the ideal angle is always at around 50$^{\circ}$ while for systems with weaker anisotropies 
$D$ (compared to the exchange interaction $J$), this angle may lie close to $90^{\circ}$.
It should be mentioned here that one would not speak of a toroidal system if the
anisotropy was not dominant.
These results show the exact opposite of what would be expected if the argument 
for superpositions with strong toroidal moments having long coherence times was right: 
The more anisotropic the system, the more stable the toroidal states should become. 
In our calculations, the configurations near $\theta=90^{\circ}$, 
which contain superpositions with high toroidal moments, are doing worse 
in terms of coherence times when the strength of the anisotropy is increased,
compare \figref{angle_D}.

\section{Gap sizes and coherence times}
\label{sec6}

\begin{figure}[ht!]
\centering
\includegraphics[width=0.48\columnwidth]{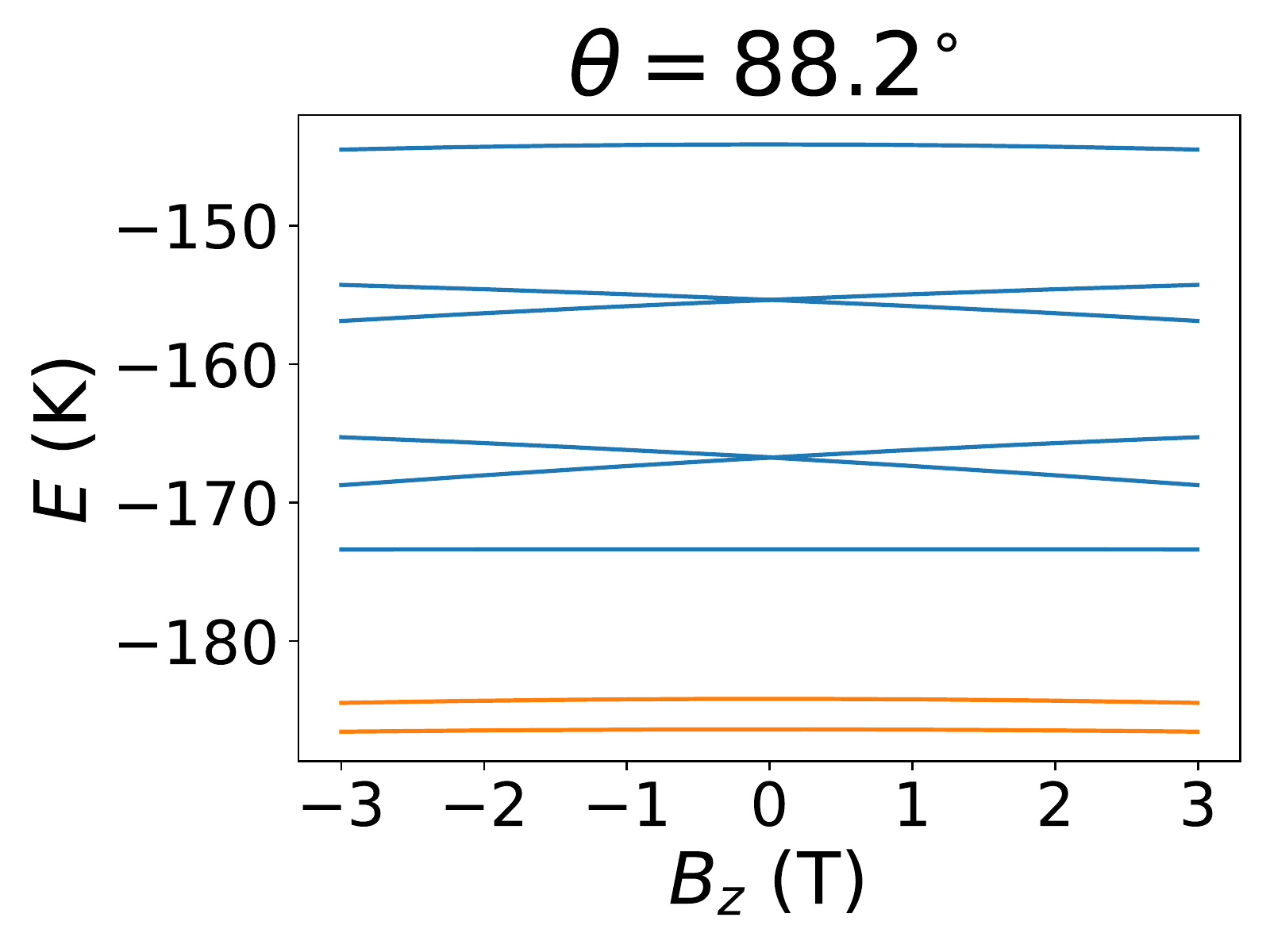}
\includegraphics[width=0.48\columnwidth]{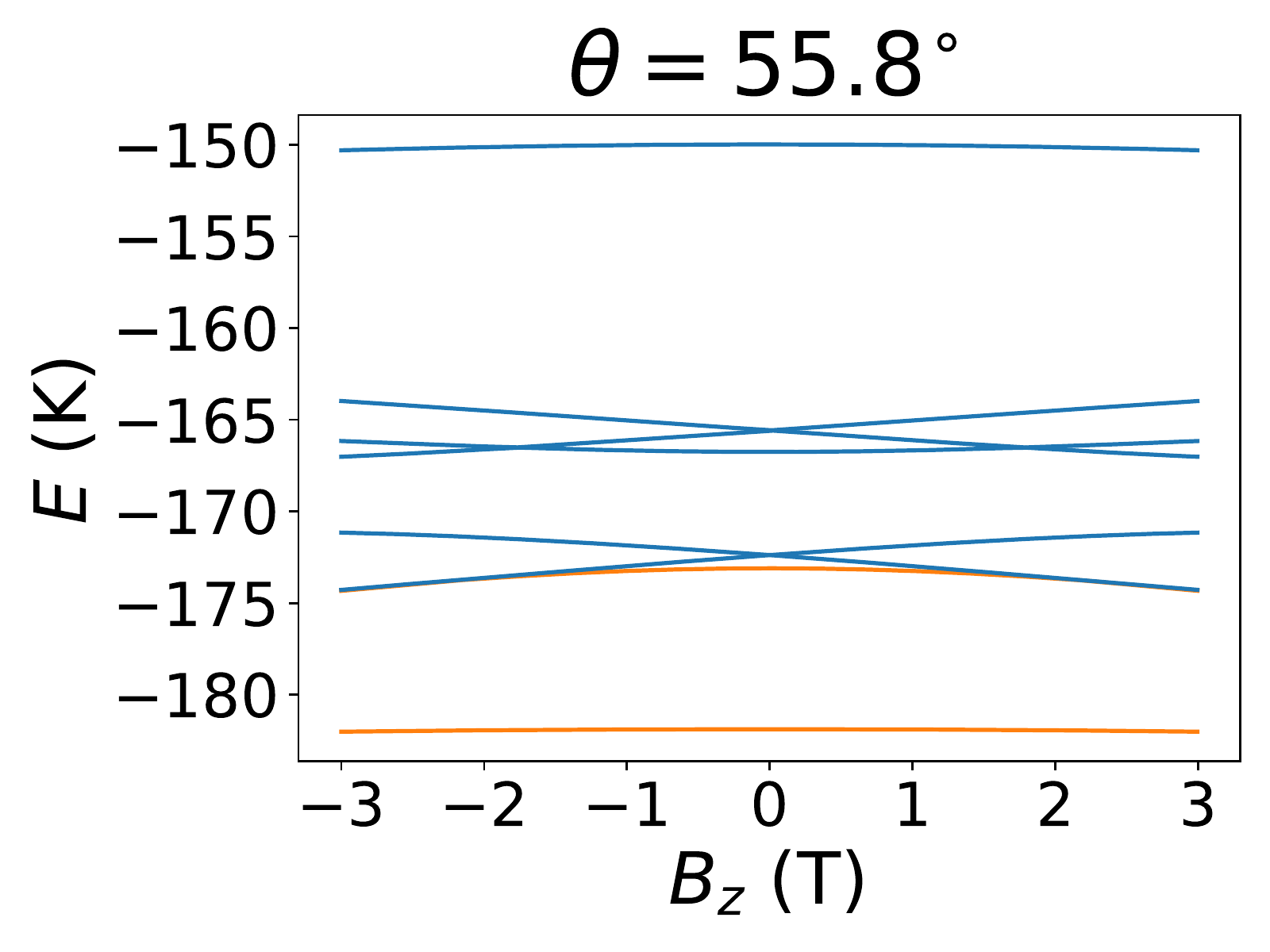}
\includegraphics[width=0.48\columnwidth]{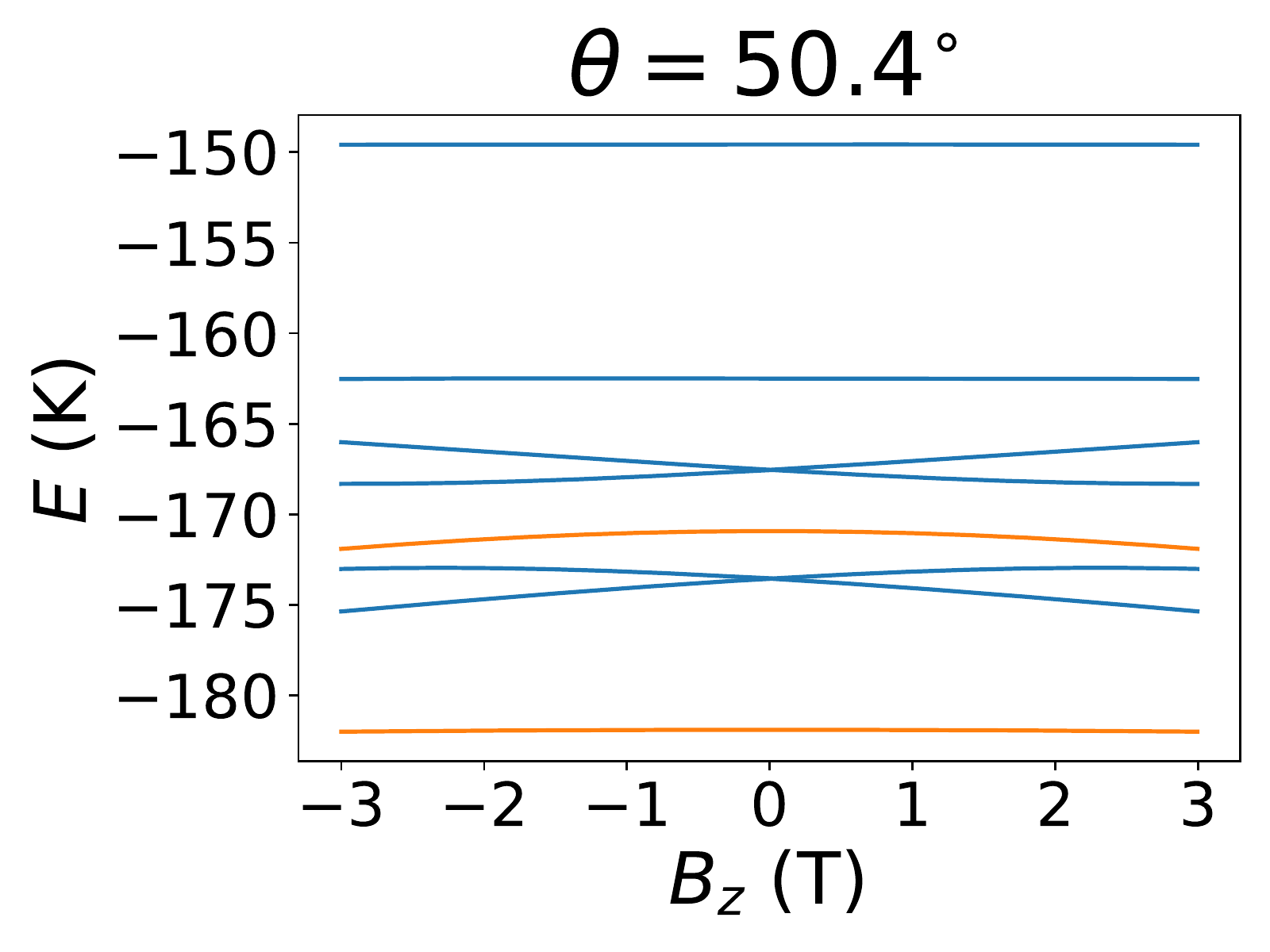}
\includegraphics[width=0.48\columnwidth]{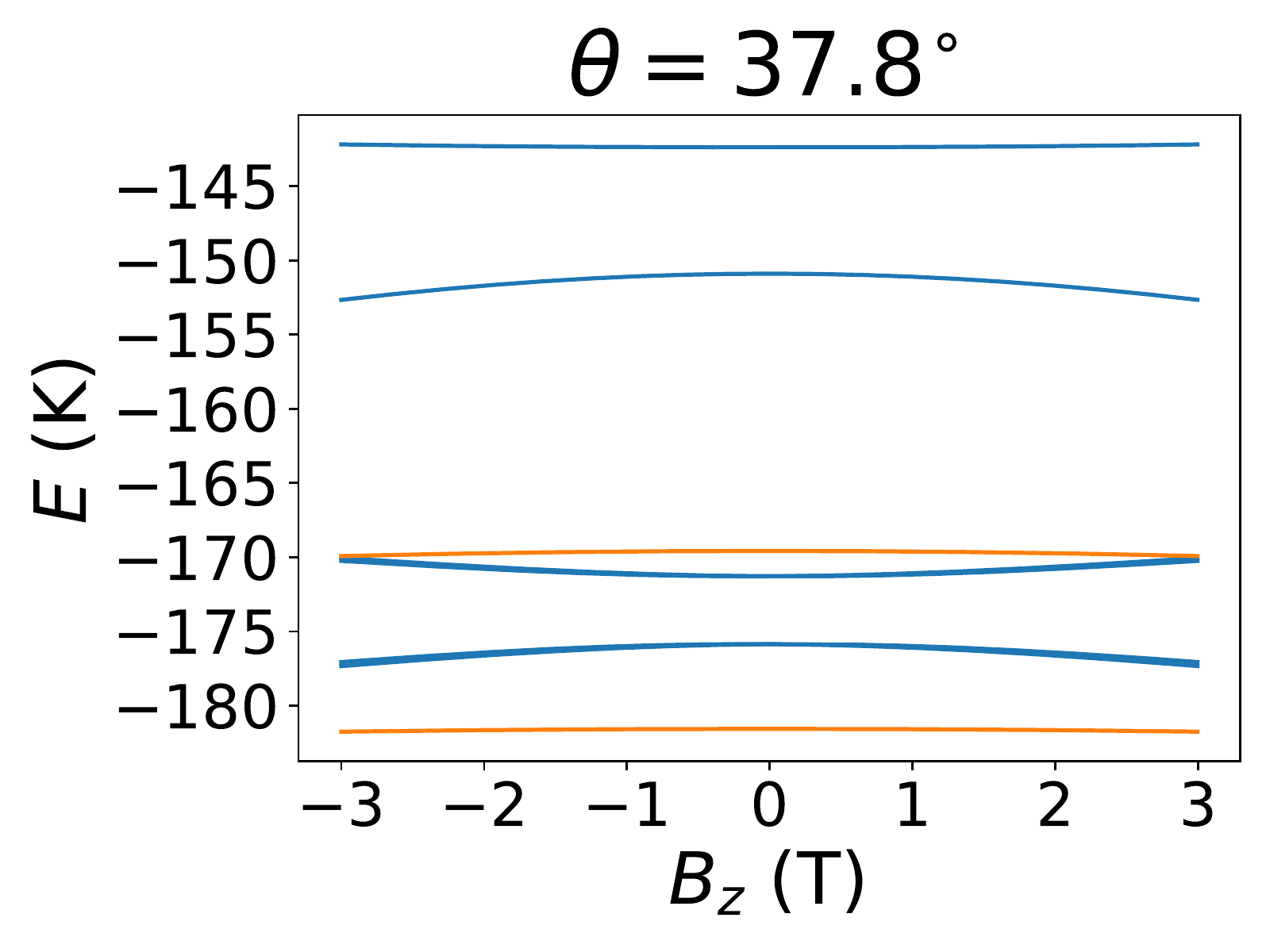}
\caption{\label{sup_gap}Zeeman diagrams of the eight lowest-lying states of the system 
without bath for some sample values of $\theta$. The states to be used for decoherence 
calculations are colored in orange. Notice the different scalings on the $y$-axes.}
\end{figure}

The size of the energy gap $\Delta E$ between the two superposed states is an 
easily measured characteristic of a clock transition. It is obvious that by scaling up 
$J$ and $D$, the spacing of the energy levels and therefore $\Delta E$ will increase. 
It is also obvious that this will lead to longer coherence times when we leave the 
coupling to the bath $A_1$ unchanged as this is akin to reducing $A_1$ while keeping 
$J$ and $D$ unchanged which is a situation which was already investigated to some extend in see Sec.~\xref{sec2}
and is also covered in the appendix.
Therefore one might arrive at the conclusion that large energy gaps $\Delta E$ always lead to longer coherence times. 
To test if this is true, we investigate how coherence times depend on $\Delta E$ by changing 
the tilting angle $\theta$ instead of $J$ and $D$. The clock transition of the two lowest-lying states 
in the almost ``toroidal" flat triangle ($\theta=88.2^{\circ}$) configuration is chosen as an example. 
As $\theta$ is decreased, the upper superposition state sometimes changes order with other states 
as can be inferred from \figref{sup_gap} while the lower state always stays the ground state. 

\begin{figure}[ht!]
\centering
\includegraphics[width=0.675\columnwidth]{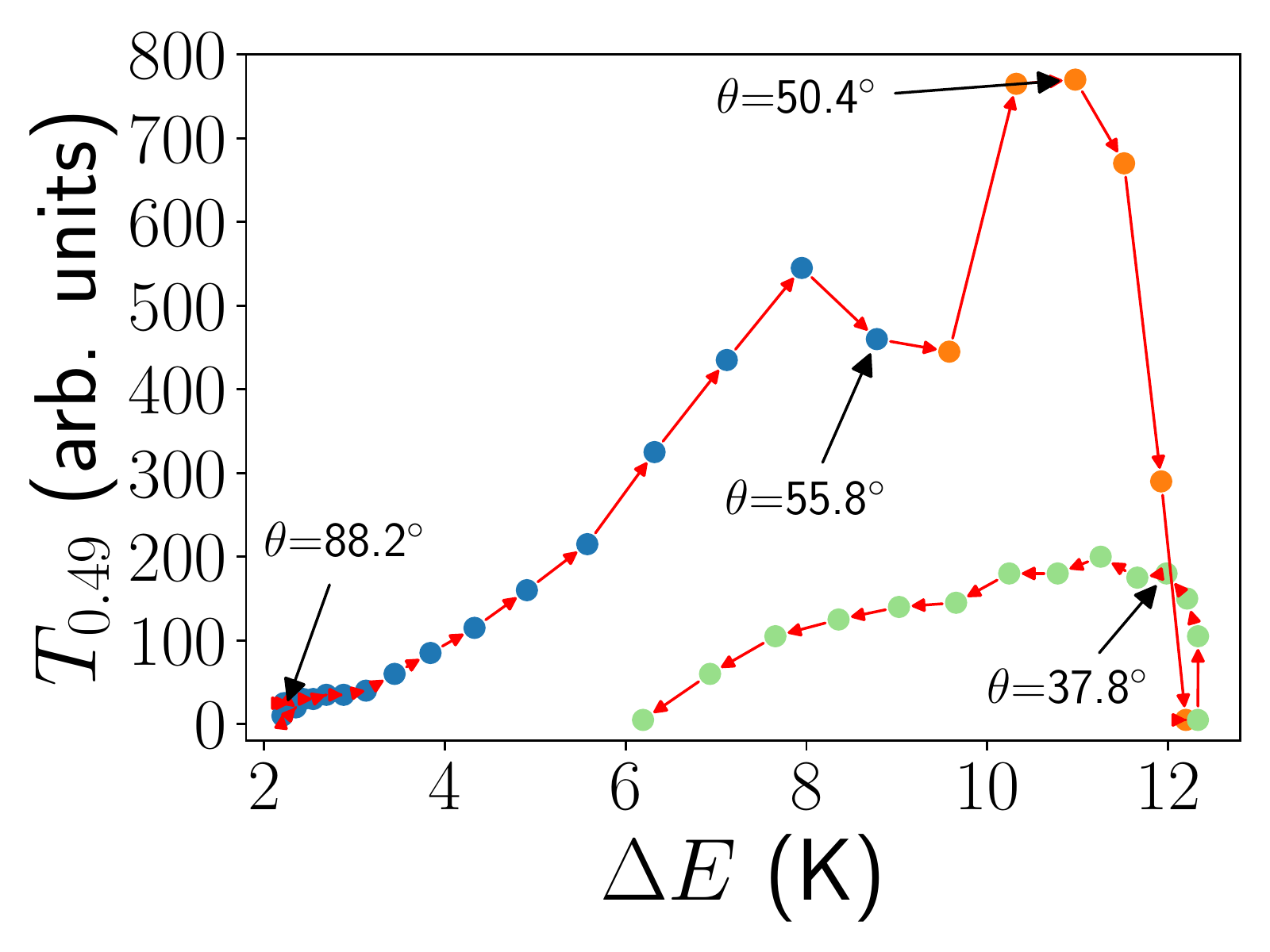}
\caption{\label{gap_break}Coherence times vs. gap size $\Delta E$ for the superposition indicated 
by the orange states in \figref{sup_gap} for different values of $\theta$ for a single bath with $n_{\text{bath}}=8$. 
The dots are colored according to the legend displayed in \figref{legend} and indicate the calculated coherence times. Red arrows connect consecutive values of $\theta$; 
they start in the almost ``toroidal" configuration at $\theta=88.2^{\circ}$ and always point 
to the next configuration with $\theta$ decreased by $1.8^{\circ}$ until $\theta=19.8^{\circ}$. 
Blue arrows indicate data points corresponding to the systems displayed in \figref{sup_gap}.}
\end{figure}

The results for the decoherence calculations of these systems are displayed in \figref{gap_break}. Here, the idea is to show decoherence time, gap size and tilting angle all in one figure without having to resort to 3D plots. The colored dots indicate decoherence time vs. gap size while, starting in the left-hand lower corner with $\theta=88.2^{\circ}$ (almost toroidal configuration), 
the red arrows always point to the next configuration with the tilting angle decreased by $1.8^{\circ}$.  

There seems to be a strong positive relationship between gap size and coherence time between $\theta=88.2^{\circ}$ and $\theta=50.4^{\circ}$ interrupted only by a brief decrease in coherence times between $\theta=57.6^{\circ}$ and $\theta=54^{\circ}$ as one of the superposition states gets energetically close to other eigenstates with which it can now be connected via interactions with the spin bath and therefore decoheres. In this range of tilting angles, the upper superposition state is either the energetically second or fourth lowest state. When decreasing the tilting angle further, it again gets energetically close to other states which causes the coherence time to collapse almost to zero. From approximately $\theta=45^{\circ}$ on, the upper superposition state is now the sixth lowest eigenstate and there is no clear correlation between gap size and coherence times. The reason for this is not well understood and warrants further investigation.

\section{Impact of a symmetry-breaking dipolar interaction}
\label{sec7}

\begin{figure}[H]
\centering
% \includegraphics*[clip,width=0.45\columnwidth]{blender_environment.png}
% \qquad
% \includegraphics*[clip,width=0.45\columnwidth]{blender90.png}
\includegraphics[width=0.48\columnwidth]{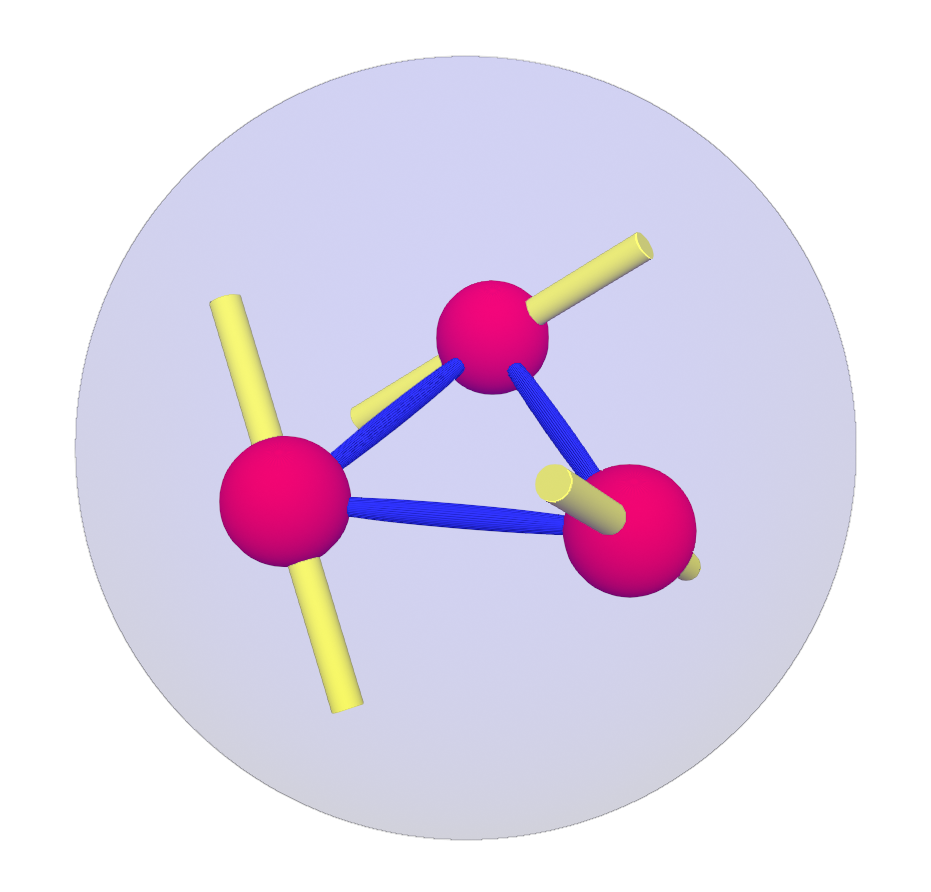}
\includegraphics[width=0.48\columnwidth]{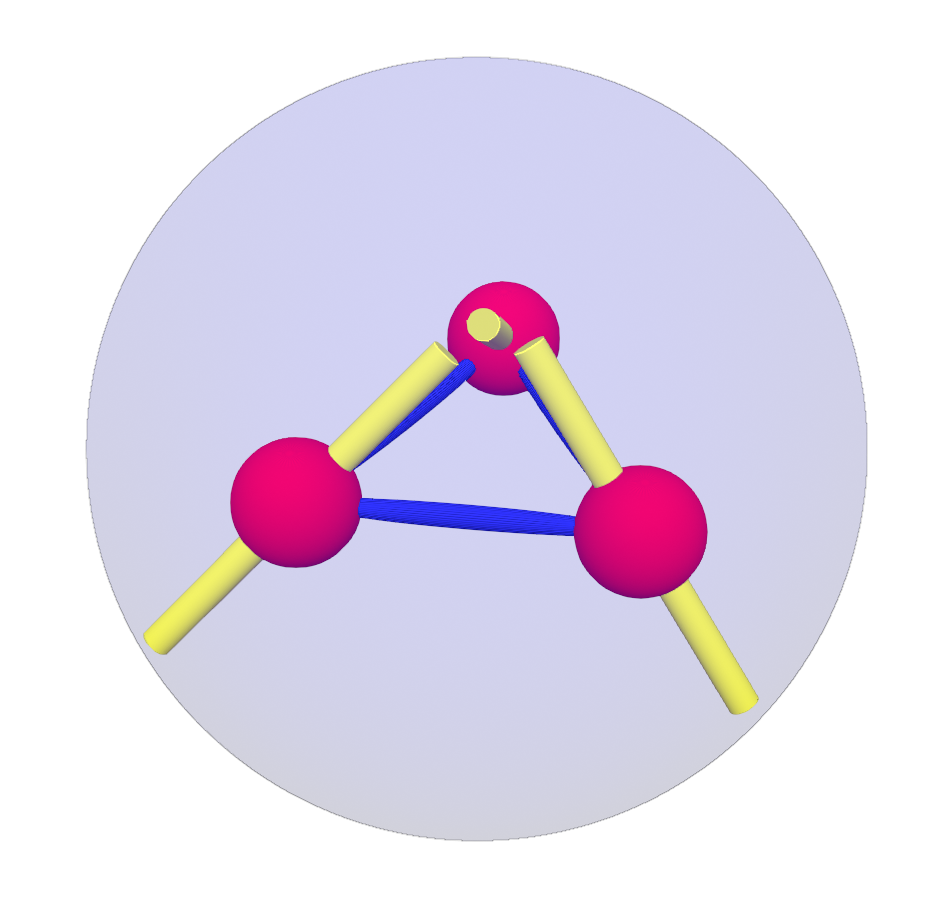}
\caption{3D representation of the spin system from \figref{system} (left) and with anisotropy axes rotated by $\varphi=-\pi/2$ (right).} 
\label{system_90}
\end{figure}

As discussed in Ref.~\cite{PIW:PRR22}, a Hamiltonian consisting of a Heisenberg term
and single-ion easy anisotropy axes is invariant under a collective rotation 
of all anisotropy axes (and the field direction) by a common angle.
Therefore, a rotation by e.g. $\varphi=-\pi/2$, which rotates all anisotropy axes 
to point inwards, see \figref{system_90}, yields the same energy spectrum and leaves 
(many) observables unchanged. The superpositions, however, have zero toroidal moment now.

Out of curiosity, we introduced a symmetry-breaking dipole-dipole interaction between the system spins 
(we could have also chosen a Dzyaloshinskii-Moriya interaction) to see if there would be any effect 
on the coherence times if the above mentioned symmetry is lost and the toroidal character of the system 
even stabilized \cite{PIW:PRR22}. The new system Hamiltonian then reads 
\begin{align}
\op{H}_{\text{S}}^{\text{N}}=\  &\op{H}_{\text{Heisenberg}}+\op{H}_{\text{anisotropy}}+\op{H}_{\text{Zeeman}}\\ \notag
&+\sum_{\substack{i=0,\\ j=i+1}}^{2}\frac{A_3}{r_{ij}^3}\left(\vecops{i} \cdot\vecops{j}-\frac{3\left(\vecops{i}\cdot\vec{r}_{ij}\right)\left(\vecops{j}\cdot\vec{r}_{ij}\right)}{r_{ij}^2}\right) 
\
\end{align}
with the third spin being again the zeroth spin so that all system spins are coupled to each other.

When choosing $A_3$ to be very small, the symmetry is broken but numerically, 
the difference in coherence times is also very small; 
the superpositions with large toroidal moments do not perform significantly better than before.

 \begin{figure}[ht!]
 \centering
\includegraphics[width=0.48\columnwidth]{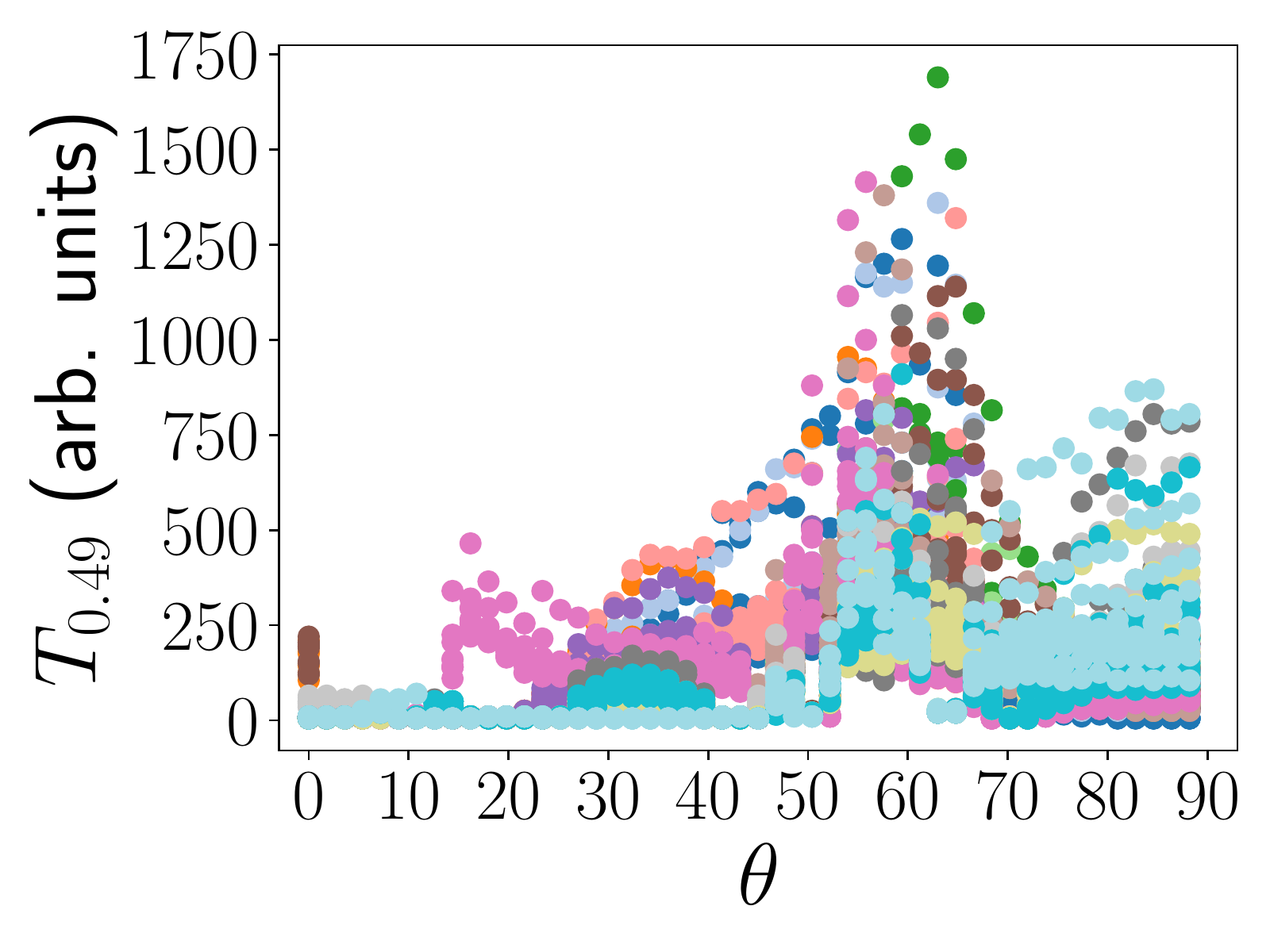}
\includegraphics[width=0.48\columnwidth]{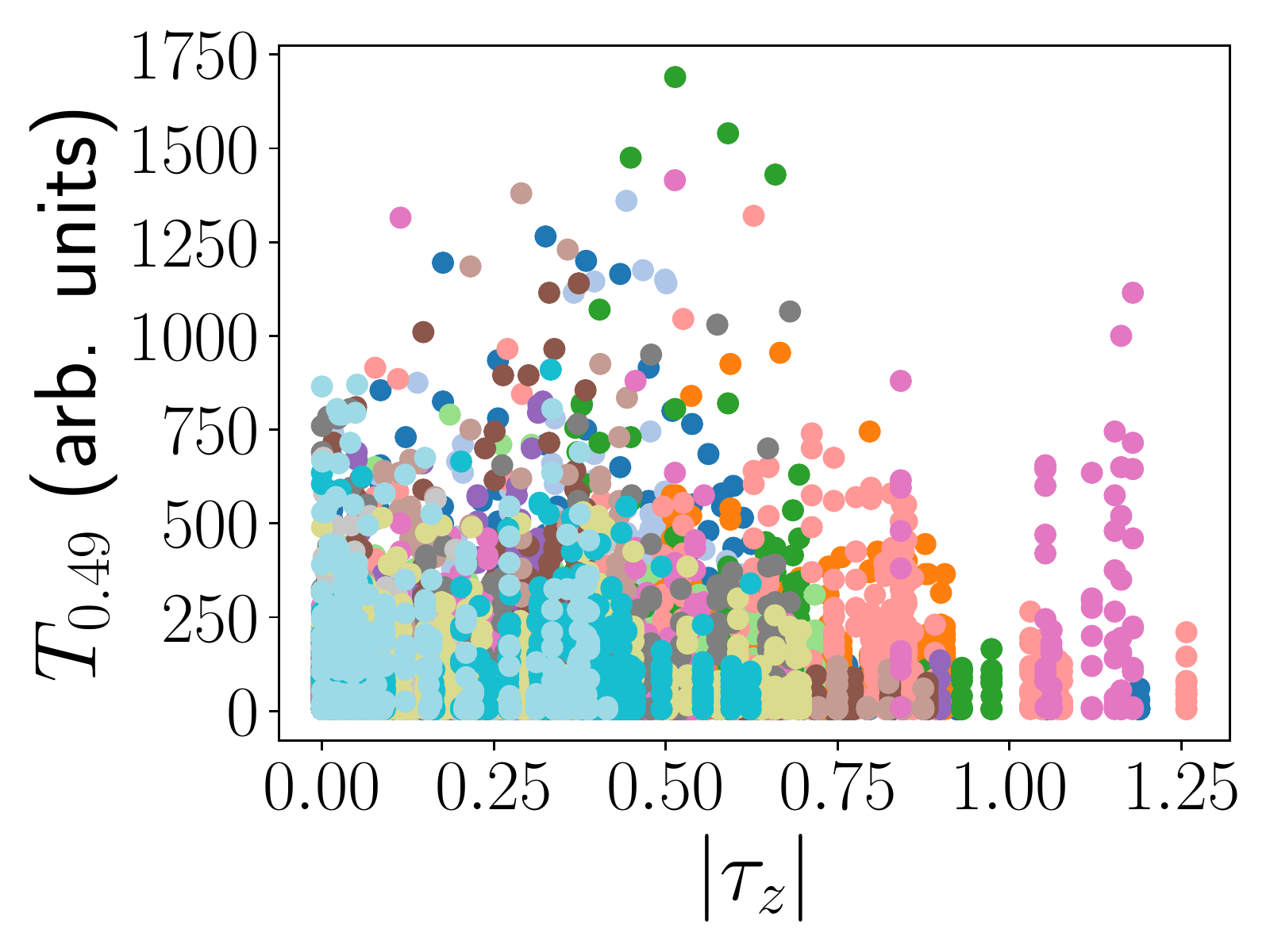}
  \caption{Coherence times vs. tilting angle (left) and absolute value of the $z$-component of the toroidal moment of superpositions (right, see \eqref{tau}) with additional dipole interactions between central spins with $A_3=|J|=10$ K for 50 values of $\theta$ between $0^{\circ}$ and $88.2^{\circ}$ at $B_z=0.05$~T for ten different random baths with $n_{\text{bath}}=8$. Again, best coherence times are observed for mid-sized angles and toroidal moments although the latter now have a smaller range compared to the original system, see \figref{anglevsbreak}. Legend is displayed in \figref{legend}.} 
 \label{hdtauz}
 \end{figure}

Choosing an $A_3$ with the same magnitude as $J$, \figref{hdtauz} shows there are even 
fewer states with large toroidal moments displaying long coherence times. 
We repeated the calculation for ten different random baths to again eliminate 
the possibility of choosing a non-typical bath. Otherwise, there are no significant new findings; 
the system behaves very similarly to the one without dipolar interactions 
between the system spins, see \figref{anglevsbreak}.

%\vspace*{5mm}
%%%%%%%%%%%%%%%%%%%%%%%%%%%%%%%%%%%%%%%%%%%%%%%%%%%%%%%%%%%%%%%%%%%%%%%%
\section{Discussion and conclusions}
\label{sec8}

On the basis of our calculations, we can state that systems with maximally ``toroidal"-oriented anisotropy 
axes do not necessarily exhibit long coherence times. There are a number of factors at play, 
mainly the need for $|M_p-M_n|$ to be small, 
while other factors such as the energy gap and perhaps the toroidal moment enter in a complex way 
and cannot be considered independently of each other. 
It was shown that, for the example of a 
spin-1 triangle, superpositions with large toroidal moments 
can exhibit even very short coherence times. 
All in all, we found no evidence that toroidicity should be a desirable characteristic 
when designing, e.g., a qubit with long coherence times. 
To our surprise, rather, non-collinear tilted anisotropy axes 
that are almost mutually orthogonal 
seem to be most promising in many cases.
The fundamental advantage of these systems is given by 
the presence of clock transitions amongst the energetically 
low-lying states which is a feature not easily inferred from the 
geometry or other characteristics of the system.

The present paper mainly states numerical findings about decoherence
properties of a triangular arrangement of spins with $C_3$-symmetric anisotropy axes.
The very interesting question \emph{why} certain superpositions decohere more
slowly and how these effects are influenced by e.g.\ the size of energy gaps of the 
system compared to the spectral width of the bath remains open and thus subject
to further studies. From previous studies we know that decoherence is intimately 
related to entanglement between system and bath \cite{VoS:PRB20} which suggests 
that some initial states entangle more easily and quickly than others for a
given Hamiltonian of the total system.

%\vspace*{5mm}
%%%%%%%%%%%%%%%%%%%%%%%%%%%%%%%%%%%%%%%%%%%%%%%%%%%%%%%%%%%%%%%%%%%%%%%%
\section*{Acknowledgment}

This work was supported by the Deutsche Forschungsgemeinschaft DFG
(355031190 (FOR~2692); 397300368 (SCHN~615/25-2)). 
We acknowledge support for the publication costs by the Open Access Publication 
Fund of Bielefeld University and the Deutsche Forschungsgemeinschaft (DFG).

%%%%%%%%%%%%%%%%%%%%%%%%%%%%%%%%%%%%%%%%%%%%%%%%%%%%%%%%%%%%%%%%%%%%%%%%
\appendix

\begin{figure}[ht!]
\centering
\includegraphics[width=0.675\columnwidth]{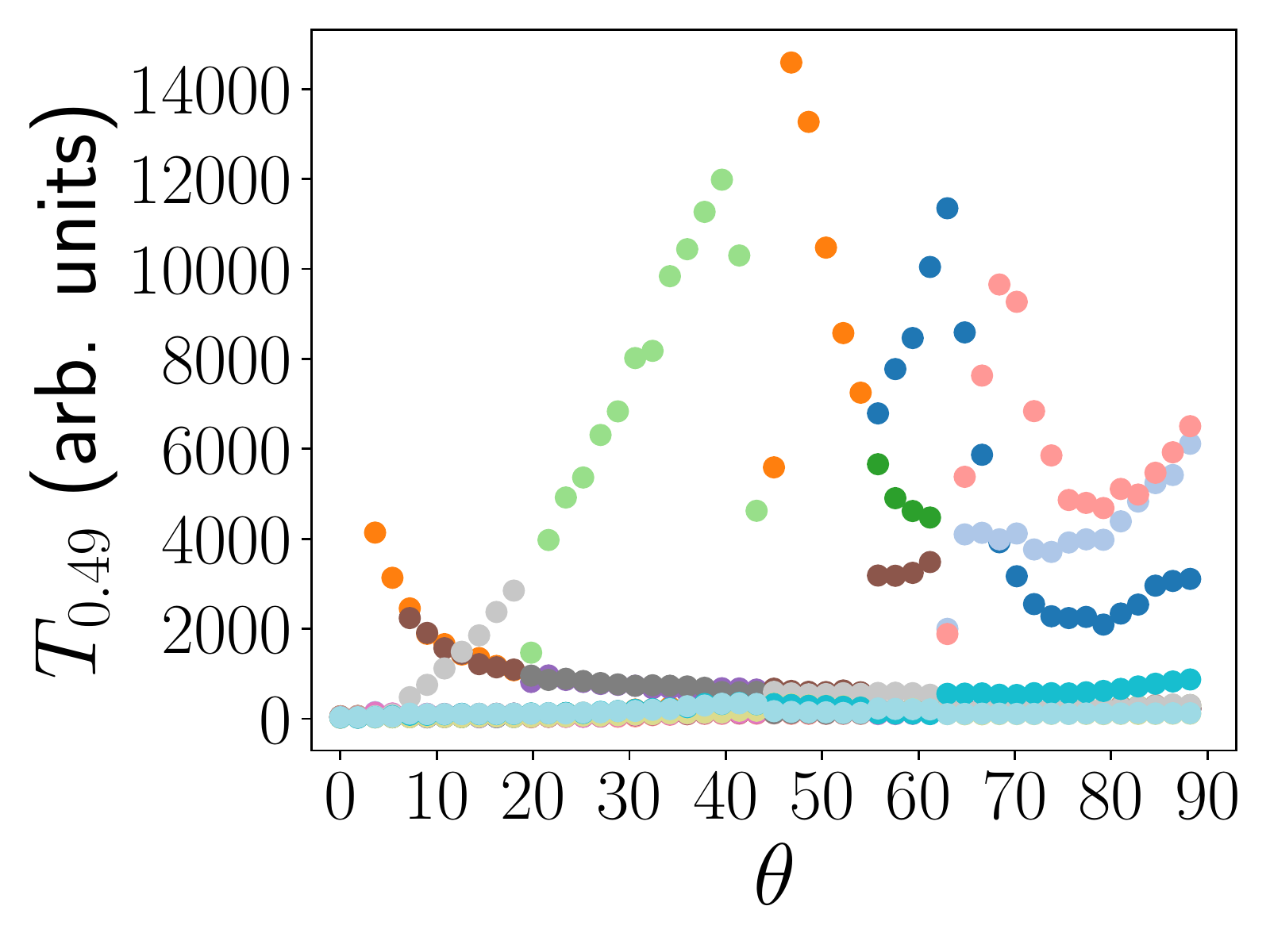}
\caption{Coherence times vs$.$ tilting angle for 50 values of $\theta$ between $0^{\circ}$ and $88.2^{\circ}$ at $B_z=0.05$~T for a random bath
with $n_{\text{bath}}=8$, $A_1=0.01$ K, and $A_2=0.001$ K. Observations align with those for the system with $A_1=A_2=0.1$ K shown in \figref{anglevsbreak}. Legend is displayed in \figref{legend}.} 
\label{A1001A20001}
\end{figure}

\begin{figure}[ht!]
\centering
\includegraphics[width=0.675\columnwidth]{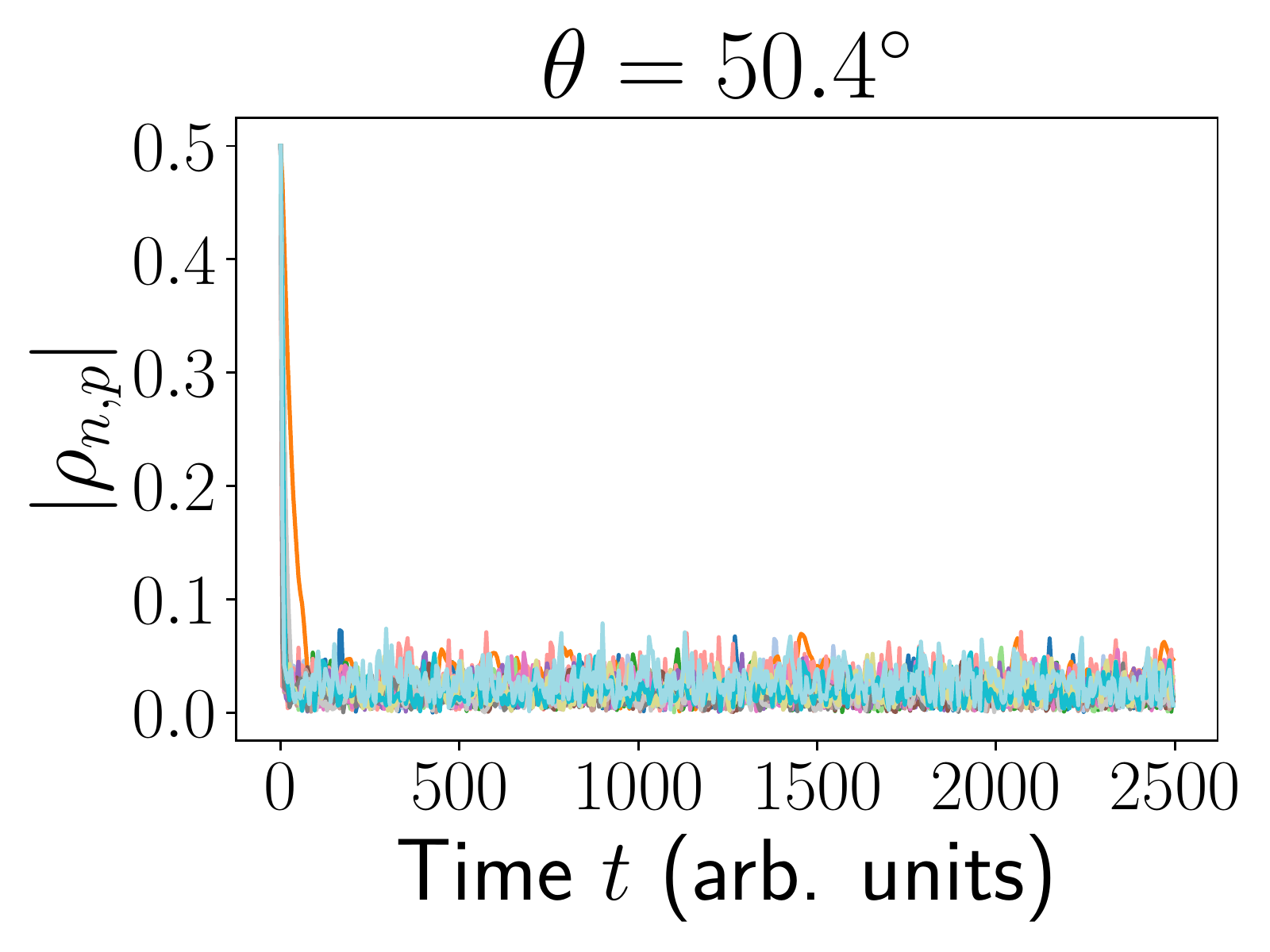}
\includegraphics[width=0.675\columnwidth]{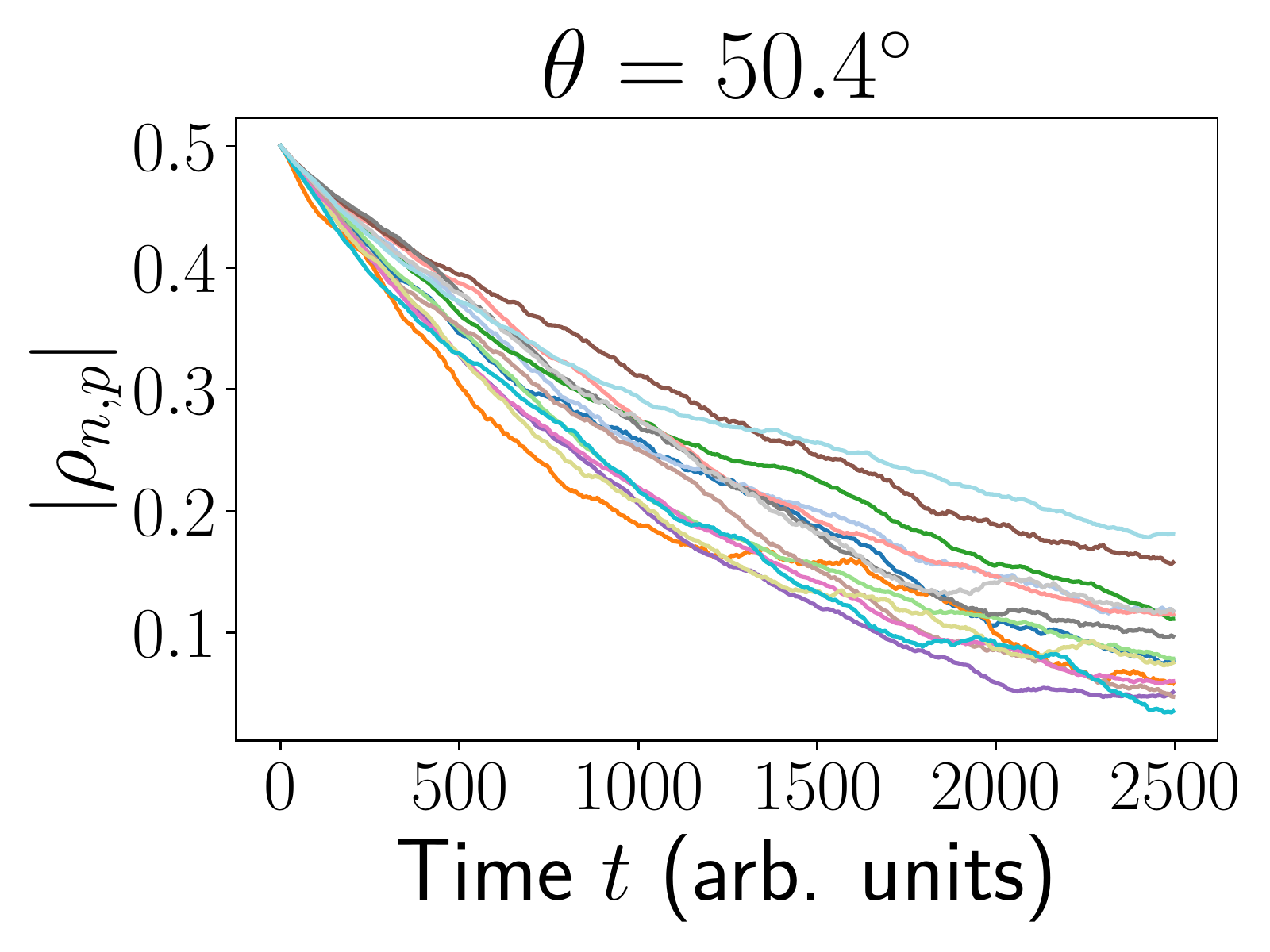}
\caption{Decoherence over time of all two-state superpositions of the six lowest-lying energy eigenstates for a random bath with $n_{\text{bath}}=8$ at $B_z=0.05$ and $\theta=50.4^{\circ}$ with $A_1=1.0$ K and $A_2=0.1$ K (top) or $A_1=0.1$ K and $A_2=100.0$ K (bottom). At these extremes, all superpositions perform equally well and there are no advantages in coherence times for clock transitions. Large $A_1$ lead to virtually instantaneous decoherence, while large $A_2$ lead to neither very quick nor very slow decoherence. Legend is displayed in \figref{legend}.} 
\label{extreme}
\end{figure}

\section{Investigations for other parameters of the system}
Following a suggestion by a referee, we considered different values for the system-bath coupling strength $A_1$ and the bath-bath coupling strength $A_2$ as choosing them to be the same indicates an electronic spin bath which is maybe too special. Furthermore, we also looked at a system with $J=-1$ K and $D=-10$ K to work with numbers more typical for dysprosium
triangles. Our aim here is to show that our main qualitative findings hold for a wide range of parameters.

\subsection{Variation of $A_1$ and $A_2$ in the standard system described in the paper}

Figure \xref{A1001A20001} shows the coherence times of all superpositions considered vs.\
the tilting angle $\theta$ for the standard system with $J=-10$ K and $D=-50$ K 
but with $A_1=0.01$ K and $A_2=0.001$ K instead of $A_1=A_2=0.1$ K. 
This could e.g.\ represent a bath of protons. As expected, the coherence times 
increase substantially, as the coupling between system and bath is now an order of magnitude weaker. 
Other than that, however, the findings align with the main qualitative observations 
of the standard configuration shown in \figref{anglevsbreak}.

This behaviour only breaks down when considering very strong $A_1$ and/or $A_2$, see \figref{extreme} for an example. If $A_1$ is chosen too large, no superposition has a significant coherence time and all decohere apparently instantly. We believe that the reason for this behaviour is that the Zeeman levels of the original system entangle strongly with the bath and are thus deformed so extremely that it does not matter if they formed a clock transition in the original system.

If $A_2$ is chosen very large, all superpositions show similar, mid-sized coherence times. We believe this to be caused by the fact that the states of each superposition can, when combined with the energetically now very broad spectrum of the environment, be energetically connected to a multitude of other states of the original system. On the other hand, the density of bath states is significantly reduced, so that the process of decoherence may be hindered to some extend as suggested in \cite{VoS:PRB20}.

\subsection{More realistic values of $J$ and $D$ to approximately represent dysprosium triangles}

\begin{figure}[H]
\centering
\vspace{1cm}
\includegraphics[width=0.675\columnwidth]{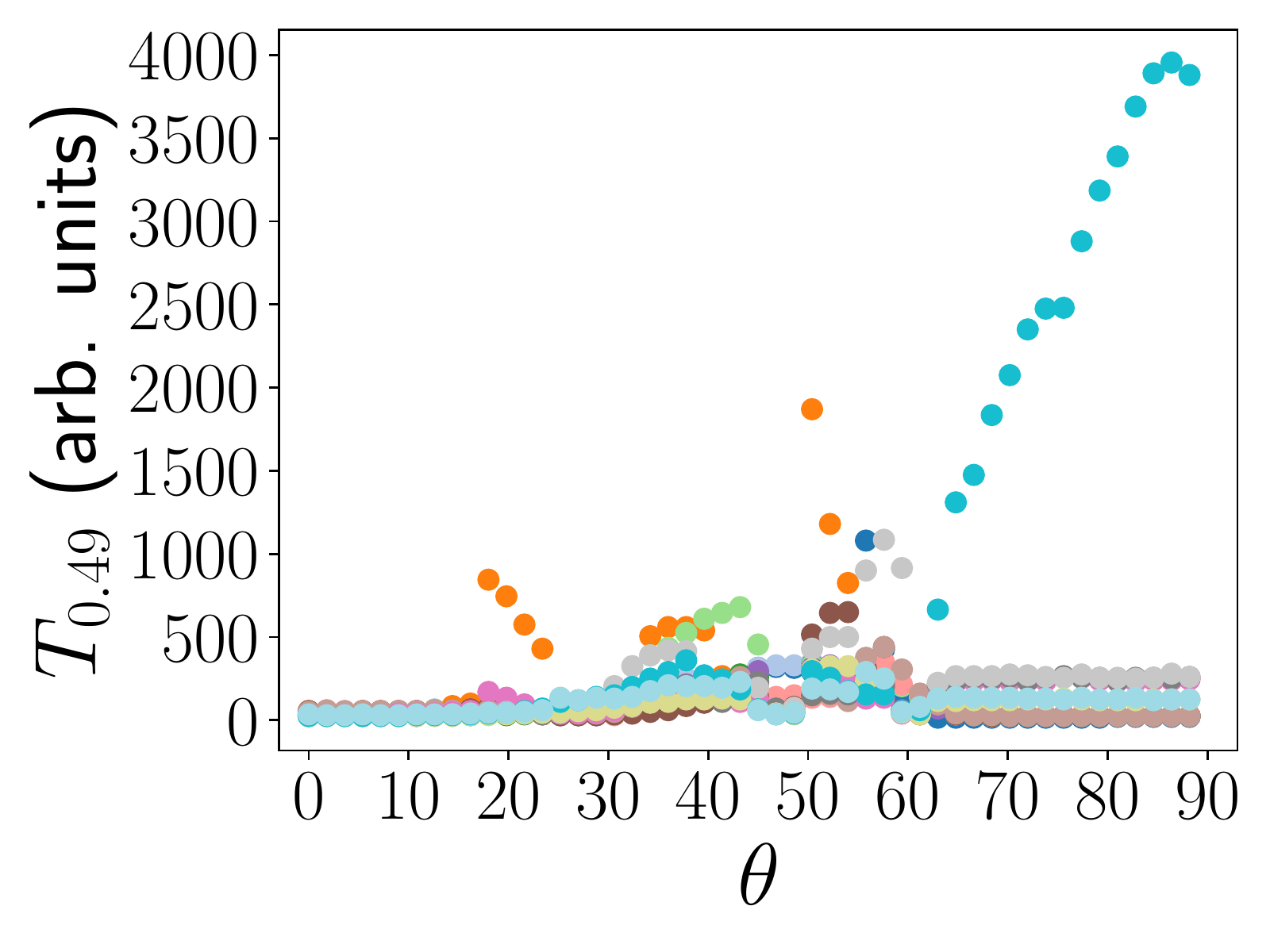}
\includegraphics[width=0.675\columnwidth]{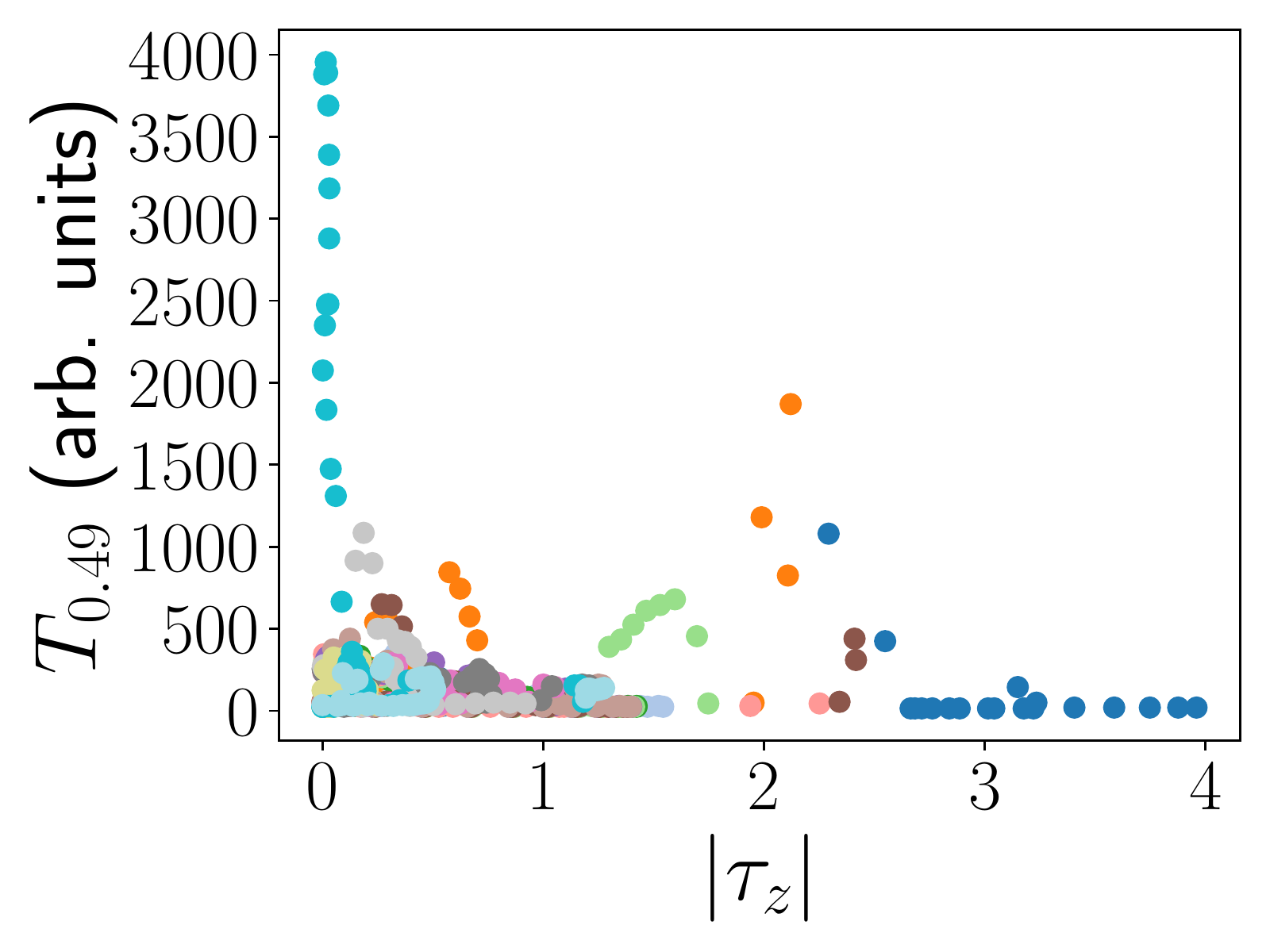}
\caption{Coherence times vs$.$ tilting angle (top) and absolute value of the $z$-component 
of the toroidal moment of superpositions (bottom, see \eqref{tau}) for 50 values of $\theta$ between $0^{\circ}$ and $88.2^{\circ}$ at $B_z=0.05$~T for a random bath
with $n_{\text{bath}}=8$, $J=-1$ K, and $D=-10$ K, $A_1=0.01$ K, $A_1=0.001$ K. The coherence times are again very similar to those shown for the system with $J=-10$ K, $D=-50$ K, see \figref{anglevsbreak} with the exception of the superposition of the third and fifth excited states at angles $\theta \gtrapprox 60^{\circ}$. However, these superpositions have a toroidal moment close to zero so that they are of no significance to the main findings of this paper. Legend is displayed in \figref{legend}.} 
\label{dysprosium}
\end{figure}

Papers claiming toroidal moments to be promising candidates for quantum technologies often consider triangles of dysprosium
\cite{THM:ACIE06,CUS:ACIE08}. Therefore, we choose $J=-1$ K and $D=-10$ K in order 
to work with more realistic values for the Heisenberg interaction and strength of the anisotropy. 
We find, as shown in \figref{dysprosium}, that there are again some superpositions with 
long coherence times around $\theta=50^{\circ}$. However, the best-performing superpositions are made up of the 
third and fifth excited states similar to those in \figref{angle_J} (top) and \figref{angle_D} (bottom).
These have almost zero toroidal moment and therefore do not disprove the main statement of this paper.

%%%%%%%%%%%%%%%%%%%%%%%%%%%%%%%%%%%%%%%%%%%%%%%%%%%%%%%%%%%%%%%%%%%%%%%%
%\bibliographystyle{/home/schnack/tex/sty/revtex4-1/revtex4-1/bibtex/bst/revtex/apsrev4-2}
%\bibliography{/home/schnack/tex/bibtex/js-own.bib,/home/schnack/tex/bibtex/js-other.bib}
%\bibliography{js-own.bib,js-other.bib}


\end{document}